\newif\ifconfver
\confverfalse      

\newif\ifcutshort      
\cutshorttrue

\newif\ifcutshortlvltwo  
\cutshortlvltwofalse

\ifconfver
    \documentclass[10pt,twocolumn,twoside]{IEEEtran}
\else
    \documentclass[11pt,draftcls,onecolumn]{IEEEtran}
\fi

\usepackage{array}
\usepackage{cite}
\usepackage{calc}
\usepackage{caption}
\usepackage{graphicx}
\usepackage{psfrag}
\usepackage[tight,footnotesize]{subfigure}
\usepackage{stfloats}
\usepackage{url}
\usepackage{subfig}
\usepackage{longtable}
\usepackage{amsmath,amsfonts,amssymb,amsbsy,bm,paralist,theorem,ifthen,color}
\usepackage{algorithmic,algorithm}
\usepackage[dvips]{epsfig}
\usepackage[dvips]{graphicx}

\newcommand\Gc{\ensuremath{\mathcal{G}}}

\newcommand\Nc{\ensuremath{{\mathcal{N}}}}
\newcommand\xb{\ensuremath{{\bm x}}}

\newcommand\yb{\ensuremath{{\bm y}}}

\newcommand\ssb{\ensuremath{{\bm s}}}
\newcommand\ub{\ensuremath{{\bm u}}}

\newcommand\Ab{\ensuremath{{\bm A}}}
\newcommand\ab{\ensuremath{{\bm a}}}

\newcommand\bb{\ensuremath{{\bm b}}}

\newcommand\Db{\ensuremath{{\bm D}}}
\newcommand\eb{\ensuremath{{\bm e}}}
\newcommand\Eb{\ensuremath{{\bf E}}}
\newcommand\fb{\ensuremath{{\bm f}}}
\newcommand\Gb{\ensuremath{{\bm G}}}

\newcommand\Ib{\ensuremath{{\bm I}}}
\newcommand\Lb{\ensuremath{{\bm L}}}
\newcommand\Pb{\ensuremath{{\bm P}}}
\newcommand\pb{\ensuremath{{\bm p}}}

\newcommand\Mb{\ensuremath{{\bf M}}}

\newcommand\rb{\ensuremath{{\bm r}}}
\newcommand\tb{\ensuremath{{\bm t}}}
\newcommand\Qb{\ensuremath{{\bm Q}}}
\newcommand\qb{\ensuremath{{\bm q}}}

\newcommand\vb{\ensuremath{{\bm v}}}
\newcommand\Wb{\ensuremath{{\bm W}}}

\newcommand\zb{\ensuremath{{\bm z}}}

\newcommand\nub{\ensuremath{{\bm \nu}}}

\newcommand\Upsilonb{\ensuremath{{\bm \Upsilon}}}
\newcommand\zerob{\ensuremath{{\bm 0}}}

\newcommand\blkdiag{\ensuremath{{\rm blkdiag}}}
\newcommand\diag{\ensuremath{{\rm diag}}}

\newcommand\oneb{\ensuremath{{\bf 1}}}

\newcommand\Xc{\ensuremath{{\mathcal{X}}}}

\newtheorem{Theorem}{Theorem}

\newtheorem{Remark}{Remark}
\newtheorem{Assumption}{Assumption}

\ifconfver
\def\blue{\color{black}}
\else

\def\blue{\color{black}}
\fi

\begin{document}

\bibliographystyle{IEEEtran}

\title{Multi-Agent Distributed Optimization via Inexact Consensus ADMM}

\ifconfver \else {\linespread{1.1} \rm \fi

\author{\vspace{0.8cm}Tsung-Hui Chang$^\star$, Mingyi Hong$^\dag$ and Xiangfeng Wang$^\ddag$\\
\thanks{
The work of Tsung-Hui Chang is supported by National Science Council, Taiwan (R.O.C.), under Grant
NSC 102-2221-E-011-005-MY3. Part of this work is presented in IEEE ICASSP 2014.}
\thanks{$^\star$Tsung-Hui Chang is the corresponding author. Address:
Department of Electronic and Computer Engineering, National Taiwan University of Science and Technology, Taipei 10607, Taiwan, (R.O.C.). E-mail:
tsunghui.chang@ieee.org. }
\thanks{$^\dag$Mingyi Hong is with Department of Electrical and Computer Engineering, University of Minnesota, Minneapolis, MN 55455, USA, E-mail: mhong@umn.edu}
\thanks{$^\ddag$Xiangfeng Wang is with Shanghai Key Lab for Trustworthy Computing, Software Engineering Institute, East China Normal University, Shanghai, 200062, China, E-mail: xfwang@sei.ecnu.edu.cn}}

 \maketitle

\begin{abstract}
Multi-agent distributed consensus optimization problems arise in many signal processing
applications. Recently, the alternating direction method of multipliers (ADMM) has been used for solving this family of problems. ADMM based distributed optimization method is shown to have faster convergence rate compared with classic methods based on consensus subgradient, but can be computationally expensive, especially for problems with complicated structures or large dimensions. In this paper, we propose low-complexity algorithms that can reduce the overall computational cost of consensus ADMM by an order of magnitude for certain large-scale problems. Central to the proposed algorithms is the use of an inexact step for each ADMM update, which enables the agents to perform cheap computation at each iteration. Our convergence analyses show that the proposed methods
converge well {\blue under some convexity assumptions.} {\blue Numerical results show that the proposed algorithms offer considerably lower computational complexity than the standard ADMM based distributed optimization methods.} 
\\\\
\noindent {\bfseries EDICS}:  OPT-DOPT, MLR-DIST, NET-DISP, SPC-APPL.
\end{abstract}

\ifconfver \else
\newpage
\fi

\ifconfver \else \IEEEpeerreviewmaketitle} \fi

\vspace{-0.3cm}
\section{Introduction}\label{sec: intro}
{We consider a network with multiple agents, for example a sensor network,
a data cloud network or a communication network.
The agents seek to collaborate to accomplish certain task.}
For example, distributed database servers may cooperate for data mining or for parameter learning
in order to fully exploit the data collected from individual servers \cite{Foster2008}.
Another example arises from {large-scale machine learning} applications \cite{BK:Bekkerman12}, where a computation task may be executed by collaborative microprocessors with individual memories and storage spaces \cite{BK:Bekkerman12,BK:Andrews99,BK:Ghosh07}.
{Distributed optimization becomes favorable as it is not always efficient to pool all the local information for centralized computation, due to large size of problem dimension, a large amount of local data, energy constraints and/or privacy issues \cite{Angelia2010,Johansson08,MZhu2012,Chen_Sayed2012}.}
Many of the distributed optimization tasks, such as those described above, can be cast as an optimization problem of the following form
\begin{align}\label{consensus problem}
  {\sf (P1)}~~\min_{\yb \in \mathbb{R}^K}~ &\sum_{i=1}^N \phi_i(\yb)
\end{align}
{where $\yb \in \mathbb{R}^K$ is the decision variable} and {\blue $\phi_i:\mathbb{R}^{K} \rightarrow \mathbb{R}\cup\{\infty\}$} is the cost function associated with agent $i$. Here the function $\phi_i$ is composed of a smooth component {\blue $f_i:\mathbb{R}^{M} \rightarrow \mathbb{R}\cup\{\infty\}$ (possibly with extended values)} and a non-smooth component {\blue $g_i:\mathbb{R}^{K} \rightarrow  \mathbb{R}\cup\{\infty\}$}, i.e.,
\begin{align}\label{eqn: phi}
  \phi_i(\yb)= f_i(\Ab_i\yb) + g_i(\yb),
\end{align}
where $\Ab_i\in \mathbb{R}^{M\times K}$ is some data matrix not necessarily of full rank. Such model is common in practice: the smooth component usually represents the cost function to be minimized, {while the non-smooth component is often {\blue used as a regularization function} \cite{Elad2009Book} or an indicator function representing that $\yb$ is subject to a constraint set\footnote{For example, if $\yb\in \Xc \subseteq \mathbb{R}^K$ for some set $\Xc$, then this can be implicitly included in the nonsmooth component $g_i$ by letting \cite[Section 5]{BoydADMM11}
\begin{align}\label{eqn: indicator function}
  g_i(\yb)=\left\{
   \begin{array}{ll}
    0 ~&  \text{if}~\yb\in \Xc\\
    \infty ~& \text{otherwise}.
  \end{array}
  \right.
\end{align} }.}

In the setting of distributed optimization, it is commonly assumed that each agent $i$ only has knowledge about the local information $f_i$, $g_i$ and $\Ab_i$. The challenge is to obtain, for each agent in the system, the optimal $\xb$ of {\sf (P1)} using only local information and messages exchanged with neighbors \cite{Angelia2010,Johansson08,MZhu2012,Chen_Sayed2012}.


{In addition to {\sf (P1)}, another common problem formulation has the following form }
\begin{align}\label{consensus problem 2}
  {\sf (P2)}~~ \min_{\xb_1,\ldots,\xb_N \in \mathbb{R}^K}~ &\sum_{i=1}^N \phi_i(\xb_i)
  ~~  \text{s.t.}~~ \sum_{i=1}^N\Eb_i\xb_i =\qb, 
\end{align}
{where $\Eb_i\in \mathbb{R}^{M \times K}$, $\qb\in \mathbb{R}^M$ and $\phi_i$ is given as in \eqref{eqn: phi}.}
Unlike {\sf (P1)}, in {\sf (P2)}, each agent $i$ owns a local control variable\footnote{\blue Here we let all $\xb_i$'s have the same dimension without loss of generality.} {\blue $\xb_i \in \mathbb{R}^K$}, and these variables are
coupled together through the linear constraint.  Examples of
{\sf (P2)} include the basis pursuit (BP) problem \cite{ChenBP98,Mota2012}, the network flow control problem \cite{BK:Bersekas_netowork} and
interference management problem in communication networks \cite{ShenTSP2012}.
To relate {\sf (P2)} with {\sf (P1)}, let $\nub \in \mathbb{R}^M$ be the Lagrange dual variable associated with the linear constraint
$\sum_{i=1}^N\Eb_i\xb_i =\qb$. 
The Lagrange dual problem of {\sf (P2)} can be {equivalently} written as

\vspace{-0.1cm}
{\small \begin{align}\label{eq: dual of consensus problem 2}
  \min_{\nub \in \mathbb{R}^M} ~\sum_{i=1}^N \bigg(\varphi_i(\nub) + \frac{1}{N}\nub^T\qb\bigg)
\end{align}}\hspace{-0.1cm}
where
\begin{align}\label{eq: varphi}
  \varphi_i(\nub) = \max_{\xb_i} \bigg\{ -\phi_i(\xb_i)  - \nub^T\Eb_i\xb_i\bigg\},~i=1,\ldots,N.
\end{align}
Problem \eqref{eq: dual of consensus problem 2} thus has the same form as {\sf (P1)}.
Given the optimal $\nub$ of \eqref{eq: dual of consensus problem 2} and assuming that {\sf (P2)} has a zero duality gap \cite{BK:BoydV04},
each agent $i$ can obtain the associated optimal variable $\xb_i$ by solving \eqref{eq: varphi}.
Therefore, a distributed optimization method that can solve
{\sf (P1)} may also be used for {\sf (P2)} through solving \eqref{eq: dual of consensus problem 2}.

There is an extensive literature on distributed consensus optimization methods, such as the consensus subgradient methods; 
see \cite{Angelia2010,Johansson08} and the recent developments
in \cite{Chen_Sayed2012,Ram2012,MZhu2012,ChangTAC2013}. The consensus subgradient methods are appealing
 owing to  their simplicity and the ability to handle a wide range of problems. However, the convergence of the consensus subgradient methods are usually slow.

Recently, the alternating direction method of multipliers (ADMM) \cite{BertsekasADMM,BoydADMM11} has become popular for solving problems with forms of {\sf (P1)} and {\sf (P2)} in a distributed fashion. In \cite{ShenTSP2012}, distributed transmission designs for multi-cellular wireless communications were developed based on ADMM.
In \cite{Mateos2010}, several ADMM based distributed optimization algorithms were developed for solving the sparse LASSO problem \cite{Tibshirani96}. In \cite{Mota2012}, using a different consensus formulation from \cite{Mateos2010} and assuming the availability of a certain coloring scheme for the graph, ADMM is applied to solving the BP problem \cite{ChenBP98} for both row partitioned and column partitioned data models \cite{Ram2012}. 
In \cite{MotaDADMM2013}, the methodologies proposed in \cite{Mota2012} are extended to handling a more general class of problems with forms of {\sf (P1)} and {\sf (P2)}. {In \cite{ErminWeiSIP13}, a distributed ADMM with a sequential update rule is proposed; while in \cite{ErminWeiCDC12}, the method is extended and can be implemented asynchronously.}
The fast practical performance of ADMM is corroborated by its nice theoretical property. In particular, ADMM was found to converge linearly for a large class of problems \cite{DengYin2013J,HongLuo2013}, meaning a certain optimality measure can decrease by a constant fraction in each iteration of the algorithm. {In \cite{ShiLing2013J,Jakovetic2013arxiv}, such fast convergence rate has also been built for distributed optimization.}

It is important to note that existing ADMM based algorithms can be readily used  to solve problems {\sf (P1)} and {\sf (P2)}. For example, by applying the consensus formulation proposed in \cite{Mateos2010} and ADMM to {\sf (P1)}, a fully parallelized distributed optimization algorithm can be obtained (where the agents update their variables in a fully parallel manner), which we refer to as the consensus ADMM (C-ADMM). To solve {\sf (P2)}, the same consensus formulation and ADMM can be used on its Lagrange dual problem in \eqref{eq: dual of consensus problem 2}, 
referred to as the dual consensus ADMM (DC-ADMM).
The main drawback of these algorithms lies in the fact that each agent needs to repeatedly solve certain subproblems to {\it global optimality}. This can be computationally demanding, especially when the cost functions $f_i$'s have complicated structures or when the problem size is large \cite{BK:Bekkerman12}. If a low-accuracy suboptimal solution is used for these subproblems instead, the convergence is no longer guaranteed.


The main objective of this paper is to study algorithms that can significantly reduce the computational burden for the agents. In particular, we propose two algorithms, named the inexact consensus ADMM (IC-ADMM) and the inexact dual consensus ADMM (IDC-ADMM'), both of which allow the agents to perform a single proximal gradient (PG) step \cite{Nesterov2005} at each iteration. The benefit of the proposed approach lies in the fact that the PG step is usually simple, especially when $g_i$'s are structured functions \cite{Nesterov2005,Elad2009Book}.
Notably, the cheap iterations of the proposed algorithms is made possible by {\it inexactly} solving the subproblems arising in C-ADMM and DC-ADMM, in a way that is not known in the ADMM or consensus literature. For example, the proposed IC-ADMM approximates the smooth functions $f_i$'s in C-ADMM, which is very different from the known inexact ADMM methods \cite{HeYuan2013,MaSQ2013}, where only the quadratic penalty is approximated (thus does not always result in cheap PG steps).
We summarize our main contributions below.

\begin{itemize}\itemsep=-0pt
\item For {\sf (P1)}, we propose an IC-ADMM method for reducing the computational complexity of C-ADMM. Conditions for global convergence of IC-ADMM are analyzed. Moreover, we show that IC-ADMM converges linearly, under similar conditions as in \cite{ShiLing2013J}.
\item {\blue For {\sf (P2)}, we first propose a DC-ADMM method which 
    can globally solve {\sf (P2)} for any connected graph and convex $\phi_i$'s. We further propose an IDC-ADMM method for reducing the computational burden of DC-ADMM. Conditions for global (linear) convergence are presented.}
\end{itemize}
Numerical examples for solving distributed sparse logistic regression problems \cite{Liu2009} will show that the proposed IC-ADMM and IDC-ADMM methods converge much faster than the consensus subgradient method \cite{Angelia2010}. {Further, compared with the original C-ADMM and DC-ADMM, the proposed method can reduce the overall computational cost by an order of magnitude.} 


The paper is organized as follows. Section \ref{sec: problem statement} presents the applications and assumptions.
The C-ADMM and IC-ADMM are presented in Section \ref{sec: C-ADMM}; while DC-ADMM and IDC-ADMM are presented in Section \ref{sec: DC-ADMM}. Numerical results are given in Section \ref{sec: simulation} and conclusions are drawn in Section \ref{sec: conclusions}.

{\textit{\bf Notations:} 
$\Ab\succeq \zerob$ ($\succ \zerob$) means that matrix $\Ab$ is positive semidefinite (positive definite). $\Ib_K$ is the $K \times K$ identity matrix; $\oneb_K$ is the $K$-dimensional all-one vector.
$\|\ab\|_2$ denotes the Euclidean norm of vector $\ab$, and $\|\zb\|^2_\Ab\triangleq \zb^T\Ab\zb$ for some $\Ab\succeq \zerob$.
Notation $\otimes$ denotes the Kronecker product. $\diag\{a_1,\ldots,a_N\}$ is a diagonal matrix with the $i$th diagonal element being $a_i$; while $\blkdiag\{\Ab_1,\ldots,\Ab_N\}$ is a block diagonal matrix with the $i$th diagonal block matrix being $\Ab_i$. $\lambda_{\max}(\Ab)$ and $\lambda_{\min}(\Ab)$ denote the maximum and minimum eigenvalues of matrix $\Ab$, respectively.}

\section{Applications and Network Model}\label{sec: problem statement}

\subsection{Application to Data Regression}\label{subsec: application}

As discussed in Section \ref{sec: intro}, {\sf (P1)} and  {\sf (P2)} arise in many problems in sensor networks, data networks and machine learning tasks. Here let us focus on the classical regression problems. We consider a general formulation that incorporates the LASSO \cite{Mateos2010} and logistic regression (LR) \cite{Liu2009} as special instances. {Let $\Ab =[\Ab_1^T,\ldots,\Ab_N^T]^T \in \mathbb{R}^{N M \times K}$ denote a regression data matrix, where $\Ab_i\in \mathbb{R}^{M\times K}$ for all $i=1,\ldots,N$.} For a row partitioned data (RPD) model \cite[Fig. 1]{Mota2012},\cite{Ram2012}, the distributed regression problem is given by
\begin{align}\label{HPD regression}
  \min_{\yb \in \mathbb{R}^K}~ &\sum_{i=1}^N \Psi_i(\yb; \Ab_i,\bb_i),
\end{align}
where $\Psi_i(\yb; \Ab_i,\bb_i)$ is the cost function defined on {the local regression data $\Ab_i$ and a local response signal $\bb_i\in \mathbb{R}^{M}$.} For example, the LASSO problem has $\Psi_i(\yb; \Ab_i,\bb_i)=\|\bb_i-\Ab_i\yb\|_2^2 +g_i(\yb)$. 
Similarly, for the LR problem, one has
\begin{align}\label{eqn: logistic func}
\!\!\!\!\Psi_i(\yb; \Ab_i,\bb_i) =  \sum_{m=1}^{M} \log\big( 1+ \exp(-b_{im}\ab_{im}^T\yb) \big) +g_i(\yb),
\end{align}
where {$\Ab_i=[\ab_{i1},\ldots,\ab_{iM}]^T$ contains $M$ training data vectors} and $b_{im}\in \{\pm 1\}$ 
are binary labels for the training data. It is clear that \eqref{HPD regression} has the same form as {\sf (P1)}.
{\blue Here, the non-smooth function $g_i$ can be 1-norm for sparse regression, as well as
mixture with an indicator functions specifying that $\yb$ is confined in certain constraint set.}

{On the other hand, let $\Eb =[\Eb_1,\ldots,\Eb_N] \in \mathbb{R}^{M \times NK}$ denote a regression data matrix, where $\Eb_i\in \mathbb{R}^{M\times K}$ for all $i=1,\ldots,N$. Then,} for the column partitioned data (CPD) model \cite[Fig. 1]{Mota2012},\cite{Ram2012}, the distributed regression problem is formulated as
\begin{align}\label{VPD regression}
\min_{\xb_1,\ldots,\xb_N\in \mathbb{R}^K}~ &\sum_{i=1}^N \Psi_i(\xb_i; \Eb_i,\bb),
\end{align}
where the response signal $\bb$ is known to all agents while each agent $i$ has a local regression variable {\blue $\xb_i \in \mathbb{R}^{K}$ and local regression data matrix $\Eb_i=[\eb_{i1},\ldots,\eb_{iM}]^T \in \mathbb{R}^{M \times K}$.} For example, the LR problem has
{\begin{align}\label{eqn: logistic func2}
\Psi_i(\xb_i; \Eb_i,\bb) =  \sum_{m=1}^{ M} \log\big( 1+ \exp(-b_{m}\sum_{i=1}^N \eb_{im}^T\xb_i) \big) +g_i(\xb_i).
\end{align}} \hspace{-0.2cm} By introducing a slack variable {\blue $\zb=[z_1,\ldots,z_M]^T\triangleq \sum_{i=1}^N \Eb_i\xb_i$}, the CPD LR problem can be reformulated as
\begin{align}\label{VPD logstic regression}
  \min_{\substack{\xb_1,\ldots,\xb_N \in \mathbb{R}^K, \\ \zb \in \mathbb{R}^{M}}}~ & \bigg\{\sum_{m=1}^{M} \log\big( 1+ \exp(-b_{m}z_m) \big) + \sum_{i=1}^N g_i(\xb_i) \bigg\}
   \notag \\
   ~~  \text{s.t.}~~ &\textstyle{ \sum_{i=1}^N\Eb_i\xb_i -\zb =\zerob,}
\end{align}
which is an instance of {\sf (P2)}.
In Section \ref{sec: simulation}, we will primarily test our algorithms on the RPD and CPD regression problems.

\vspace{-0.2cm}
\subsection{Network Model and Assumptions}

Let {\blue an undirected} graph $\mathcal{G}$ denote a multi-agent network, which contains a node set $V=\{1,\ldots,N\}$ and an edge set $\mathcal{E}$. An edge $(i,j)\in \mathcal{E}$ if and only if agent $i$ and agent $j$ can communicate with each other (i.e., neighbors). The edge set $\mathcal{E}$ defines an adjacency matrix $\Wb \in \{0,1\}^{N\times N}$, where $[\Wb]_{i,j}=1$ if $(i,j)\in \mathcal{E}$ and $[\Wb]_{i,j}=0$ otherwise. In addition, one can define an index subset $\mathcal{N}_i = \{j\in V \mid (i,j)\in \mathcal{E}\}$ for the neighbors
of each agent $i$, and a degree matrix $\Db={\rm diag}\{|\Nc_1|,\ldots,|\Nc_N|\}$ (a diagonal matrix).
{With $\Wb$ and $\Db$, the Laplacian matrix of $\mathcal{G}$ is given by $\Lb = \Db-\Wb$ which is a positive semidefinite matrix (i.e., $\Lb\succeq\zerob$) and satisfies $\Lb\oneb_N=\zerob$ \cite{BK:Chung96}.}

We make the following assumptions on  $\mathcal{G}$  and problems {\sf (P1)} and {\sf (P2)}.

\begin{Assumption} \label{assumption connected graph} The {undirected} graph $\mathcal{G}$ is connected.
\end{Assumption}
Assumption \ref{assumption connected graph} implies that any two agents in the network can always influence each other in the long run. We also have the following assumptions on problems {\sf (P1)} and {\sf (P2)}.

\begin{Assumption} \label{assumption convex prob}
\begin{enumerate}[(a)]
\item {\blue In {\sf (P1)}, the functions $\phi_i:\mathbb{R}^K \rightarrow \mathbb{R}\cup\{\infty\}$ are proper closed convex functions; at every $\yb$ for which both $f_i(\Ab_i\yb)$ and $g_i(\yb)$ are well defined and $\phi_i(\yb)<\infty$, there exists at least one bounded subgradient $\partial \phi_i(\yb) \in \mathbb{R}^K$ such that
    $\phi_i(\xb) \geq \phi_i(\yb) + (\partial \phi_i(\yb))^T(\xb-\yb)~\forall \xb \in \mathbb{R}^K$.
    Moreover, the minimum of {\sf (P1)} can be attained.
}
\item {\blue In {\sf (P2)}, the functions $\phi_i:\mathbb{R}^K \rightarrow \mathbb{R}\cup\{\infty\}$ are proper closed convex functions; $\phi_i$ has at least one bounded subgradient at every $\xb_i$ for which both $f_i(\Ab_i\xb_i)$ and $g_i(\xb_i)$ are well defined and $\phi_i(\xb_i)<\infty$;} the minimum of {\sf (P2)} is attained and so is its optimal dual value; moreover, strong duality holds for {\sf (P2)}.
    \end{enumerate}
\end{Assumption}
\begin{Assumption} \label{assumption strongly convex fi} For all $i\in V$, {the smooth function $f_i$ in \eqref{eqn: phi}} is strongly convex, i.e., there exists some $\sigma_{f,i}^2 >0$ such that
\begin{align*}
    (\nabla f_i(\yb) - \nabla f_i(\xb))^T(\yb-\xb) \geq &\sigma_{f,i}^2 \|\yb-\xb\|^2_2
    \; \forall \yb,\xb\in \mathbb{R}^{M}.
\end{align*}
Moreover, $f_i$ has Lipschitz continuous gradients, i.e., there exists some $L_{f,i}>0$ such that
\begin{align}\label{eq: lipschitz gradient of f}
  \|\nabla f_i(\yb) - \nabla f_i(\xb)\|_2 \leq L_{f,i} \|\yb-\xb\|_2~~\forall \yb,\xb\in \mathbb{R}^{M}.
\end{align}
\end{Assumption}

Note that, even under Assumption \ref{assumption strongly convex fi}, $\phi_i(\xb)=f_i(\Ab_i\xb)+g_i(\xb)$ is not necessarily strongly convex in $\xb$ since the matrix $\Ab_i$ can be fat and rank deficient.
Both the LASSO problem \cite{Mateos2010} and the LR function
in \eqref{eqn: logistic func} satisfy Assumption \ref{assumption strongly convex fi}
\footnote{The logistic regression function $\log(1+\exp(-x)$ is
strongly convex given that $x$ lies in a compact set.}.

\section{Distributed Consensus ADMM}\label{sec: C-ADMM}

In Section \ref{subsec: review of C-ADMM}, we briefly review the original C-ADMM \cite{Mateos2010} for solving {\sf (P1)}. 
In Section \ref{subsec: IC-ADMM}, we propose a computationally efficient inexact C-ADMM method.

\vspace{-0.2cm}
\subsection{Review of C-ADMM}\label{subsec: review of C-ADMM}

Under Assumption \ref{assumption connected graph}, {\sf (P1)} can be equivalently written as
\begin{subequations}\label{consensus problem equi}
\begin{align}
  \min_{\substack{\yb_1,\ldots,\yb_N,\\\{\tb_{ij}\}}}~ &\sum_{i=1}^N \phi_i(\yb_i)
  \label{consensus problem equi C1}\\
  \text{s.t.}~ & \yb_i = \tb_{ij}~ \forall\; j\in \mathcal{N}_i,~ i\in V ,
  \label{consensus problem equi C2} \\
  & \yb_j = \tb_{ij}~\forall\; j\in \mathcal{N}_i,~ i\in V ,\label{consensus problem equi C3}
\end{align}
\end{subequations}
where $ \{\tb_{ij}\}$ are slack variables. According to \eqref{consensus problem equi}, each agent $i$ can optimize
its local function $f_i(\Ab_i\yb_i) + g_i(\yb_i)$ with respect to a local copy of $\yb$, i.e,
$\yb_i$, under the consensus constraints in \eqref{consensus problem equi C2} and \eqref{consensus problem equi C3}.
In \cite{Mateos2010}, ADMM is employed to solve \eqref{consensus problem equi} in a
distributed manner. 
Let $\{\ub_{ij}\}$ and $\{\vb_{ij}\}$ denote the Lagrange dual variables associated with constraints \eqref{consensus problem equi C2} and \eqref{consensus problem equi C3}, respectively. According to \cite{Mateos2010},
ADMM leads to the following iterative updates at each iteration $k$:
\begin{subequations}\label{eqn: admm steps for P1 2}
\begin{align}
\ub_{ij}^{(k)}&\!=\!\ub_{ij}^{(k-1)}+\frac{c}{2}(\yb_i^{(k-1)}\!-\!\yb_j^{(k-1)}) ~\forall j\in \Nc_i, i\in V,  \label{eqn: admm steps for P1 u} \\
\vb_{ij}^{(k)}&\!=\!\vb_{ij}^{(k-1)}+\frac{c}{2}(\yb_j^{(k-1)}\!-\!\yb_i^{(k-1)})
~\forall j\in \Nc_i, i\in V, \label{eqn: admm steps for P1 v}
\\
\label{eqn: admm steps for P1 22}
\yb_i^{(k)} &\!=\arg\min_{\yb_i}~
\bigg\{\phi_i(\yb_i)+\textstyle{\sum_{j\in \mathcal{N}_i} (\ub_{ij}^{(k)}+ \vb_{ji}^{(k)})^T\yb_i } \notag \\
&~~+ \textstyle{c \sum_{j\in \mathcal{N}_i}\big\|\yb_i-\frac{\yb_i^{(k-1)}+\yb_j^{(k-1)}}{2}\big\|_2^2\bigg\} ~\forall  i\in V},
\end{align}
\end{subequations}
where $c>0$ is a penalty parameter and $\ub_{ij}^{(0)}+\vb_{ij}^{(0)}=\zerob~\forall i,j$.
{Note that variables $\{\tb_{ij}^{(k)}\}$ are not shown in \eqref{eqn: admm steps for P1 2} as they can be expressed by variables $\{\yb_i^{(k-1)}\}$; see \cite{Mateos2010} for the details.}

{The updates in \eqref{eqn: admm steps for P1 2} are useful for convergence analysis. For practical implementation, we define $\pb_i^{(k)}\triangleq \sum_{j\in \Nc_i} ( \ub_{ij}^{(k)} + \vb_{ji}^{(k)} ),$ $ i\in V $.
Then, \eqref{eqn: admm steps for P1 2} boils down to Algorithm \ref{table: C-ADMM}.}

\begin{algorithm}[h!]
\caption{C-ADMM for solving {\sf (P1)}}
\begin{algorithmic}[1]\label{table: C-ADMM}
\STATE {\bf Given} initial variables
$\yb_i^{(0)}\in \mathbb{R}^{K}$ and $\pb_{i}^{(0)}=\zerob$ for each agent $i$,  $i\in V$. Set $k=1.$
\REPEAT
\STATE  For all  $i\in V$ (in parallel), 
\ifconfver 
{\small\begin{align}\label{eqn: subproblem of pc admm pp}
 \textstyle \pb_{i}^{(k)}=&\textstyle \pb_{i}^{(k-1)}+{c}\sum_{j \in \Nc_i}(\yb_i^{(k-1)}-\yb_j^{(k-1)}). \\
 \label{eqn: subproblem of pc admm}
 \yb_i^{(k)} =&\arg~\min_{\yb_i}~\bigg\{f_i(\Ab_i\yb_i) + g_i(\yb_i)+\yb_i^T\pb_i^{(k)} \notag \\
 &~~+ \textstyle c \sum_{j\in \mathcal{N}_i}\big\|\yb_i-\frac{\yb_i^{(k-1)}+\yb_j^{(k-1)}}{2}\big\|^2_2\bigg\}.
\end{align}
}
\else
$~~~~~~~~~\pb_{i}^{(k)}=\pb_{i}^{(k-1)}+{c}\sum_{j \in \Nc_i}(\yb_i^{(k-1)}-\yb_j^{(k-1)}),$
\begin{align}
 \label{eqn: subproblem of pc admm}
 \yb_i^{(k)} =&\arg~\min_{\yb_i}~\bigg\{f_i(\Ab_i\yb_i) + g_i(\yb_i)+\yb_i^T\pb_i^{(k)} + \textstyle c \sum_{j\in \mathcal{N}_i}\big\|\yb_i-\frac{\yb_i^{(k-1)}+\yb_j^{(k-1)}}{2}\big\|^2_2\bigg\}.
\end{align}
\fi
\STATE {\bf Set} $k=k+1.$
\UNTIL a predefined stopping criterion (e.g., a maximum iteration number) is satisfied.%
\end{algorithmic}
\end{algorithm}

It is important to note from Step 4 and Step 5 of Algorithm \ref{table: C-ADMM} {that, except for the parameter $c$ which has to be universally known,} each agent $i$ updates the variables $(\yb_i^{(k)},\pb_i^{(k)})$ in a fully parallel manner, by only using the local function $\phi_i$ and
messages $\{\yb_j^{(k-1)}\}_{j\in \Nc_i}$, which come from its direct neighbors. It has been shown in \cite{Mateos2010} that, under Assumptions \ref{assumption connected graph} and \ref{assumption convex prob}, C-ADMM is guaranteed to converge for any $c>0$\footnote{\blue In general, the parameter $c$ is chosen empirically. Only for some special instance, optimal $c$ may be analytically found; e.g., see \cite{Ghadimi2013arxiv}. }:
\begin{align}
  \lim_{k\rightarrow \infty } \yb_i^{(k)} = \yb^\star,~ 
  \lim_{k\rightarrow \infty } (\ub_{ij}^{(k)},\vb_{ij}^{(k)}) = (\ub_{ij}^\star,\vb_{ij}^\star),\; \forall j,i, 
\end{align}
where {\blue $\yb^\star\triangleq \yb_1^\star =\cdots=\yb_N^\star$} and $\{\ub_{ij}^\star,\vb_{ij}^\star\}$ denote
a pair of optimal primal and dual solutions to problem \eqref{consensus problem equi}, and $\yb^\star$ is optimal to {\sf (P1)}. {It is also shown that C-ADMM can converge linearly when $\phi_i$'s are purely smooth (i.e., $g_i(\yb_i)=0~\forall i$) and strongly convex with respect to $\yb_i$ \cite{ShiLing2013J}.} 

One key issue about C-ADMM is that the subproblem in \eqref{eqn: subproblem of pc admm}
is not always easy to solve. 
For instance,
for the LR function in \eqref{eqn: logistic func}, 
the associated subproblem \eqref{eqn: subproblem of pc admm} is given by
\begin{align}\label{eqn: subprob logistic func}
&\yb_i^{(k)} =\arg~\min_{\yb_i}~\bigg\{\sum_{m=1}^{M} \log\big( 1+ \exp(-b_{im}\ab_{im}^T\yb_i) \big) +g_i(\yb_i) \notag \\
 &~~~~~~+\yb_i^T\pb_i^{(k)}+  c \sum_{j\in \mathcal{N}_i}\big\|\yb_i-\frac{\yb_i^{(k-1)}+\yb_j^{(k-1)}}{2}\big\|^2_2\bigg\}.
\end{align}
As seen, due to the complicated LR cost, problem \eqref{eqn: subprob logistic func} cannot yield simple solutions,
and a numerical solver has to be employed. Clearly, obtaining a high-accuracy solution
of \eqref{eqn: subprob logistic func} can be computationally expensive, especially when the problem dimension or the number of training data is large. While a low-accuracy solution to \eqref{eqn: subprob logistic func}
can be adopted for complexity reduction, it may destroy the convergence behavior of C-ADMM, as will be shown in Section \ref{sec: simulation}.

\vspace{-0.3cm}
\subsection{Proposed Inexact C-ADMM}\label{subsec: IC-ADMM}

To reduce the complexity of C-ADMM, instead of solving
subproblem \eqref{eqn: subproblem of pc admm} directly, we consider the following update:
\ifconfver
\begin{align}\label{eqn: inexact ADMM step xi 2}
    &\yb_i^{(k)} =\arg~\min_{\yb_i}~ \bigg\{\nabla f_i(\Ab_i\yb_i^{(k-1)})^T\Ab_i (\yb_i - \yb_i^{(k-1)})  \notag \\
 &~~~~~~~~~~~~~~~~~+ \frac{\beta_i}{2}\|\yb_i-\yb_i^{(k-1)}\|^2_2+g_i(\yb_i) + \yb_i^T\pb_{i}^{(k)} \notag \\
 &~~~~~~~~~~~~~~~~~\textstyle + c \sum_{j\in \mathcal{N}_i}\big\|\yb_i-\frac{\yb_i^{(k-1)}+\yb_j^{(k-1)}}{2}\big\|^2_2\bigg\}.
\end{align}
\else
\begin{align}\label{eqn: inexact ADMM step xi 2}
    &\yb_i^{(k)} =\arg~\min_{\yb_i}~ \bigg\{\nabla f_i(\Ab_i\yb_i^{(k-1)})^T\Ab_i (\yb_i - \yb_i^{(k-1)})  \notag \\
 &~~~~~~~~~~~~~~~~~~~~+ \frac{\beta_i}{2}\|\yb_i-\yb_i^{(k-1)}\|^2_2+g_i(\yb_i) + \yb_i^T\pb_{i}^{(k)} \textstyle + c \sum_{j\in \mathcal{N}_i}\big\|\yb_i-\frac{\yb_i^{(k-1)}+\yb_j^{(k-1)}}{2}\big\|^2_2\bigg\}.
\end{align}
\fi
In \eqref{eqn: inexact ADMM step xi 2} we have replaced the smooth cost function $f_i(\Ab_i\yb_i)$ in \eqref{eqn: subproblem of pc admm} with {a proximal first-order approximation around $\yb_i^{(k-1)}$}: $$\nabla f_i(\Ab_i\yb_i^{(k-1)})^T\Ab_i (\yb_i - \yb_i^{(k-1)})+ \frac{\beta_i}{2}\|\yb_i-\yb_i^{(k-1)}\|^2_2,$$
where $\beta_i > 0$ is {a penalty parameter of the proximal quadratic term.}
To obtain a concise representation of $\yb_i^{(k)}$, let us define the {\it proximity operator} for the non-smooth function $g_i$ at a given point $\ssb\in\mathbb{R}^K$ as \cite{Nesterov2005}
\begin{align}\label{eqn: proximal operator}
 {\rm prox}_{g_i}^{\gamma_i}[\ssb]\triangleq \arg~\min_{\yb}~ \bigg\{g_i(\yb) + \frac{\gamma_i}{2}\|\yb - \ssb\|^2_2\bigg\},
\end{align}
where $\gamma_i= \beta_i + 2c|\Nc_i|$. Clearly, using this definition, {\eqref{eqn: inexact ADMM step xi 2} can be expressed more compactly as
\begin{align}\label{eqn: inexact ADMM step xi}
 \textstyle  &\yb_i^{(k)} =\arg\min_{\yb_i}
  \bigg\{g_i(\yb) + \frac{\gamma_i}{2}\bigg\|\yb_i - \frac{1}{\gamma_i}\big( \beta_i\yb_i^{(k-1)}- \pb_i^{(k)}\! \notag \\
   & \textstyle~~~ -\! \Ab_i^T
   \nabla f_i(\Ab_i\yb_i^{(k-1)})   + c \sum_{j \in \Nc_i} (\yb_i^{(k-1)} + \yb_j^{(k-1)})\big)\bigg\|^2_2\bigg\}
   \notag \\
   &~~~~~={\rm prox}_{g_i}^{\gamma_i}\bigg[\frac{1}{\gamma_i}\bigg( \beta_i\yb_i^{(k-1)}\! - \pb_i^{(k)}-\! \Ab_i^T
   \nabla f_i(\Ab_i\yb_i^{(k-1)})  \notag \\
   &~~~~~~~~~~~~~~~~~~~~~+ \textstyle c \sum_{j \in \Nc_i} (\yb_i^{(k-1)} + \yb_j^{(k-1)})\bigg) \bigg],
\end{align}
which is a proximal gradient (PG) update.}

The PG updates like \eqref{eqn: inexact ADMM step xi} often admit closed-form expression, especially when $g_i$'s {\blue are functions} including the $\ell_1$ norm, Euclidean norm, infinity norm and matrix nuclear norm \cite{Combettes2009}. For example, when $g_i(\yb)=\|\yb\|_1$, \eqref{eqn: proximal operator} has a closed-form solution known as the soft thresholding operator \cite{Nesterov2005,Combettes2009}:
\begin{align}\label{soft thresholding}
 \textstyle   \mathcal{S}\left[\ssb,\frac{1}{\gamma_i}\right] = \left( \ssb - \frac{1}{\gamma_i}\mathbf{1}_K\right)^+
    +\left( -\ssb - \frac{1}{\gamma_i}\mathbf{1}_K\right)^+,
\end{align}
where $(x)^+ \triangleq \max\{x,0\}$. 
The IC-ADMM is presented in Algorithm \ref{table: IC-ADMM}.
\begin{algorithm}[h!]
\caption{Proposed IC-ADMM for solving {\sf (P1)}}
\begin{algorithmic}[1]\label{table: IC-ADMM}{
\STATE
{\bf Given} initial variables
$\yb_i^{(0)}\in \mathbb{R}^{K}$ and $\pb_{i}^{(0)}=\zerob$ for each agent $i$,  $i\in V$. Set $k=1.$
\REPEAT
\STATE  For all  $i\in V$ (in parallel), 

{\small~~~ $\pb_{i}^{(k)}=\pb_{i}^{(k-1)}+{c}\sum_{j \in \Nc_i}(\yb_i^{(k-1)}-\yb_j^{(k-1)}),$
\begin{align}\label{eqn: inexact ADMM step xi2}
 \textstyle  &\yb_i^{(k)} ={\rm prox}_{g_i}^{\gamma_i}\bigg[\frac{1}{\gamma_i}\bigg( \beta_i\yb_i^{(k-1)}\! -\! \Ab_i^T \nabla f_i(\Ab_i\yb_i^{(k-1)})  \notag \\
   &~~~~~~~~ \textstyle - \pb_i^{(k)} + c \sum_{j \in \Nc_i} (\yb_i^{(k-1)} + \yb_j^{(k-1)})\bigg) \bigg].
\end{align}}
\STATE {\bf Set} $k=k+1.$
\UNTIL {a predefined stopping criterion (e.g., a maximum iteration number) is satisfied.}%
}
\end{algorithmic}
\end{algorithm}

Although the idea of ``inexact ADMM" is not new, our approach is significantly different from the existing
methods \cite{HeYuan2013,MaSQ2013}, where the inexact
update is obtained by approximating the quadratic penalization term only.
It can be seen that problem \eqref{eqn: subprob logistic func} is still difficult to solve even the inexact update in \cite{HeYuan2013,MaSQ2013} is applied. {Two notable exceptions are the algorithms proposed in \cite{HePengWang2011} and \cite{MaS2013arxiv} where the cost function is also linearized. However, an additional back substitution step and two extragradient steps are required in \cite{HePengWang2011} and \cite{MaS2013arxiv}, respectively, which is not suited for distributed optimization.}

The convergence properties of IC-ADMM is characterized by the following theorem.
\begin{Theorem} \label{thm: conv of inexact pc admm} Suppose that Assumptions \ref{assumption connected graph}, \ref{assumption convex prob}(a) and \ref{assumption strongly convex fi} hold. Let
\begin{align}\label{eq: conv condition beta}
\!\!\! \beta_i > \frac{L_{f,i}^2 }{\sigma_{f,i}^2} \lambda_{\max}(\Ab^T_i\Ab_i)\! -\! c \lambda_{\min}(\Db+\Wb)>0
 ~\forall i\in V,
\end{align}
{\blue and let $\yb^\star\triangleq \yb_1^\star =\cdots=\yb_N^\star$ and $\{\ub_{ij}^\star,\vb_{ij}^\star\}$ denote
a pair of optimal primal and dual solutions to problem \eqref{consensus problem equi} (i.e., {\sf (P1)}).}

\begin{itemize}
\item [(a)] {\blue For Algorithm \ref{table: IC-ADMM}, $\yb_1^{(k)},\ldots,\yb_N^{(k)}$ converge to a common point $\yb^\star$.} 

\item [(b)] 
{If $\phi_i(\yb)=f_i(\Ab_i\yb)$, where $\Ab_i$ has full column rank, for all $i \in V$,}
then we have
\begin{align*}
& \blue  \lim_{k\to\infty}\|\yb^{(k)}-\oneb_N\otimes \yb^\star\|_{\frac{1}{2}\Gb+\alpha\Mb}^2 \notag \\
 &~~~~~~~~~~\blue+ \frac{1}{c}\|\ub^{(k+1)} - \ub^\star\|^2_2 =0\; \mbox{linearly},
\end{align*}
where $\yb^{(k)}=[(\yb^{(k)}_1)^T, \ldots, (\yb^{(k)}_N)^T]^T$;
$\ub^{(k)}_i \in \mathbb{R}^{K|\Nc_i|}$ {\blue ($\ub_i^\star$)} is a vector that stacks $\ub^{(k)}_{ij}$ {\blue ($\ub_{ij}^\star$)}~$\forall j\in \Nc_i$; 
$\ub^{(k)}\in \mathbb{R}^{K\sum_{i=1}^N|\Nc_i|}$ {\blue ($\ub^\star$)} stacks $\ub^{(k)}_{i}$ {\blue ($\ub_{i}^\star$)} $\forall i=1,\ldots,N$.
and
\begin{align}
  \Gb &\triangleq \Db_{\beta} + c ((\Db+\Wb)\otimes \Ib_K) \succ \zerob, \label{eq: G}\\
  \Mb &\triangleq \tilde \Ab^T(\Db_{\sigma_f}-\frac{1}{2}\Db_{\rho})\tilde \Ab \succ \zerob, \label{eq: M}
\end{align}
for some $0<\alpha<1$ and $\rho>0$. {Here,} 
{$\tilde \Ab=\blkdiag\{\Ab_1,\ldots,\Ab_N\}$; $\Db_{\beta}=\diag\{\beta_1,\ldots,\beta_N\}\otimes \Ib_K$;
$\Db_{\sigma_f}=\diag\{\sigma_{f,1}^2,\ldots,\sigma_{f,N}^2\}\otimes \Ib_K$;}
{and $\Db_{\rho}=\diag\{\rho_1,\ldots,\rho_N\}\otimes \Ib_K$}. 
\end{itemize}
\end{Theorem}
The proof is presented in Appendix \ref{proof of conv of inexact pc-admm}.
Theorem \ref{thm: conv of inexact pc admm} implies that, given sufficiently large $\beta_i$'s, IC-ADMM not only achieves consensus and optimality, but also converges linearly provided that {$\phi_i$ is purely smooth and strongly convex.}
{\blue Note that, to ensure \eqref{eq: conv condition beta}, the global knowledge of $\lambda_{\min}(\Db+\Wb)$ is required by all agents.}
{As a parallel work, we should mention that a concurrent result similar as Theorem \ref{thm: conv of inexact pc admm}(b) is presented in \cite{LinICASSP14}.}

\begin{Remark}\label{rmk: topology}{\rm
  We remark that the convergence condition in \eqref{eq: conv condition beta} depends on the network topology.
  Let $\Lb = \Db-\Wb$ denote the Laplacian matrix of $\mathcal{G}$. Then $\Db+\Wb = 2\Db - \Lb.$
  By the graph theory \cite{BK:Chung96}, the normalized Laplacian matrix, i.e., $\tilde\Lb=\Db^{-\frac{1}{2}}\Lb\Db^{-\frac{1}{2}}$, must have
  $\lambda_{\max}(\tilde\Lb)\leq 2$. Further, $\lambda_{\max}(\tilde\Lb)< 2$ if and only if the connected graph $\mathcal{G}$ is not bipartite.
  Thus, we have
  $
      \lambda_{\min}(\Db+\Wb) = \lambda_{\min}(\Db^{\frac{1}{2}}(2\Ib_N- \tilde\Lb )\Db^{\frac{1}{2}}) \geq 0,
  $  and $\lambda_{\min}(\Db+\Wb)>0$ whenever $\mathcal{G}$ is non-bipartite.
  }
\end{Remark}

\vspace{-0.2cm}
\section{Distributed Dual Consensus ADMM}\label{sec: DC-ADMM}

In this section, we turn the focus to {\sf (P2)}. In Section \ref{subsec: DC-ADMM}, we present a DC-ADMM method for solving {\sf (P2)}.
In Section \ref{subsec: inexact DC-ADMM}, an inexact DC-ADMM method is proposed.

\vspace{-0.2cm}
\subsection{Proposed DC-ADMM} \label{subsec: DC-ADMM}
The DC-ADMM is obtained by applying the C-ADMM (Algorithm \ref{table: C-ADMM}) to problem \eqref{eq: dual of consensus problem 2} which {is equivalent to the Lagrange dual of {\sf (P2)}.}
Firstly, similar to \eqref{consensus problem equi}, we write problem \eqref{eq: dual of consensus problem 2} as
\begin{subequations}\label{consensus problem equi dual}
\begin{align}
  \min_{\substack{\nub_1,\ldots,\nub_N\\ \{\tb_{ij}\}} }~ &\sum_{i=1}^N \bigg(\varphi_i(\nub_i) + \frac{1}{N}\nub_i^T\qb\bigg)
  \label{consensus problem equi C1 dual}\\
  \text{s.t.}~ & \nub_i = \tb_{ij},~ \nub_j = \tb_{ij}~\forall\; j\in \mathcal{N}_i,~ i\in V ,
  \label{consensus problem equi C2 dual} 
\end{align}
\end{subequations}
where $\nub_i\in \mathbb{R}^{M}$ is the $i$th agent's local copy of the dual variable $\nub$ and $\varphi_i$ is given in \eqref{eq: varphi}.
Following a similar argument as in deriving Algorithm \ref{table: C-ADMM}, we obtain the following update steps at each iteration $k${
\begin{subequations}\label{eqn: admm steps for P2 2}
\begin{align}
\!\!\!\!\pb_{i}^{(k)}&=\textstyle \pb_{i}^{(k-1)}+{c}\sum_{j \in \Nc_i}(\nub_i^{(k-1)}-\nub_j^{(k-1)}),\\
\nub_i^{(k)} &=\arg\min_{\nub_i\in \mathbb{R}^{M}} \bigg\{\varphi_i(\nub_i) + \frac{1}{N}\nub^T_i\qb+\nub_i^T\pb_{i}^{(k)} \notag \\
 &\textstyle~~~+ c
\sum_{j\in \mathcal{N}_i}\big\|\nub_i-\frac{\nub_i^{(k-1)}+\nub_j^{(k-1)}}{2}\big\|^2_2\bigg\}~\forall~ i\in V,\label{eqn: dc admm steps yi}
\end{align}
\end{subequations}
where, with a slight abuse of notation,
\begin{align}\label{eqn: pi}
 \pb_i^{(k)}=\textstyle \sum_{j\in \Nc_i} ( \ub_{ij}^{(k)} + \vb_{ji}^{(k)} ),
\end{align}
in which $\{\ub_{ij}\}$ and $\{\vb_{ij}\}$
are dual variables associated with the two constraints in \eqref{consensus problem equi C2 dual} and are updated in a similar fashion as in \eqref{eqn: admm steps for P1 u} and \eqref{eqn: admm steps for P1 v}, i.e.,}
\begin{subequations}\label{eq: inexact ADMM u v}
\begin{align}
\ub_{ij}^{(k)}&=\ub_{ij}^{(k-1)}+\frac{c}{2}(\nub_i^{(k-1)}\!-\!\nub_j^{(k-1)}) ~\forall j\in \Nc_i, i\in V,
\label{eq: inexact ADMM ub} \\
\vb_{ij}^{(k)}&=\vb_{ij}^{(k-1)}+\frac{c}{2}(\nub_j^{(k-1)}\!-\!\nub_i^{(k-1)})
~\forall j\in \Nc_i, i\in V. \label{eq: inexact ADMM vb}
\end{align}
\end{subequations}

In general, subproblem \eqref{eqn: dc admm steps yi} is not easy to handle because $\varphi_i$ is implicit and \eqref{eqn: dc admm steps yi} is in fact a min-max optimization problem {given by
\begin{align}\label{eqn: minmax0}
\nub_i^{(k)} &=\arg \min_{\nub_i} \max_{\xb_i}\bigg\{ -\phi_i(\xb_i)  - \nub_i^T\Eb_i\xb_i
+ \frac{1}{N}\nub^T_i\qb \notag \\
 &\textstyle~~+\nub_i^T\pb_{i}^{(k)}+ c
\sum_{j\in \mathcal{N}_i}\big\|\nub_i-\frac{\nub_i^{(k-1)}+\nub_j^{(k-1)}}{2}\big\|^2_2\bigg\}.
\end{align}
Fortunately,
since the objective function in \eqref{eqn: minmax0} is convex in $\nub_i$ for any $\xb_i$ and is concave in $\xb_i$ for any $\nub_i$,
the minimax theorem \cite[Proposition 2.6.2]{BK:Bertsekas2003_analysis} can be applied
so that the min-max problem \eqref{eqn: minmax0} can be equivalently solved by considering its max-min counterpart and saddle point exists. Specifically, the max-min counterpart of \eqref{eqn: minmax0} is given by
\begin{align}\label{eqn: fenchel duality}
&\max_{\xb_i} \min_{\nub_i} \bigg\{ -\phi_i(\xb_i)  - \nub_i^T\Eb_i\xb_i
+ \frac{1}{N}\nub^T_i\qb+\nub_i^T\pb_{i}^{(k)} \notag \\
 &\textstyle~~+ c
\sum_{j\in \mathcal{N}_i}\big\|\nub_i-\frac{\nub_i^{(k-1)}+\nub_j^{(k-1)}}{2}\big\|^2_2\bigg\}
 \\
=& \max_{\xb_i} \min_{\nub_i} \bigg\{ \textstyle -\phi_i(\xb_i)
+(c|\mathcal{N}_i|) \bigg\|  \nub_i
- \frac{1}{2|\mathcal{N}_i|}\big[
    \sum_{j\in \mathcal{N}_i} (\nub_i^{(k-1)}
    \notag \\
    &~~~~~~~~\textstyle  +\nub_j^{(k-1)})- \frac{1}{c} \pb_{i}^{(k)} +\frac{1}{c}(\Eb_i\xb_i -\frac{1}{N}\qb )
    \big]
\bigg\|^2_2 \notag \\
 &\textstyle ~~~~~~~~-\frac{c}{4|\mathcal{N}_i|}\textstyle
    \bigg\|\frac{1}{c}(\Eb_i\xb_i-\frac{1}{N}\qb) \notag \\
    &~~~~~~~~\textstyle - \frac{1}{c}\pb_{i}^{(k)} + \sum_{j\in \mathcal{N}_i} (\nub_i^{(k-1)}+\nub_j^{(k-1)})
    \bigg\|^2_2 \bigg\}\label{eqn: fenchel duality2}
\end{align}
where the equality is obtained by completing the quadratic term of $\nub_i$.
Let $\xb_i^{(k)}$ be an inner maximizer of \eqref{eqn: minmax0} so that $(\nub_i^{(k)},\xb_i^{(k)})$ is a saddle point of
\eqref{eqn: minmax0}. Then, $(\xb_i^{(k)},\nub_i^{(k)})$ is {\blue a pair of outer-inner} solution to \eqref{eqn: fenchel duality} and \eqref{eqn: fenchel duality2} \cite[Proposition 2.6.1]{BK:Bertsekas2003_analysis}.
From \eqref{eqn: fenchel duality2}, the {\blue inner minimizer $\nub_i^{(k)}$} can be uniquely determined by
\begin{align}\label{eq: dual ADMM lambda 2}
    \nub_i^{(k)} =  &\textstyle\frac{1}{2|\mathcal{N}_i|}\big[
    \sum_{j\in \mathcal{N}_i} (\nub_i^{(k-1)}+\nub_j^{(k-1)})
    - \frac{1}{c} \pb_{i}^{(k)} \notag \\
    &~~~~~~~~\textstyle  +\frac{1}{c}(\Eb_i\xb_i^{(k)} -\frac{1}{N}\qb )
    \big],
\end{align}
and that the outer maximizer is given by}
\begin{align}\label{eq: dual ADMM x}
    &\xb_i^{(k)}=\arg~\min_{\xb_i}~\bigg\{\phi_i(\xb_i)+\frac{c}{4|\mathcal{N}_i|}\textstyle
    \big\|\frac{1}{c}(\Eb_i\xb_i-\frac{1}{N}\qb) \notag \\
    &~~~~~~~~\textstyle - \frac{1}{c}\pb_{i}^{(k)} + \sum_{j\in \mathcal{N}_i} (\nub_i^{(k-1)}+\nub_j^{(k-1)})
    \big\|^2_2\bigg\}.
\end{align}
As a result, the min-max subproblem \eqref{eqn: dc admm steps yi} can actually be obtained {by first solving the subproblem
\eqref{eq: dual ADMM x} with respect to the primal variable $\xb_i$} followed by evaluating $\nub_i^{(k)}$ using the { close-form} in \eqref{eq: dual ADMM lambda 2}.
The proposed DC-ADMM is summarized in Algorithm \ref{table: DC-ADMM}.

\begin{algorithm}[h!]
\caption{Proposed DC-ADMM for solving {\sf (P2)}}
\begin{algorithmic}[1]\label{table: DC-ADMM}
\STATE {\bf Given} initial variables
$\xb_i^{(0)}\in \mathbb{R}^{K}$, $\nub_i^{(0)}\in \mathbb{R}^{M}$  and $\pb_{i}^{(0)}=\zerob$ for each agent $i$,  $i\in V$. Set $k=1.$
\REPEAT
\STATE  For all  $i\in V$ (in parallel), 

{\small
$~~~~\pb_{i}^{(k)}=\pb_{i}^{(k-1)}+{c}\sum_{j \in \Nc_i}(\nub_i^{(k-1)}-\nub_j^{(k-1)})$,
\begin{align}\label{eq: dual ADMM x1}
    &\xb_i^{(k)}=\arg~\min_{\xb_i}~\bigg\{\phi_i(\xb_i)+\frac{c}{4|\mathcal{N}_i|}\textstyle
    \big\|\frac{1}{c}(\Eb_i\xb_i-\frac{1}{N}\qb) \notag \\
    &~~~~~~~\textstyle- \frac{1}{c}\pb_i^{(k)} + \sum_{j\in \mathcal{N}_i} (\nub_i^{(k-1)}+\nub_j^{(k-1)})
    \big\|^2_2\bigg\},
\end{align}
\begin{align}\label{eq: dual ADMM lambda 21}
\!\!\!\!\!\!\!\!   \nub_i^{(k)} =  &\textstyle\frac{1}{2|\mathcal{N}_i|}\big(
    \sum_{j\in \mathcal{N}_i} (\nub_i^{(k-1)}+\nub_j^{(k-1)})
    - \frac{1}{c}   \pb_{i}^{(k)} \notag \\
    &~~~~~~~~\textstyle +\frac{1}{c}(\Eb_i\xb_i^{(k)} -\frac{1}{N}\qb )
    \big).
\end{align}}
\STATE {\bf Set} $k=k+1.$
\UNTIL {a predefined stopping criterion is satisfied.} 
\end{algorithmic}
\end{algorithm}

Interestingly, while DC-ADMM handles the {equivalent} dual problem in \eqref{eq: dual of consensus problem 2}, it directly yields primal optimal solution of
{\sf (P2)}, {as we state in the following theorem.} 
\vspace{-0.2cm}
\begin{Theorem}\label{thm: conv of DADMM}
  Suppose that Assumptions \ref{assumption connected graph} and \ref{assumption convex prob}(b) hold. Then $(\nub_1^{(k)},\ldots,\nub_N^{(k)})$ converges to a common point $\nub^\star$, which is optimal to the dual problem \eqref{eq: dual of consensus problem 2}. Moreover, any limit point of $(\xb_1^{(k)},\ldots,\xb_N^{(k)})$ is primal optimal to {\sf (P2)}.
\end{Theorem}
\vspace{-0.2cm}

{\bf Proof:} Since DC-ADMM is a direct application of C-ADMM to the dual problem \eqref{eq: dual of consensus problem 2}, it follows from \cite{Mateos2010} that as $k\rightarrow \infty$,
\begin{align}\label{eq: proof of thm 2 eq0}
  \textstyle \nub_i^{(k)} \rightarrow \nub^\star,~
  {\nub_i^{(k)}-\nub_j^{(k)} \rightarrow \zerob}~\forall j\in \mathcal{N}_i,~i\in V.
\end{align}
What remains is to show that {\blue any limit point of} $(\xb_1^{(k)},\ldots,\xb_N^{(k)})$ is asymptotically optimal to {\sf (P2)}, i.e.,
{\blue as $k\to \infty$,}
\begin{align}
\blue  \partial \phi_i(\xb_i^{(k)}) + \Eb_i^T\nub^\star & \blue \to\zerob ~\forall i\in V, \label{eq: KKT of P1}\\
 \blue \textstyle \sum_{i=1}^N \Eb_i\xb_i^{(k)} -\qb & \blue \to \zerob. \label{eq: KKT of P2}
\end{align}
To show \eqref{eq: KKT of P1}, consider the optimality condition of \eqref{eq: dual ADMM x}, i.e.,
\begin{align}\label{eq: proof of thm 2 eq01}
\zerob &=  \partial\phi_i(\xb_i^{(k)}) + \frac{1}{2|\mathcal{N}_i|}\Eb_i^T\bigg( \frac{1}{c}(\Eb_i\xb_i^{(k)}-\frac{1}{N}\qb)
\notag \\
&~~~~~~~~~\textstyle  - {\frac{1}{c}\pb_i^{(k)}} + \sum_{j\in \mathcal{N}_i} (\nub_i^{(k-1)}+\nub_j^{(k-1)}) \bigg) \notag \\
&= \partial\phi_i(\xb_i^{(k)}) + \Eb_i^T\nub_i^{(k)},
\end{align}
where the second equality is obtained by \eqref{eq: dual ADMM lambda 2}.
{\blue Since \eqref{eq: proof of thm 2 eq01} holds for all $k$ and $\nub_i^{(k)} \rightarrow \nub^\star$ by \eqref{eq: proof of thm 2 eq0}, \eqref{eq: KKT of P1} is true when $k\to \infty$.}

To show \eqref{eq: KKT of P2}, rewrite \eqref{eq: dual ADMM lambda 2} as follows
\begin{align}\label{eq: proof of thm 2 eq1}
\!\!  &\zerob= 
    -(\Eb_i\xb_i^{(k)} -\frac{1}{N}\qb ) + \textstyle 2c\sum_{j\in \mathcal{N}_i}\big(\nub_i^{(k)} -
    \frac{\nub_i^{(k)}+\nub_j^{(k)}}{2}\big)
    \notag \\
    & ~~~~~~~~+ {\pb_i^{(k)}} \!+\! c\sum_{j\in \mathcal{N}_i} (\nub_i^{(k)}+\nub_j^{(k)}\!-\nub_i^{(k-1)}\!-\nub_j^{(k-1)}) \notag \\
    &~=-(\Eb_i\xb_i^{(k)} -\frac{1}{N}\qb ) +  {\pb_i^{(k+1)}}  \notag \\
    &\textstyle ~~~~~~~~~~+ c\sum_{j\in \mathcal{N}_i} (\nub_i^{(k)}+\nub_j^{(k)}-\nub_i^{(k-1)}-\nub_j^{(k-1)}),
\end{align}
{where the last equality is obtained by \eqref{eqn: pi} and \eqref{eq: inexact ADMM u v}.} 
Upon summing \eqref{eq: proof of thm 2 eq1} for $i=1,\ldots,N$, {and by the fact that
$$
  \sum_{i=1}^N \pb_i^{(k)}=\sum_{i=1}^N \sum_{j\in \Nc_i} ( \ub_{ij}^{(k)} + \vb_{ji}^{(k)} )
  =\zerob
$$
(by applying \eqref{eq: proof 7.55} and \eqref{eq: proof 7.6} in Appendix \ref{proof of conv of inexact pc-admm}),} we can obtain
\begin{align}\label{eq: proof of thm 2 eq1.5}
  \sum_{i=1}^N\Eb_i\xb_i^{(k)} -\qb = c \sum_{i=1}^N\sum_{j\in \mathcal{N}_i} (\nub_i^{(k)}+\nub_j^{(k)}-\nub_i^{(k-1)}-\nub_j^{(k-1)}).
\end{align}
{\blue Note that $\nub_i^{(k)}-\nub_i^{(k-1)}\to \zerob~\forall i\in V$ as inferred from $\nub_i^{(k)} \to \nub^\star~\forall i\in V$ in \eqref{eq: proof of thm 2 eq0}. By applying this fact to \eqref{eq: proof of thm 2 eq1.5}, we obtain that \eqref{eq: KKT of P2} is true as $k\to \infty$.}
\hfill $\blacksquare$

Interestingly, from \eqref{eq: proof of thm 2 eq1.5}, one observes that the primal feasibility of $(\xb_1^{(k)},\ldots,\xb_N^{(k)})$ to {\sf (P2)} depends on the agents' consensus on the dual variable $\nub$.

{We remark that Algorithm \ref{table: DC-ADMM} is different from the D-ADMM algorithm in \cite[Algorithm 3]{Mota2012}. Firstly, Algorithm \ref{table: DC-ADMM} can be implemented in a fully parallel manner; secondly, Algorithm \ref{table: DC-ADMM} does not involve solving a min-max subproblem at each iteration; thirdly, convergence of Algorithm \ref{table: DC-ADMM} can be achieved without the assumption that the graph $\Gc$ is bipartite.
}

\subsection{Proposed Inexact DC-ADMM}\label{subsec: inexact DC-ADMM}

In this subsection, we propose an inexact version of DC-ADMM, referred to as the IDC-ADMM.
In view of the fact that solving the subproblem in \eqref{eq: dual ADMM x1} can be expensive, we consider an inexact update of $\xb_i^{(k)}$. Specifically, since a non-trivial $\Eb_i$ can also complicate the solution\footnote{When $\Eb_i$ has orthogonal columns (e.g., $\Eb_i^T\Eb_i=\alpha\Ib_K$ for some $\alpha\in \mathbb{R}$), then it may not be necessary to approximate the quadratic term.}, we propose to approximate both $f_i(\Ab_i\xb_i)$ and the quadratic term $\frac{c}{4|\mathcal{N}_i|}
\|\frac{1}{c}(\Eb_i\xb_i-\frac{1}{N}\qb) - \frac{1}{c}\pb_i^{(k)} +
\sum_{j\in \mathcal{N}_i} (\nub_i^{(k-1)}+\nub_j^{(k-1)})\|^2_2$ in \eqref{eq: dual ADMM x1}
by a {proximal first-order approximation around $\xb_i^{(k-1)}$; this leads to the following update
\begin{align}\label{eq: dual ADMM x linearized}
 &\xb_i^{(k)}\!=\!\arg\min_{\xb_i}\textstyle \bigg\{\!  \bigg [\!\! \Ab_i^T\nabla f_i(\Ab_i\xb_i^{(k-1)})\!+\!\!
 \frac{1}{2|\mathcal{N}_i|}\Eb_i^T
    \bigg(\!\!\frac{1}{c}(\Eb_i\xb_i^{(k-1)}\! \notag \\
    &\textstyle  -\!\frac{1}{N}\qb)- \frac{1}{c}\pb_{i}^{(k)} + \sum_{j\in \mathcal{N}_i} (\nub_i^{(k-1)}+\nub_j^{(k-1)})
    \bigg)\! \bigg ]^T\!\!\!(\xb_i-\xb_i^{(k-1)})\notag \\
 &~~~~~~~~~ +\frac{\beta_i}{2}\|\xb_i-\xb_i^{(k-1)}\|_2^2 +g_i(\xb_i)\bigg\},
\end{align}
where, with a slight abuse of notation, $\beta_i>0$ is a penalty parameter.
By \eqref{eqn: proximal operator},
equation \eqref{eq: dual ADMM x linearized} can be further written as the following PG update
\begin{align}\label{eqn: inexact DADMM step xi}
\!\!   &\xb_i^{(k)}\! =\!\arg\min_{\xb_i}\textstyle \bigg\{\!
\frac{\beta_i}{2}\bigg\|\xb_i - \bigg[ \xb_i^{(k-1)}-
\frac{1}{\beta_i}\Ab_i^T\nabla f_i(\Ab_i\xb_i^{(k-1)}) \notag\\
&~~~~~~~~~~~~~~~~~\textstyle - \frac{1}{2\beta_i|\mathcal{N}_i|}
\Eb_i^T\big(\frac{1}{c}(\Eb_i\xb_i^{(k-1)}\!-\!\frac{1}{N}\qb)- \frac{1}{c}\pb_{i}^{(k)} \notag \\
&~~~~~~~~~~~~~~~~~\textstyle + \sum_{j\in \mathcal{N}_i} (\nub_i^{(k-1)}+\nub_j^{(k-1)})\big)
\bigg]
\bigg\|_2^2 + g_i(\xb_i) \bigg\}
\notag\\
&={\rm prox}_{g_i}^{\beta_i}\!\bigg[\textstyle  \xb_i^{(k-1)}-
\frac{1}{\beta_i}\Ab_i^T\nabla f_i(\Ab_i\xb_i^{(k-1)}) \notag\\
&~~~~~~~~~~~~~~~~~\textstyle - \frac{1}{2\beta_i|\mathcal{N}_i|}
\Eb_i^T\big(\frac{1}{c}(\Eb_i\xb_i^{(k-1)}\!-\!\frac{1}{N}\qb)- \frac{1}{c}\pb_{i}^{(k)} \notag \\
&~~~~~~~~~~~~~~~~~\textstyle + \sum_{j\in \mathcal{N}_i} (\nub_i^{(k-1)}+\nub_j^{(k-1)})\big)\bigg].
\end{align}}
We summarize the proposed IDC-ADMM in Algorithm \ref{table: IDC-ADMM}.

\begin{algorithm}[h!]
\caption{Proposed IDC-ADMM for solving {\sf (P2)}}
\begin{algorithmic}[1]\label{table: IDC-ADMM}{
\STATE
 {\bf Given} initial variables
$\xb_i^{(0)}\in \mathbb{R}^{K}$ and $\pb_{i}^{(0)}=\zerob$ for each agent $i$,  $i\in V$. Set $k=1.$
\REPEAT
\STATE  For all  $i\in V$ (in parallel), 

~~ {\small $\pb_{i}^{(k)}=\pb_{i}^{(k-1)}+{c}\sum_{j \in \Nc_i}(\nub_i^{(k-1)}-\nub_j^{(k-1)})$,
\begin{align}\label{eq: dual ADMM x122}
    &\xb_i^{(k)}={\rm prox}_{g_i}^{\beta_i}\!\bigg[\textstyle  \xb_i^{(k-1)}-
\frac{1}{\beta_i}\Ab_i^T\nabla f_i(\Ab_i\xb_i^{(k-1)}) \notag\\
&~~~~~~~~~~~~\textstyle - \frac{1}{2\beta_i|\mathcal{N}_i|}
\Eb_i^T\big(\frac{1}{c}(\Eb_i\xb_i^{(k-1)}\!-\!\frac{1}{N}\qb)- \frac{1}{c}\pb_{i}^{(k)} \notag \\
&~~~~~~~~~~~~\textstyle + \sum_{j\in \mathcal{N}_i} (\nub_i^{(k-1)}+\nub_j^{(k-1)})\big)\bigg],
\\
\label{eq: dual ADMM lambda 2122}
&\nub_i^{(k)} =  \textstyle\frac{1}{2|\mathcal{N}_i|}\big(
    \sum_{j\in \mathcal{N}_i} (\nub_i^{(k-1)}+\nub_j^{(k-1)})
    - \frac{1}{c}   \pb_{i}^{(k)} \notag \\
    &~~~~~~~~\textstyle +\frac{1}{c}(\Eb_i\xb_i^{(k)} -\frac{1}{N}\qb )
    \big).
\end{align}}
\STATE {\bf Set} $k=k+1.$
\UNTIL {a predefined stopping criterion (e.g., a maximum iteration number) is satisfied.}%
}
\end{algorithmic}
\end{algorithm}
The convergence property of IDC-ADMM is stated below.
\begin{Theorem} \label{thm: conv of inexact dadmm} Suppose that Assumptions \ref{assumption connected graph}, \ref{assumption convex prob}(b)
and \ref{assumption strongly convex fi} hold and
\begin{align}\label{eq: conv condition beta idcadmm}
\textstyle {\blue \beta_i} > \lambda_{\max}\bigg(\frac{L_{f,i}^2}{\sigma_{f,i}^2}\Ab_i^T\Ab_i +\frac{1}{2|\Nc_i|c}\Eb_i^T\Eb_i\bigg)~\forall i\in V.
\end{align}
{\blue Let $\xb^\star = [(\xb_1^\star)^T,\ldots,(\xb_N^\star)^T]^T$ denote an optimal solution to {\sf (P2)}, and let $\nub^\star\triangleq \nub_1^\star =\cdots=\nub_N^\star$ and $\{\ub_{ij}^\star,\vb_{ij}^\star\}$ denote
a pair of optimal primal and dual solutions to problem \eqref{consensus problem equi dual} (i.e., \eqref{eq: dual of consensus problem 2}).}

\begin{itemize}
\item [(a)] {\blue The sequence $\xb^{(k)} = [(\xb_1^{(k)})^T,\ldots,(\xb_N^{(k)})^T]^T$ generated from Algorithm \ref{table: IDC-ADMM} converges to $\xb^\star$ of {\sf (P2)} while
$\nub_1^{(k)},\ldots,\nub_N^{(k)}$ converge to a common point $\nub^\star$ of problem \eqref{eq: dual of consensus problem 2}.}

\item [(b)]
{If $\phi_i(\xb)=f_i(\Ab_i\xb)$, where $\Ab_i$ has full column rank, and
$\Eb_i$ has full row rank, for all $i \in V$,} then for some $0<\alpha<1$ and $\rho>0$, we have
\begin{align}
&\|\xb^{(k)}- \xb^\star\|_{\alpha\Mb+\frac{1}{2}\Pb}^2+ \frac{1}{c}\|\ub^{(k+1)} - \ub^\star\|^2_2\nonumber\\
& + \frac{c}{2}\|\nub^{(k)} - {\bf 1}_N\otimes \nub^\star\|^2_{(\Db+\Wb)\otimes \Ib_M}\to 0\; \mbox{linearly},
 \end{align}
where {\blue $\ub^{(k)}$ and $\ub^\star$ are defined similarly as in Theorem \ref{thm: conv of inexact pc admm},} $\Mb$ is defined in \eqref{eq: M}, 
and {
$\Pb\triangleq \Db_{\beta}-\frac{1}{2c}\blkdiag\{\frac{1}{|\Nc_1|}\Eb_1^T\Eb_1,\ldots,\frac{1}{|\Nc_N|}\Eb_N^T\Eb_N\}\succ \zerob$.}
%
%
\end{itemize}
\end{Theorem}
The proof is presented in Appendix \ref{proof of conv of inexact dadmm}.
Note that, in addition to the smooth and strongly convex objective function, IDC-ADMM also requires
matrices $\Eb_i$'s
to have full row rank in order to have a linear convergence rate.

%
%

\vspace{-0.3cm}
\section{Numerical Results}\label{sec: simulation}

In this section, we examine the numerical performance of
Algorithm 1 to 4 presented so far.
\vspace{-0.2cm}
\subsection{Performance of C-ADMM and IC-ADMM}\label{subsec: simulation C-ADMM}
To test C-ADMM (Algorithm \ref{table: C-ADMM}) and IC-ADMM (Algorithm \ref{table: IC-ADMM}), we considered the distributed RPD LR problem in \eqref{HPD regression} {with $\Psi_i(\yb; \Ab_i,\bb_i)$ in \eqref{eqn: logistic func} and $g_i(\yb)=\frac{\lambda}{N}\|\yb\|_1 + \eta(\yb)$, where $\lambda>0$ is a penalty parameter, and $\eta(\yb)$ is an indicator function specifying that the regression variables lie in a set $\Xc=\{\yb\in \mathbb{R}^K\mid |x_i|\leq a~\forall\; i\}$ for some $a>0$ (see Eqn. \eqref{eqn: indicator function}).} We considered a simple two image classification task. Specifically, we used the images D24 and D68 from the Brodatz data set (\url{http://www.ux.uis.no/~tranden/brodatz.html}) to generate the regression data matrix $\Ab$. We randomly extracted {$(NM)/2$} overlapping patches with dimension $\sqrt{K} \times \sqrt{K}$ from the two images, respectively, followed by vectorizing the $M$ patches into vectors and stacking all of them into an $M \times K$ matrix. The rows of the matrix were randomly shuffled and the resultant matrix was used as the data matrix $\Ab$. For the RPD LR problem \eqref{HPD regression}, we horizontally partitioned the matrix $\Ab$ into $N$ submatrices $\Ab_1,\ldots,\Ab_N$, each with dimension $M \times K$. These matrices were used as the training data. Note that each $\Ab_i$ contains patches from both images. The binary labels $\bb_i$'s then were generated accordingly with  $1$ for one image and $-1$ for the other. 
{The connected graph $\Gc$ was randomly generated following the same method as in \cite{YildizScag08}.}


To implement C-ADMM (Algorithm \ref{table: C-ADMM}), we employed the fast iterative shrinkage thresholding algorithm (FISTA) \cite{BeckFISTA2009,Boyd13Proximal} to solve subproblem \eqref{eqn: subproblem of pc admm} for each agent $i$. For \eqref{eqn: subproblem of pc admm}, the associated FISTA steps can be shown as
\ifconfver
\begin{subequations}\label{eqn: fista step 1 pc admm}
\begin{align}\label{eqn: fista step 1 pc admm 1}
  &\textstyle \tilde\yb_i^{(\ell)}
  =\max\bigg\{\!\!-\!a,\min\!\bigg\{\!a,\mathcal{S}\!\bigg[\zb_i^{(\ell-1)} - \rho_i^{(\ell)}\!
  \bigg[\!\Ab_i^T\nabla f_i(\Ab_i \zb_i^{(\ell-1)})  \notag \\
  &\textstyle \! +\! \pb_i^{(k)}\!+\! 2 c\!\! \sum_{j\in \Nc_i}\! (\zb_i^{(\ell-1)}\!\!-\!\frac{\yb_i^{(k-1)}+\yb_j^{(k-1)}}{2})\! \bigg],\!
  \frac{\lambda\rho^{(\ell)}_i}{N} \!\bigg]\!\bigg\}\!\bigg\},
   \\
  &\textstyle \zb_i^{(\ell)} = \tilde\yb_i^{(\ell)} + \frac{\ell-1}{\ell+2}(\tilde\yb_i^{(\ell)}-\tilde\yb_i^{(\ell-1)}),
\end{align}
\end{subequations}
\else
\begin{subequations}\label{eqn: fista step 1 pc admm}
\begin{align}\label{eqn: fista step 1 pc admm 1}
  &\textstyle \tilde\yb_i^{(\ell)}
  =\max\bigg\{\!\!-\!a,\min\!\bigg\{\!a,\mathcal{S}\!\bigg[\zb_i^{(\ell-1)} - \rho_i^{(\ell)}\!
  \bigg[\!\Ab_i^T\nabla f_i(\Ab_i \zb_i^{(\ell-1)})  \notag \\
  &~~~~~~~~~~~~~~~~~~~~\! +\! \pb_i^{(k)}\!+\! 2 c\!\! \sum_{j\in \Nc_i}\! (\zb_i^{(\ell-1)}\!\!-\!\frac{\yb_i^{(k-1)}+\yb_j^{(k-1)}}{2})\! \bigg],\!
  \frac{\lambda\rho^{(\ell)}_i}{N} \!\bigg]\!\bigg\}\!\bigg\},
   \\
  &\zb_i^{(\ell)} = \tilde\yb_i^{(\ell)} + \frac{\ell-1}{\ell+2}(\tilde\yb_i^{(\ell)}-\tilde\yb_i^{(\ell-1)}),
\end{align}
\end{subequations}
\fi
where $\ell$ denotes the inner iteration index of FISTA, $\rho^{(\ell)}_i>0$ is a step size and $\mathcal{S}$ is defined in \eqref{soft thresholding}.
The stopping criterion of \eqref{eqn: fista step 1 pc admm} was based on the PG residue ({\sf pgr})
$
 {\sf {pgr}}=\|\zb^{(\ell-1)}_i-
  \tilde \yb^{(\ell)}_i\|/(\rho^{(\ell)}_i\sqrt{K})
$ \cite{BeckFISTA2009,Boyd13Proximal}.
For obtaining a high-accuracy solution of \eqref{eqn: subproblem of pc admm}, one may set the stopping criterion as, e.g., ${\sf {\sf pgr}}<10^{-5}$. 
%
Suppose that FISTA stops at iteration $\ell_i{(k)}$. We then set $\yb_i^{(k)}=\tilde \yb_i^{(\ell_i{(k)})}$ as a solution to subproblem \eqref{eqn: subproblem of pc admm}.

For IC-ADMM (Algorithm 2), the corresponding step in \eqref{eqn: inexact ADMM step xi} 
is given by
\ifconfver
\begin{align}\label{eqn: inexact ADMM step xi logistic}
  \textstyle  \yb_i^{(k)} =&\max\bigg\{\!\!-a,\min\bigg\{\!a,\frac{1}{\gamma_i}\mathcal{S}\bigg[\beta\yb_i^{(k-1)}\!\! -\!\! \Ab_i^T\nabla f_i(\Ab_i \yb_i^{(k-1)})  \notag \\
   & ~ \textstyle \pb_i^{(k)}+ c \sum_{j\in \Nc_i} (\yb_i^{(k-1)}+\yb_j^{(k-1)}),\frac{\lambda}{N} \bigg]\bigg\}\bigg\}.
\end{align}
\else
\begin{align}\label{eqn: inexact ADMM step xi logistic}
   \yb_i^{(k)} =&\max\bigg\{\!\!-a,\min\bigg\{\!a,\frac{1}{\gamma_i}\mathcal{S}\bigg[\beta\yb_i^{(k-1)}\!\! -\!\! \Ab_i^T\nabla f_i(\Ab_i \yb_i^{(k-1)})  \notag \\
   & ~~~~~~~~~~~~~~~~~~~~~~- \pb_i^{(k)}+ c \sum_{j\in \Nc_i} (\yb_i^{(k-1)}+\yb_j^{(k-1)}),\frac{\lambda}{N} \bigg]\bigg\}\bigg\}.
\end{align}
\fi
{From
\eqref{eqn: subproblem of pc admm pp} and \eqref{eqn: fista step 1 pc admm}, the complexity of agent $i$ at iteration $k$ of C-ADMM is given by the order of $K+\ell_i{(k)}(2MK + 2K)$ if one counts only the multiplication operations; while from \eqref{eqn: subproblem of pc admm pp} and \eqref{eqn: inexact ADMM step xi logistic}, the per-iteration complexity of each agent in IC-ADMM is given by the order of $K+(2MK + 2K)$.
One can see that, for each agent $i$, the computational complexity of C-ADMM per iteration $k$ (we refer this as the ``ADMM iteration (ADMM Ite.)") is roughly $\ell_i{(k)}$ times that of IC-ADMM.}


The stopping criterion of Algorithms 1 and 2 was based on measuring the solution accuracy ${\sf acc} = ({{\sf obj}(\hat \yb^{(k)}) - {\sf obj}^\star}) /{{\sf obj}^\star}$ and
variable consensus error ${\sf cserr}={\sum_{i=1}^N\|\hat \yb^{(k)}-\yb^{(k)}_i\|^2_2}/{N}$, where  $\hat \yb^{(k)}=({\sum_{i=1}^N \yb^{(k)}_i})/{N}$, ${\sf obj}(\hat \yb^{(k)})$ denotes the objective value of \eqref{HPD regression} given $\yb=\hat \yb^{(k)}$, and ${\sf obj}^\star$ is the optimal value of \eqref{HPD regression} which was obtained by FISTA \cite{BeckFISTA2009,Boyd13Proximal} with a high solution accuracy of ${\sf {\sf pgr}}<10^{-6}$. The two algorithms were set to stop whenever {\sf acc} and {\sf cserr} are both smaller than preset target values.

In Table \ref{Table: results of pc admm2}(a), we considered a simulation example of $N=10$, $K=10,000$, $M=10$, $\lambda=0.1$ and $a=1$, and display the comparison results. {We not only present the required ADMM iterations but also the computation time per agent\footnote{The simulation was performed on a desktop computer with 8-core Intel 1.3GHz CPU and 8 GB RAM. All the algorithms were implemented by MATLAB codes.} (in second) of the two methods.}
The convergence curves of C-ADMM and IC-ADMM with respect to the ADMM iteration are also shown in Figs. \ref{fig: fig_pc-admm bb100}(a) and \ref{fig: fig_pc-admm bb100}(b). The stopping conditions are {\sf acc} $<10^{-4}$ and {\sf cserr} $<10^{-5}$. For C-ADMM, we considered two cases, one with the stopping condition of FISTA for solving subproblem \eqref{eqn: subproblem of pc admm} set to {\sf pgr} $<10^{-5}$ and the other with that set to {\sf pgr} $<10^{-4}$. The penalty parameter $c$ for C-ADMM was set to $c=0.03$ and the step size $\rho_i^{(\ell)}$ of FISTA (see \eqref{eqn: fista step 1 pc admm}) was set to a constant $\rho_i^{(\ell)}=0.1$. The penalty parameters $c$ and $\beta$ of IC-ADMM were set to $c=0.01$ and $\beta=1.2$. 
We observe from Table \ref{Table: results of pc admm2}(a) that IC-ADMM in general requires more ADMM iterations than C-ADMM; {however, the computation time is significantly smaller, as also illustrated in Figure \ref{fig: fig_pc-admm bb100}(c). Specifically, the computation time of IC-ADMM is around $44.56/2.14\approx 20.8$ times} smaller than that of C-ADMM ({\sf pgr} $<10^{-5}$). We also observe that C-ADMM ({\sf pgr} $<10^{-4}$) consumes a smaller computation time for achieving {\sf acc} $<10^{-4}$. However, the associated {\sf cserr} = $3.425\times 10^{-4}$ does not achieve the target value $10^{-5}$. In fact, C-ADMM ({\sf pgr} $<10^{-4}$) cannot reduce {\sf cserr} properly.
As one can see from Fig. \ref{fig: fig_pc-admm bb100}(b), the {\sf cserr} curve of
C-ADMM ({\sf pgr} $<10^{-4}$) keeps relatively high and does not decrease along the iterations.
{In Fig. \ref{fig: fig_pc-admm bb100}(a) and Fig. \ref{fig: fig_pc-admm bb100}(b),} we also plot the convergence curves of the consensus subgradient method in \cite{Angelia2010}, where the diminishing step size $10/k$ was used. As one can see, the consensus
subgradient method converges much slower than IC-ADMM.

In Table \ref{Table: results of pc admm2}(b), we considered another example with the network size increased to $N=50$. We set $c=0.004$ for C-ADMM and $\rho_i^{(\ell)}=0.1$ for FISTA; while for IC-ADMM, we set $c=0.008$ and $\beta=1.2$. {The computation times of C-ADMM and IC-ADMM under this setting are also shown in Fig. \ref{fig: fig_pc-admm bb100}(c).} We can observe similar comparison results from Table \ref{Table: results of pc admm2}(b) {and Fig. \ref{fig: fig_pc-admm bb100}(c).} Specifically, {the computation time of IC-ADMM is around 8.75 times} smaller than C-ADMM ({\sf pgr} $<10^{-5}$).
When considering a lower accuracy of {\sf pgr} $<10^{-4}$, it is found that C-ADMM cannot properly converge.




{To corroborating the linear convergence behavior of C-ADMM and IC-ADMM as claimed in Theorem \ref{thm: conv of inexact pc admm}(b)), we consider a problem instance of \eqref{HPD regression} with $\lambda=0$, $N=10$, $K=25$, $M=1,000$ and $a=10$.
We set $c=0.2$ for C-ADMM and $\rho_i^{(\ell)}=0.01$ and {\sf pgr} $<10^{-5}$ for FISTA; while for IC-ADMM, we set $c=1.2$ and $\beta=10$. The convergence curves are shown in Figure \ref{fig: fig_pc-admm linear}. One can see from this figure that both algorithms converge linearly under this setting.
}
\begin{table}[t]
  \caption{Comparison of C-ADMM and IC-ADMM}\vspace{-0.2cm}
    \begin{center}
    {\bf (a)} $N=10$, $ K=10,000$, $ M=10$, $\lambda=0.1$, $a=1$.\\
    \begin{tabular}{|c|c|c|c|}
    \hline
          & & & \\
       & \footnotesize {\bf C-ADMM}  & \footnotesize {\bf C-ADMM} &  \footnotesize {\bf IC-ADMM}\\
       & \scriptsize ({\sf pgr} $<10^{-5}$)  & \scriptsize ({\sf pgr} $<10^{-4}$) & \footnotesize  \\
    \hline\hline
    \scriptsize{\sf ADMM Ite.} & \scriptsize 810 &\scriptsize 675 & \scriptsize 2973 \\
    \hline
    {\scriptsize{\sf Compt. Time (sec)} }& {\scriptsize {\bf 44.56}} &\scriptsize {{\bf 17.86} }&{\scriptsize {\bf 2.14} }\\
    \hline
    \scriptsize{\sf acc}$<10^{-4}$ &\scriptsize $9.982\times 10^{-5}$ &\scriptsize $9.91\times 10^{-5}$ & \scriptsize$9.99\times 10^{-5}$ \\
    \hline
    \scriptsize{\sf cserr}$<10^{-5}$&\scriptsize $1.53\times 10^{-6}$ &  \scriptsize ${\bf3.425\times 10^{-4}}$ &\scriptsize $ {\bf 3.859\times 10^{-9}}$ \\
    \hline
    \end{tabular}

    \end{center}

%
    \begin{center}
    {\bf (b)}  $N=50$, $K=10,000$, $M=10$, $\lambda=0.15$, $a=1$.\\
    \begin{tabular}{|c|c|c|c|}
    \hline
       & & & \\
       & \footnotesize {\bf C-ADMM}  & \footnotesize {\bf C-ADMM} & \footnotesize  {\bf IC-ADMM} \\
       & \scriptsize ({\sf pgr} $<10^{-5}$)  & \scriptsize ({\sf pgr} $<10^{-4}$) & \footnotesize  \\
    \hline\hline
    \scriptsize{\sf ADMM Ite.} & \scriptsize 952 &\scriptsize N/A & \scriptsize 7,251 \\
    \hline
     {\scriptsize{\sf Compt. Time (sec)} }& {\scriptsize {\bf 81.72}} &\scriptsize N/A &{\scriptsize {\bf 9.33} }\\
    \hline
    \scriptsize{\sf acc}$<10^{-4}$ &\scriptsize $9.99\times 10^{-5}$ &\scriptsize N/A &\scriptsize $9.999\times 10^{-5}$ \\
    \hline
    \scriptsize{\sf cserr}$<10^{-5}$& \scriptsize$1.305\times 10^{-7}$ &\scriptsize N/A & \scriptsize$ {\bf 1.169\times 10^{-10}}$ \\
    \hline
    \end{tabular}

    \end{center}

\label{Table: results of pc admm2}\vspace{-0.1cm}
\end{table}

\ifconfver
\begin{figure}[!t]
\begin{center}
{\subfigure[][]{\resizebox{.38\textwidth}{!}
{\includegraphics{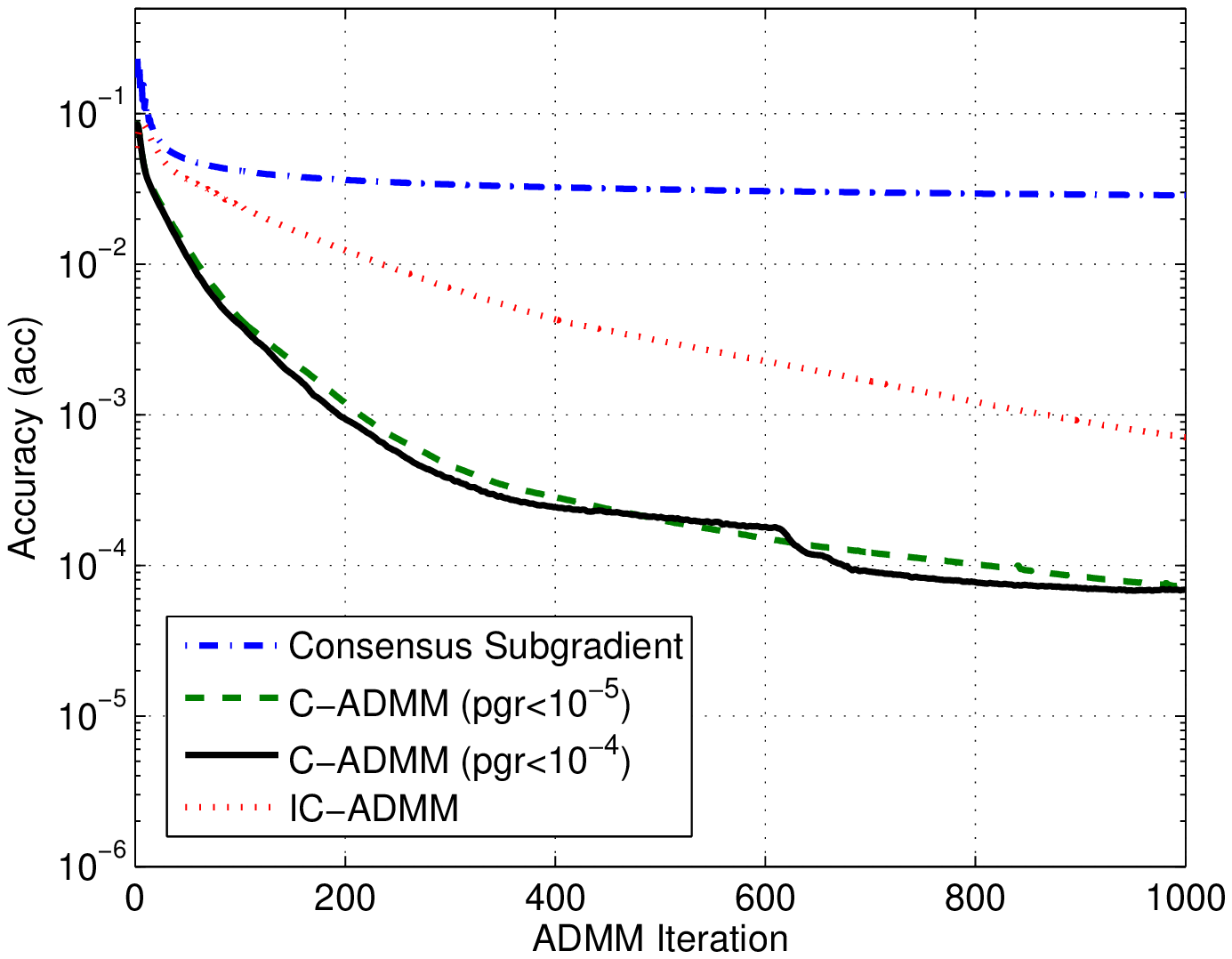}}}
}
\hspace{-1pc}
{\subfigure[][]{\resizebox{.38\textwidth}{!}{\includegraphics{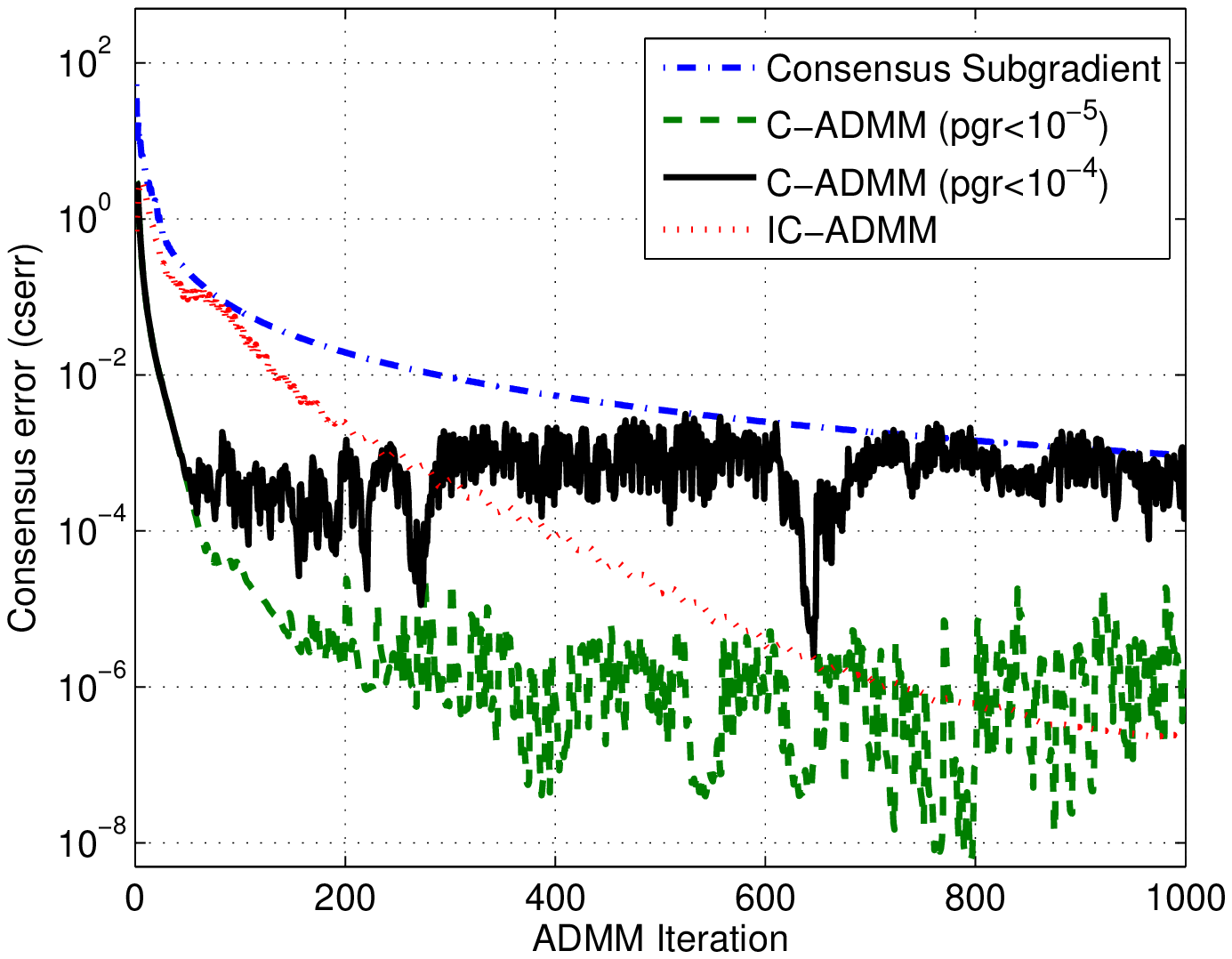}}}
 }
 \hspace{-1pc}
{
\subfigure[][]{\resizebox{.38\textwidth}{!}{\includegraphics{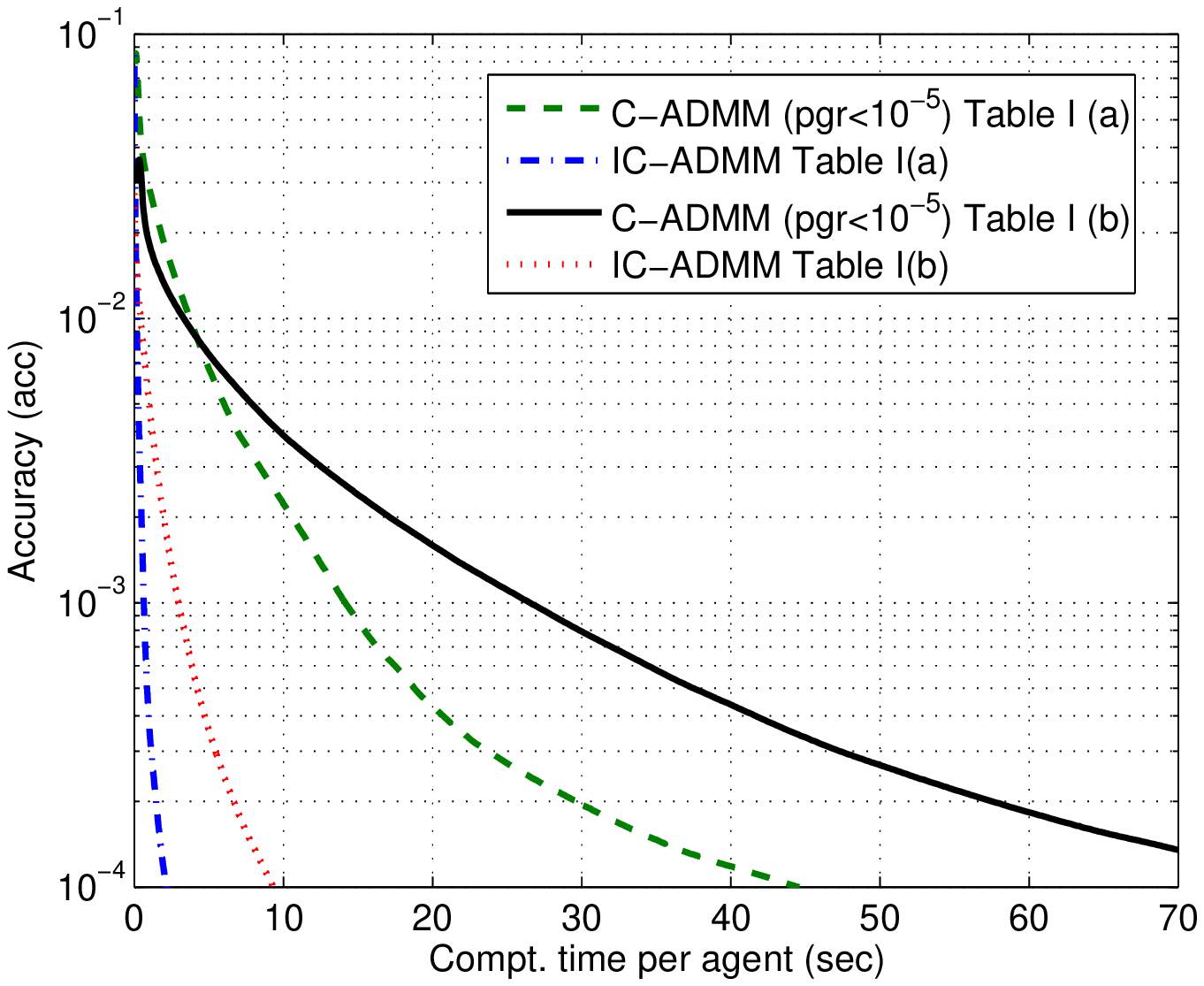}}}
 }
\end{center}\vspace{-0.4cm}
\caption{Convergence curves of C-ADMM and IC-ADMM.}
\vspace{-0.6cm}\label{fig: fig_pc-admm bb100}
\end{figure}

\begin{figure}[!t]
\begin{center}
{\resizebox{.38\textwidth}{!}
{\includegraphics{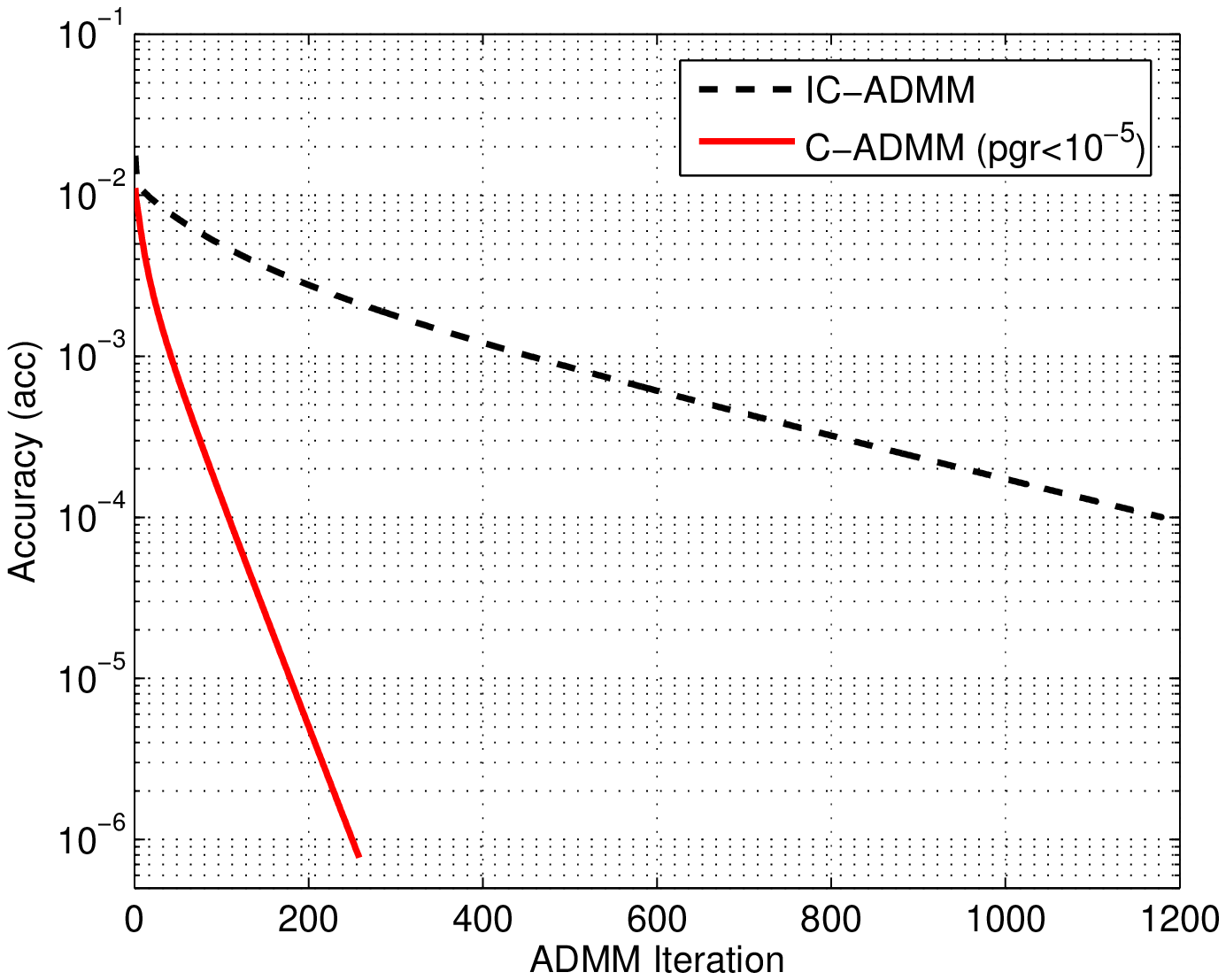}}}
\end{center}\vspace{-0.2cm}
\caption{Convergence curves of C-ADMM and IC-ADMM.}
\vspace{-0.2cm}\label{fig: fig_pc-admm linear}
\end{figure}

\else
\begin{figure}[!t]
\begin{center}
{\subfigure[][]{\resizebox{.47\textwidth}{!}
{\includegraphics{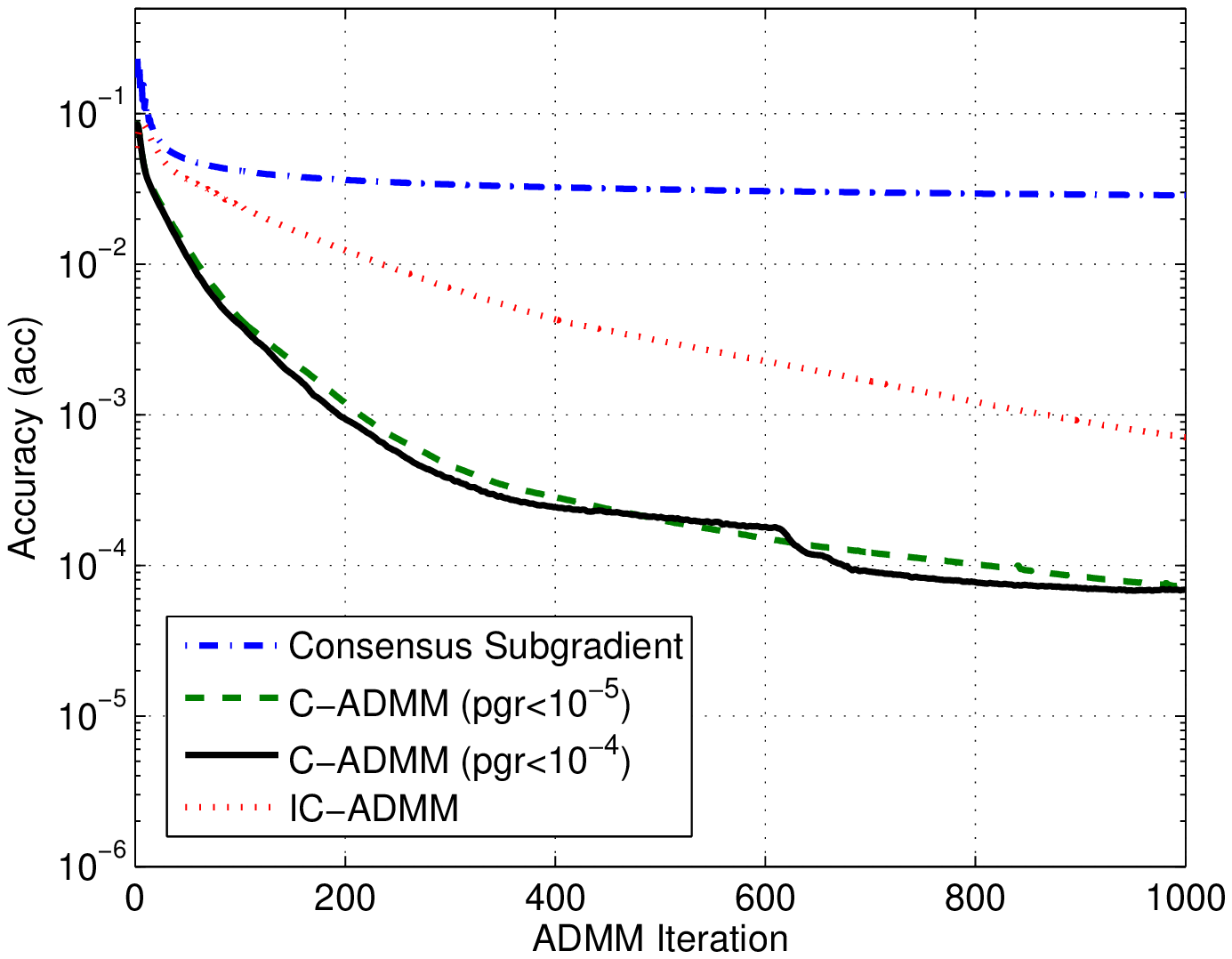}}}
}
\hspace{-1pc}
{\subfigure[][]{\resizebox{.47\textwidth}{!}{\includegraphics{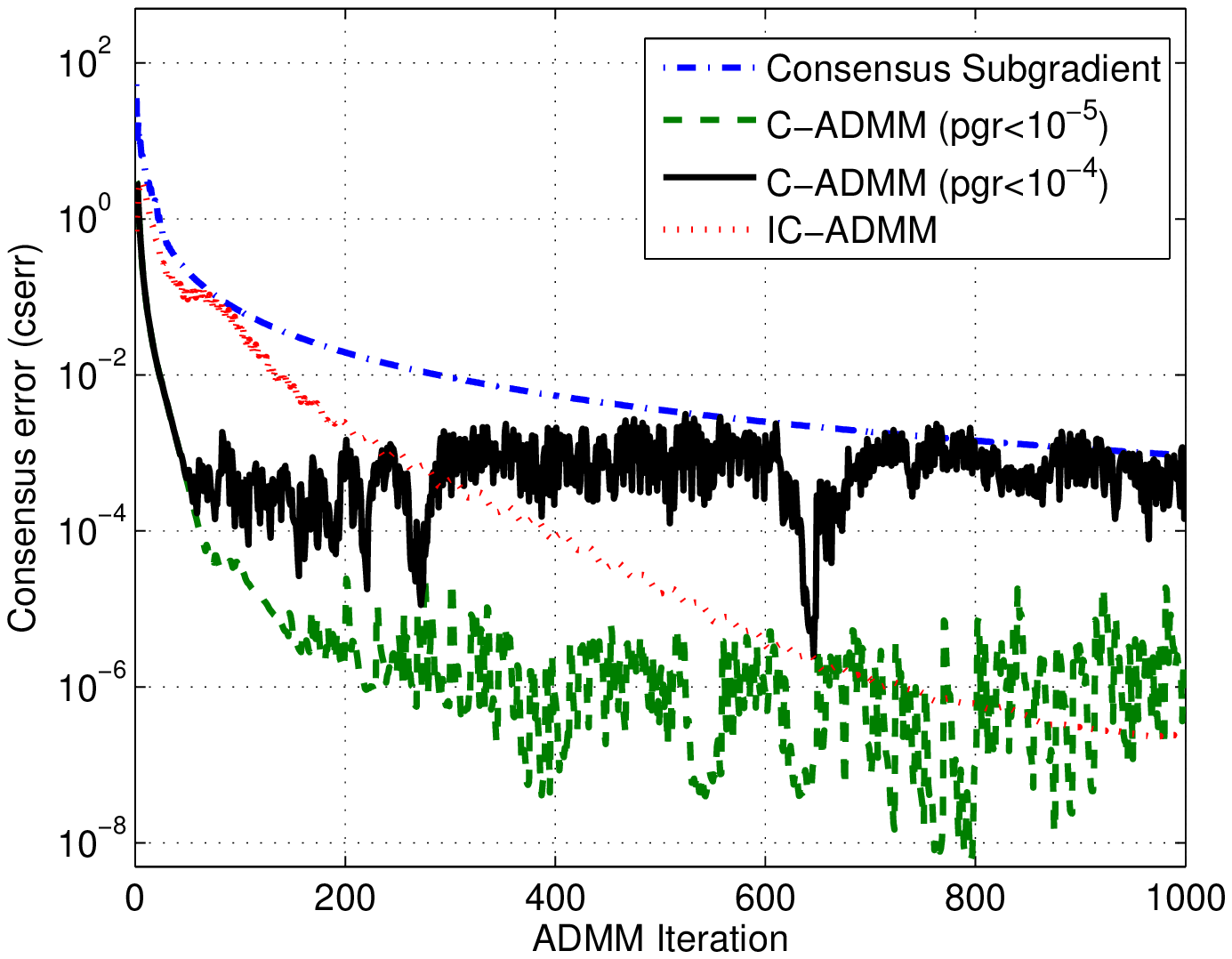}}}
 }
 \hspace{-1pc}
{
\subfigure[][]{\resizebox{.47\textwidth}{!}{\includegraphics{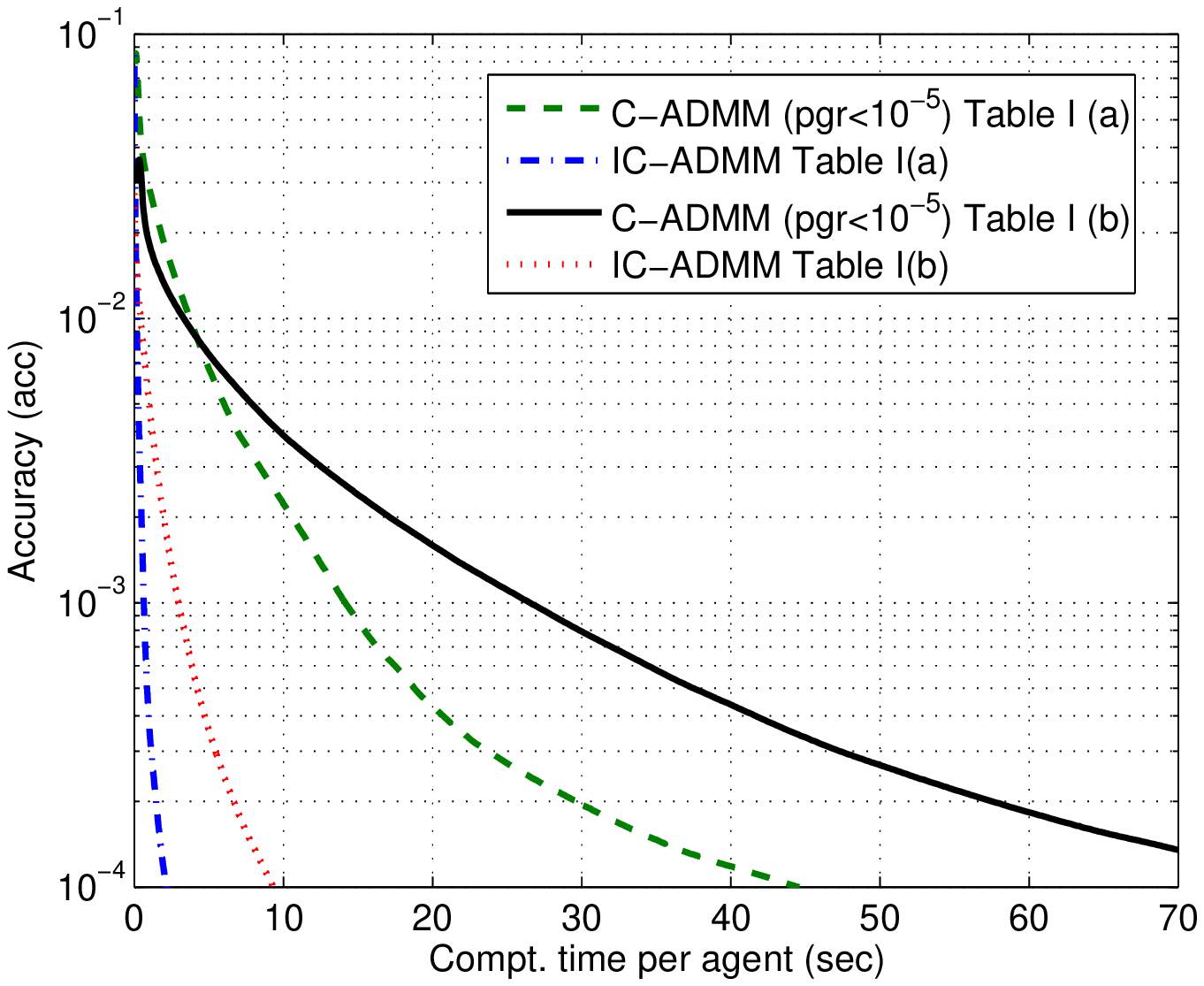}}}
 }
\end{center}\vspace{-0.4cm}
\caption{Convergence curves of C-ADMM and IC-ADMM.}
\vspace{-0.6cm}\label{fig: fig_pc-admm bb100}
\end{figure}

\begin{figure}[!t]
\begin{center}
{\resizebox{.47\textwidth}{!}
{\includegraphics{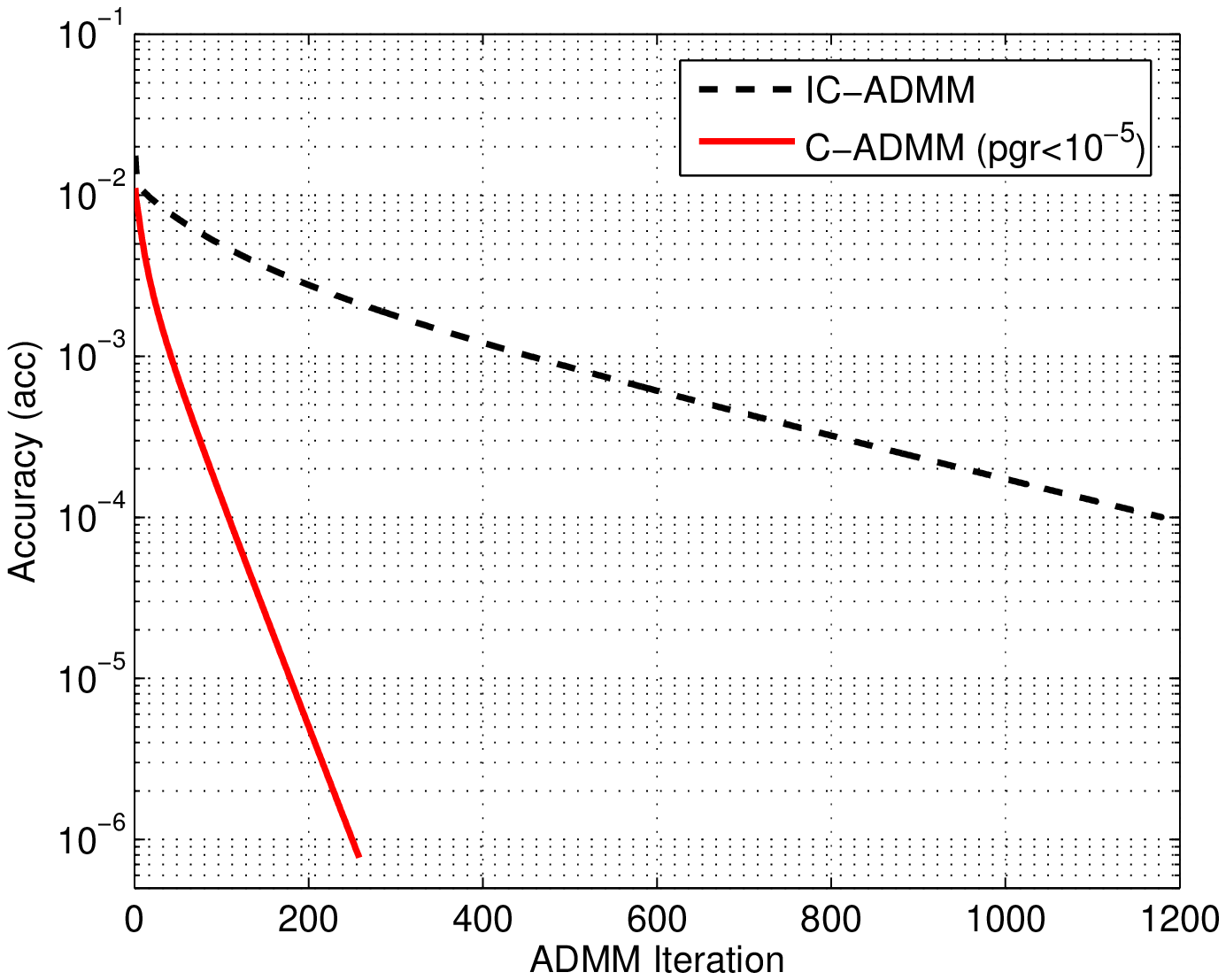}}}
\end{center}\vspace{-0.2cm}
\caption{Convergence curves of C-ADMM and IC-ADMM.}
\vspace{-0.4cm}\label{fig: fig_pc-admm linear}
\end{figure}
\fi
\vspace{-0.3cm}
\subsection{Performance of DC-ADMM and IDC-ADMM}\label{subsec: simulation DC-ADMM}

We examine the performance of DC-ADMM (Algorithm \ref{table: DC-ADMM}) and IDC-ADMM (Algorithm \ref{table: IDC-ADMM}) by considering the distributed CPD LR problem in \eqref{VPD regression}, {with $\Psi_i(\xb_i; \Eb_i,\bb)$ in \eqref{eqn: logistic func2} and $g_i(\xb_i)=\lambda\|\xb_i\|_1$. Each variable $\xb_i$ is subject to the constraint set $ \Xc_i=\{\xb_i\in {\blue \mathbb{R}^{K/N}}~|~|[\xb_{i}]_j|\leq a~\forall j\}$ for some $a>0$.}
DC-ADMM and IDC-ADMM were applied to handle the associated problem \eqref{VPD logstic regression}.
The regression data matrix {$\Eb=[\Eb_1,\ldots,\Eb_N]$ was generated following the same way as generating $\Ab$} in Section \ref{subsec: simulation C-ADMM}. {To implement DC-ADMM, we employed FISTA \cite{BeckFISTA2009,Boyd13Proximal} to solve subproblem \eqref{eq: dual ADMM x1} and the solution accuracy was measured by the PG residue 
of FISTA.} %


In Table \ref{Table: results of dadmm}(a), we show the comparison results for an example of $N=50$, $K=200$, $M=100$, $\lambda=0.05$ and $a=10$. The convergence curves are also shown in Figs. \ref{fig: fig_pc-admm bb16}(a) to \ref{fig: fig_pc-admm bb16}(c). It was set $c=0.05$ for DC-ADMM and the step size of FISTA $\rho_i^{(\ell)}$ was determined based on a line search rule \cite{Boyd13Proximal}.
We see from Table \ref{Table: results of dadmm}(a) that, for achieving {\sf acc} $<10^{-4}$, DC-ADMM ({\sf pgr} $<10^{-5}$) took 329 ADMM iterations whereas IDC-ADMM took 10,814 iterations. {However, the computation time of
DC-ADMM ({\sf pgr} $<10^{-5}$) is around $42.78/1.92\approx 22.28$ times} higher than IDC-ADMM. When one reduce the solution accuracy of FISTA for solving subproblem \eqref{eq: dual ADMM x1} to
{\sf pgr} $<10^{-4}$, DC-ADMM cannot reach the high accuracy of {\sf acc} $<10^{-4}$, as observed in Fig. \ref{fig: fig_pc-admm bb16}(a). From Fig. \ref{fig: fig_pc-admm bb16}(b), one can see that DC-ADMM converges much faster than IDC-ADMM with respect to the ADMM iterations. However, as shown from Fig. \ref{fig: fig_pc-admm bb16}(c), the comparison result is reversed when one counts {the computation times.}

In Table \ref{Table: results of dadmm}(b), we considered another example with {\blue $K$ increased to $800$.} We set $c=0.05$ for DC-ADMM, and set $c=0.08$ and $\beta=5$ for IDC-ADMM. From Table \ref{Table: results of dadmm}(b) and Figs. \ref{fig: fig_pc-admm bb16}(b) and \ref{fig: fig_pc-admm bb16}(c), one can observe similar results. 


\begin{table}[t]
  \caption{Comparison of DC-ADMM and IDC-ADMM}\vspace{-0.2cm}
    \begin{center}
   {\bf (a)} $N=50$, $K=200$, $M=100$, $\lambda=0.05$, $a=10$.\\
    \begin{tabular}{|c|c|c|c|}
    \hline
       & & & \\
       & \footnotesize {\bf DC-ADMM}  & \footnotesize {\bf DC-ADMM} & \footnotesize  {\bf IDC-ADMM} \\
       & \scriptsize ({\sf pgr} $<10^{-5}$)  & \scriptsize ({\sf pgr} $<10^{-4}$) & \footnotesize  \\
    \hline
    \hline
    \scriptsize{\sf ADMM Ite.} & \scriptsize 329 &\scriptsize N/A & \scriptsize 10814 \\
    \hline
     {\scriptsize{\sf Compt. Time (sec)} }& {\scriptsize {\bf 42.78}} &\scriptsize N/A &{\scriptsize {\bf 1.92} }\\
    \hline
    \scriptsize{\sf acc}$<10^{-4}$ &\scriptsize $9.928\times 10^{-5}$ &\scriptsize N/A &\scriptsize $9.997\times 10^{-5}$\\
    \hline
    \end{tabular}

    \end{center}

    \begin{center}
   {\bf (b)} $N=50$, $K=800$, $M=100$, $\lambda=0.01$, $a=20$.\\
    \begin{tabular}{|c|c|c|c|}
    \hline
       & & & \\
       & \footnotesize {\bf DC-ADMM}  & \footnotesize {\bf DC-ADMM} & \footnotesize  {\bf IDC-ADMM} \\
       & \scriptsize ({\sf pgr} $<10^{-5}$)  & \scriptsize ({\sf pgr} $<10^{-4}$) & \footnotesize  \\
    \hline
    \hline
    \scriptsize{\sf ADMM Ite.} & \scriptsize 475 &\scriptsize N/A & \scriptsize 38728 \\
    \hline
     {\scriptsize{\sf Compt. Time (sec)} }& {\scriptsize {\bf 427.73}} &\scriptsize N/A &{\scriptsize {\bf 18.07} }\\
    \hline
    \scriptsize{\sf acc}$<10^{-4}$ &\scriptsize $9.777\times 10^{-5}$ &\scriptsize N/A &\scriptsize $9.999\times 10^{-5}$\\
    \hline
    \end{tabular}

    \end{center}
\label{Table: results of dadmm}\vspace{-0.5cm}
\end{table}


\vspace{-0.2cm}
\section{Conclusions}\label{sec: conclusions}
\vspace{-0.1cm}
In this paper, we have presented ADMM based distributed optimization methods for solving problems {\sf (P1)} and {\sf (P2)} in multi-agent networks.
In particular, aiming at reducing the computational complexity of C-ADMM for solving large-scale instances of {\sf (P1)} with complicated objective functions, we have proposed the IC-ADMM method (Algorithm \ref{table: IC-ADMM}) where agents perform one PG update only at each iteration. For {\sf (P2)}, we have proposed the DC-ADMM method (Algorithm \ref{table: DC-ADMM}) and its complexity reduced counterpart IDC-ADMM (Algorithm \ref{table: IDC-ADMM}). Preliminary numerical results based on the distributed LR problems \eqref{HPD regression} and \eqref{VPD logstic regression} have shown that {\blue the proposed methods converge faster than the consensus subgradient method. Moreover, both IC-ADMM and IDC-ADMM require more ADMM iterations than C-ADMM and DC-ADMM, but the traded computational complexity reduction is significant.} 

\ifconfver
\begin{figure}[!t]
\begin{center}
{\subfigure[][]{\resizebox{.38\textwidth}{!}
{\includegraphics{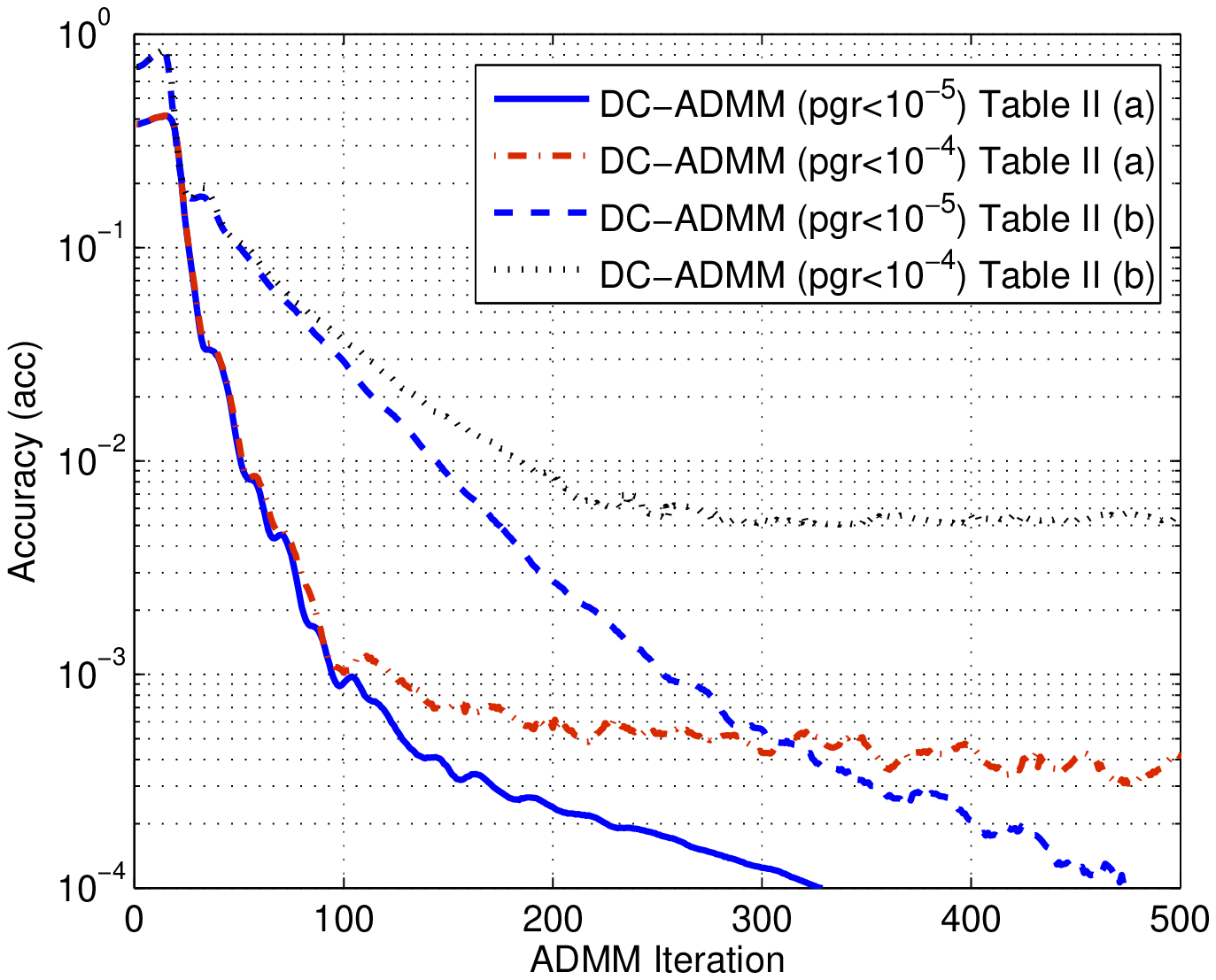}}}
}
\hspace{-1pc}
{\subfigure[][]{\resizebox{.38\textwidth}{!}{\includegraphics{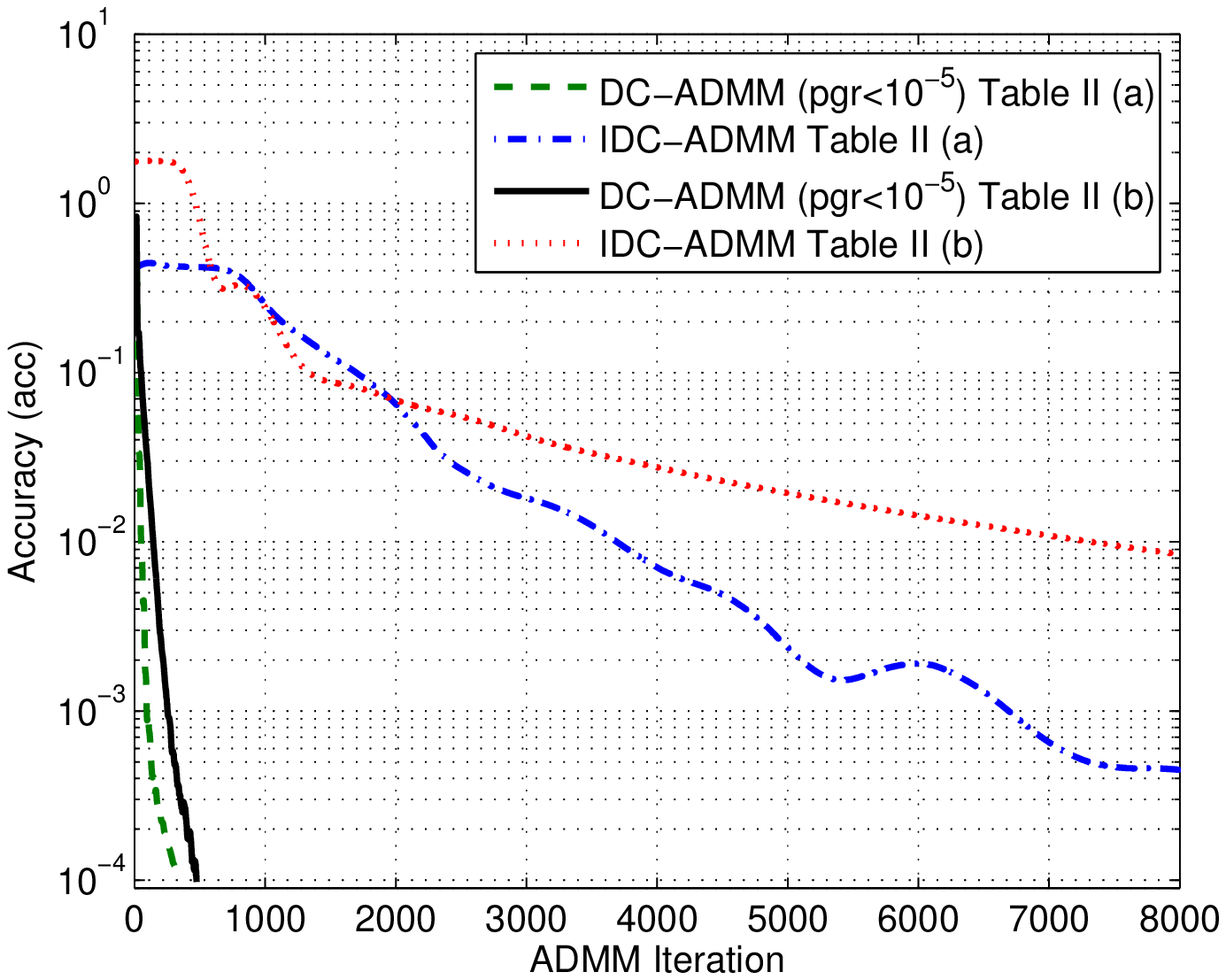}}}
 }
 \hspace{-1pc}
{\subfigure[][]{\resizebox{.38\textwidth}{!}{\includegraphics{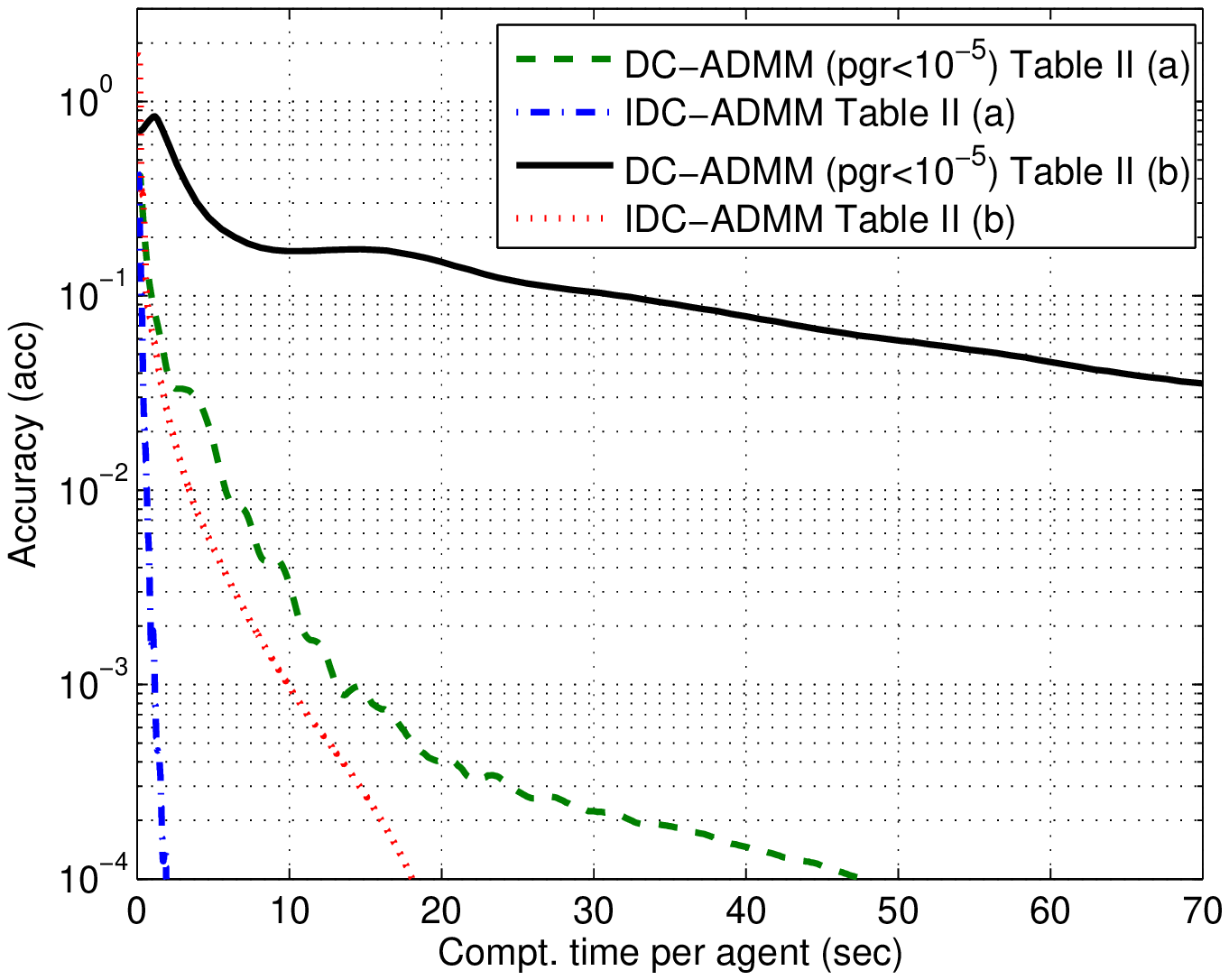}}}
 }
\end{center}\vspace{-0.4cm}
\caption{Convergence curves of DC-ADMM and IDC-ADMM.}
\vspace{-0.6cm}\label{fig: fig_pc-admm bb16}
\end{figure}
\else
\begin{figure}[!t]
\begin{center}
{\subfigure[][]{\resizebox{.47\textwidth}{!}
{\includegraphics{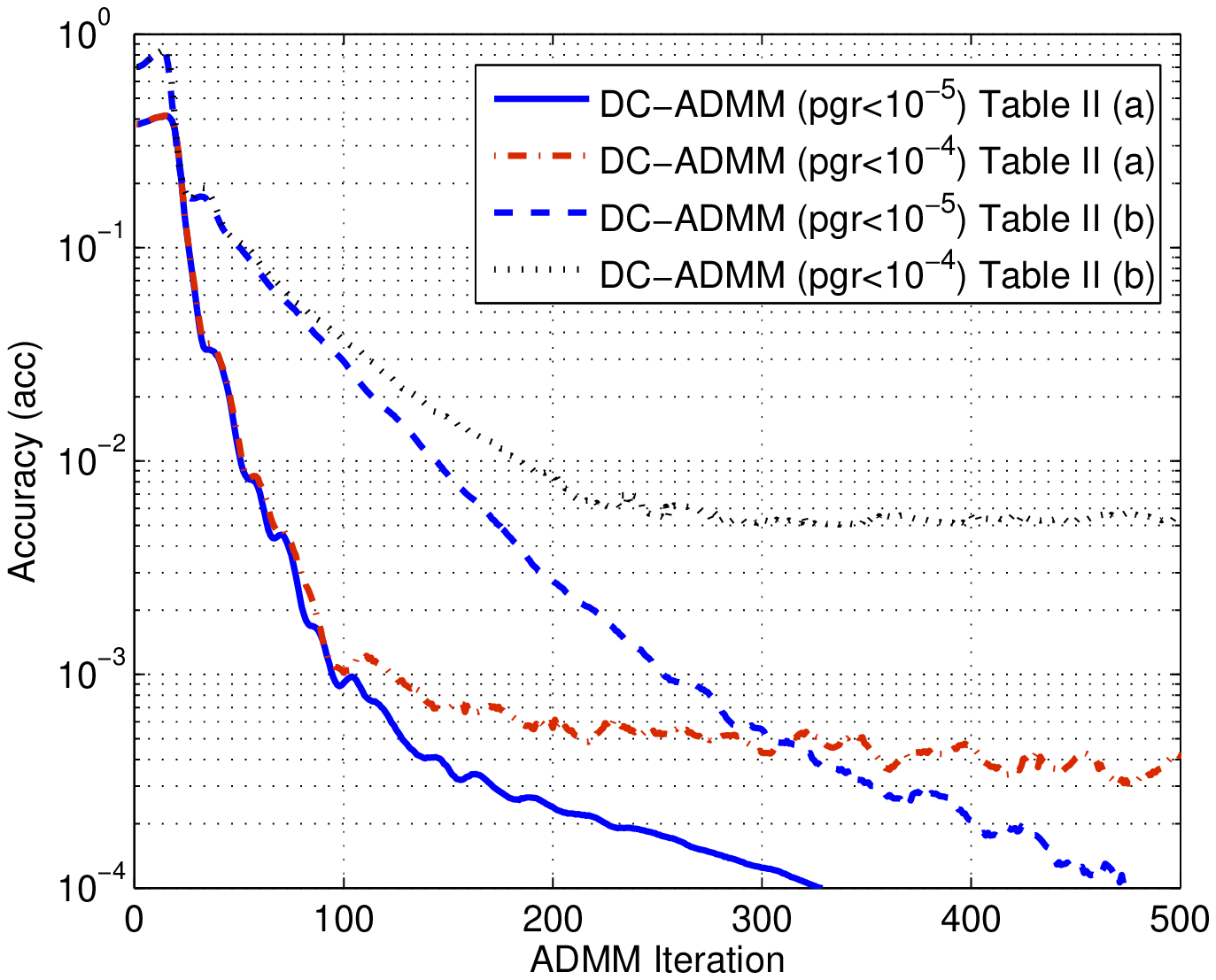}}}
}
\hspace{-1pc}
{\subfigure[][]{\resizebox{.47\textwidth}{!}{\includegraphics{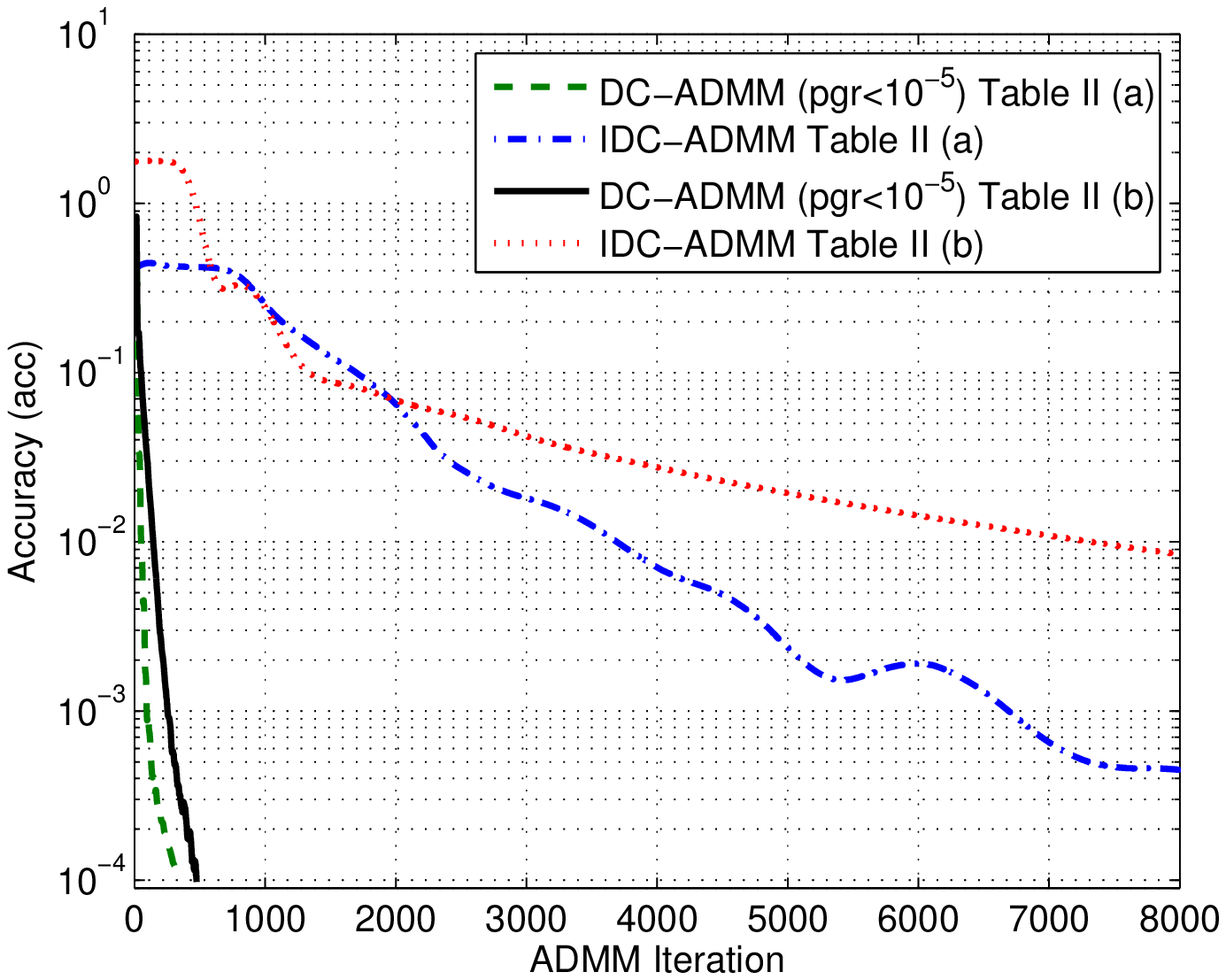}}}
 }
 \hspace{-1pc}
{\subfigure[][]{\resizebox{.47\textwidth}{!}{\includegraphics{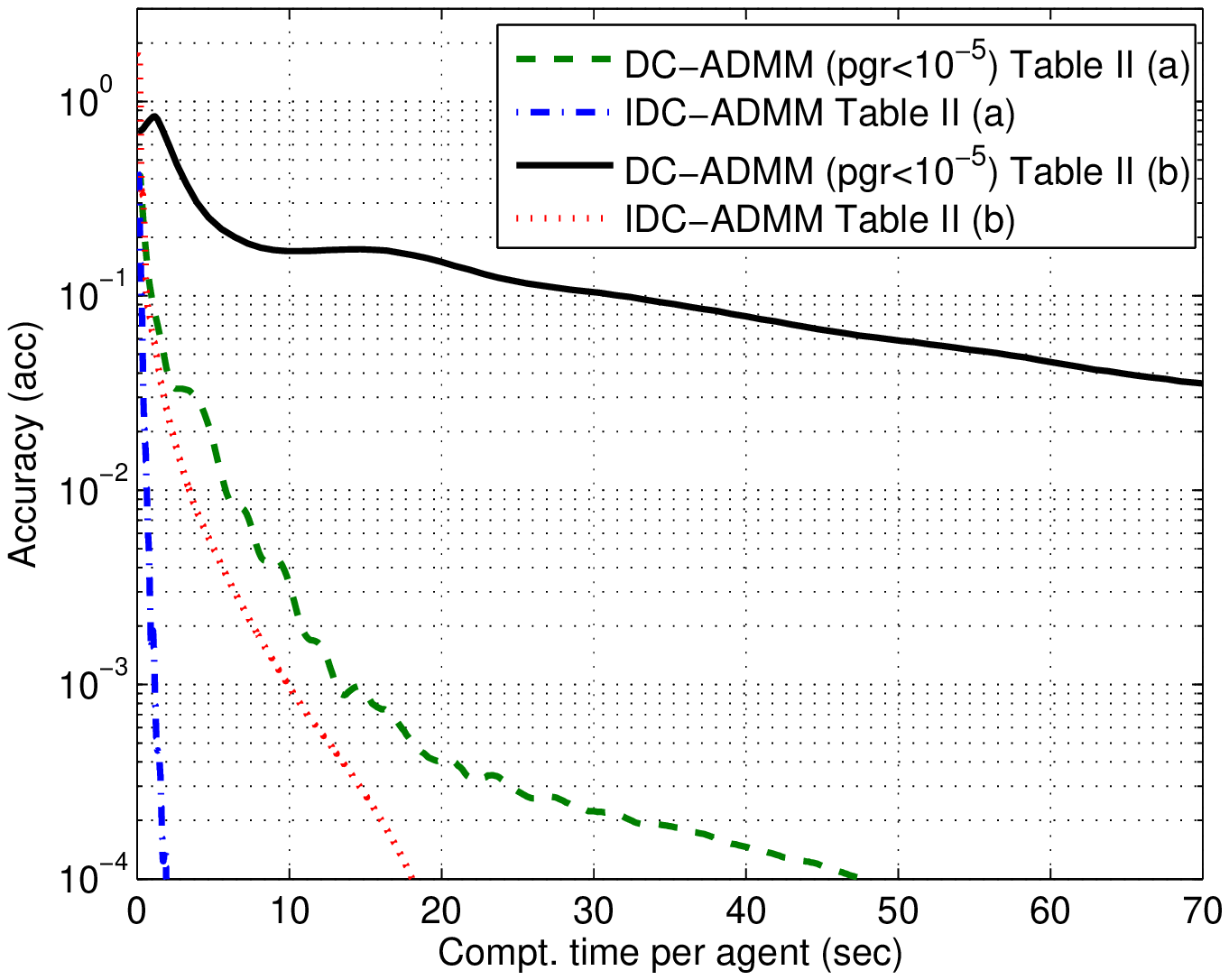}}}
 }
\end{center}\vspace{-0.4cm}
\caption{Convergence curves of DC-ADMM and IDC-ADMM.}
\vspace{-0.6cm}\label{fig: fig_pc-admm bb16}
\end{figure}
\fi


\vspace{-0.2cm}
\appendices {\setcounter{equation}{0}
\renewcommand{\theequation}{A.\arabic{equation}}

\section{Proof of Theorem \ref{thm: conv of inexact pc admm}}
\label{proof of conv of inexact pc-admm}
{\bf Proof of Theorem \ref{thm: conv of inexact pc admm}(a):}
Let {$\tilde \yb^\star \triangleq  [(\yb_1^\star)^T,\ldots,(\yb_N^\star)^T]^T$} and $\{\ub_{ij}^\star,\vb_{ij}^\star,j\in \mathcal{N}_i\}_{i=1}^N$ be a pair of
optimal primal and dual solutions to problem \eqref{consensus problem equi}. Then they satisfy the following Karush-Kuhn-Tucker (KKT) conditions: $\forall i\in V$,
\begin{align}
  &\textstyle \Ab_i^T\nabla f_i(\Ab_i\yb_i^\star) + \partial g_i(\yb_i^\star) +
  \sum_{j\in \mathcal{N}_i} (\ub_{ij}^\star + \vb_{ji}^\star) =\zerob, \label{eq: KKT1}\\
  & \yb_i^\star = \yb_j^\star~\forall j\in \mathcal{N}_i, \label{eq: KKT2}\\
  & \ub_{ij}^\star + \vb_{ij}^\star =\zerob,~\forall j\in \mathcal{N}_i, \label{eq: KKT3}
\end{align}
{where $\partial g_i(\yb_i^\star)$ denotes the subgradient of $g_i$ at $\yb_i^\star$.} Under Assumption 1, \eqref{eq: KKT2} implies that $\yb^\star\triangleq  \yb_1^\star=\cdots=\yb_N^\star$ {and $\tilde \yb^\star = \mathbf{1}_N \otimes \yb^\star$,} i.e., consensus among agents is reached, and thus
$\yb^\star$ is optimal to the original problem {\sf (P1)}.

{By recalling that $\pb_i^{(k)}= \sum_{j\in \Nc_i} ( \ub_{ij}^{(k)} + \vb_{ji}^{(k)} )$ $\forall i\in V $,
and by the optimality condition of \eqref{eqn: inexact ADMM step xi 2} \cite{BK:BoydV04}, we have that
\begin{align}
  \zerob&=\Ab_i^T\nabla f_i(\Ab_i\yb_i^{(k-1)})+ \beta_i (\yb_i^{(k)} - \yb_i^{(k-1)})
   +\partial g_i(\yb_i^{(k)})
   \notag
   \\
  &~~~~~\textstyle  +
  \sum_{j\in \mathcal{N}_i} (\ub_{ij}^{(k)} + \vb_{ji}^{(k)} ) \notag \\
  &~~~~~ \textstyle + 2c \sum_{j\in \mathcal{N}_i}\big(\yb_i^{(k)}  - \frac{\yb_i^{(k-1)} + \yb_j^{(k-1)}  }{2} \big).
  \label{eq: proof 1.0}
\end{align}
By combining \eqref{eq: proof 1.0} with \eqref{eq: KKT1}, one obtains
\begin{align}
  &\Ab_i^T\nabla f_i(\Ab_i\yb_i^{(k-1)}) - \Ab_i^T\nabla f_i(\Ab_i\yb^\star) + \beta_i (\yb_i^{(k)} - \yb_i^{(k-1)})
     \notag
     \\
  &~\textstyle +\partial g_i(\yb_i^{(k)}) -\partial g_i(\yb^\star)  +
  \sum_{j\in \mathcal{N}_i} (\ub_{ij}^{(k)} + \vb_{ji}^{(k)} -\ub_{ij}^\star - \vb_{ji}^\star ) \notag \\
  &~~~~~ \textstyle + 2c \sum_{j\in \mathcal{N}_i}\big(\yb_i^{(k)}  - \frac{\yb_i^{(k-1)} + \yb_j^{(k-1)}  }{2} \big)=\zerob.
  \label{eq: proof 1}
\end{align}}
\ifconfver
Adding and subtracting $\Ab_i^T\nabla f_i(\Ab_i\yb_i^{(k)})$ in the left hand side (LHS) of \eqref{eq: proof 1} followed by multiplying $(\yb_i^{(k)}- \yb^\star)$ on both sides yields \eqref{eq: proof 2} on the top of the next page.
\begin{figure*}[t]
\begin{align}\label{eq: proof 2}
 &(\nabla f_i(\Ab_i\yb_i^{(k-1)})-\nabla f_i(\Ab_i\yb_i^{(k)}))^T\Ab_i(\yb_i^{(k)}-\yb^\star)+
  \beta_i (\yb_i^{(k)} - \yb_i^{(k-1)})^T(\yb_i^{(k)}-\yb^\star)+(\nabla f_i(\Ab_i\yb_i^{(k)})
   \notag \\
  &~-\nabla f_i(\Ab_i\yb^\star))^T\Ab_i(\yb_i^{(k)}-\yb^\star)+(\partial g_i(\yb_i^{(k)}) - \partial g_i(\yb^\star))^T(\yb_i^{(k)}-\yb^\star)
  \notag \\
  &~+
  \textstyle\sum_{j\in \mathcal{N}_i} (\ub_{ij}^{(k)} - \ub_{ij}^\star+ \vb_{ji}^{(k)} - \vb_{ji}^\star)^T(\yb_i^{(k)}-\yb^\star)+ 2c \sum_{j\in \mathcal{N}_i}\bigg(\yb_i^{(k)}  - \frac{\yb_i^{(k-1)} + \yb_j^{(k-1)}  }{2} \bigg)^T(\yb_i^{(k)}\!-\yb^\star) =\zerob.
\end{align}
\hrulefill
\end{figure*}
\else
Adding and subtracting $\Ab_i^T\nabla f_i(\Ab_i\yb_i^{(k)})$ in the left hand side (LHS) of \eqref{eq: proof 1} followed by multiplying $(\yb_i^{(k)}- \yb^\star)$ on both sides yields 
\begin{align}\label{eq: proof 2}
 &(\nabla f_i(\Ab_i\yb_i^{(k-1)})-\nabla f_i(\Ab_i\yb_i^{(k)}))^T\Ab_i(\yb_i^{(k)}-\yb^\star)+
  \beta_i (\yb_i^{(k)} - \yb_i^{(k-1)})^T(\yb_i^{(k)}-\yb^\star)\notag \\
  &~+(\nabla f_i(\Ab_i\yb_i^{(k)})-\nabla f_i(\Ab_i\yb^\star))^T\Ab_i(\yb_i^{(k)}-\yb^\star)
   \notag \\
  &~+(\partial g_i(\yb_i^{(k)}) - \partial g_i(\yb^\star))^T(\yb_i^{(k)}-\yb^\star)
  +
  \textstyle\sum_{j\in \mathcal{N}_i} (\ub_{ij}^{(k)} - \ub_{ij}^\star+ \vb_{ji}^{(k)} - \vb_{ji}^\star)^T(\yb_i^{(k)}-\yb^\star)
  \notag \\
  &~\textstyle+ 2c \sum_{j\in \mathcal{N}_i}\bigg(\yb_i^{(k)}  - \frac{\yb_i^{(k-1)} + \yb_j^{(k-1)}  }{2} \bigg)^T(\yb_i^{(k)}\!-\yb^\star) =\zerob.
\end{align}
\fi
Note that the first term on the LHS of \eqref{eq: proof 2} can be lower bounded as
\begin{align}\label{eq: proof 3}
\!\!\!\! &\textstyle (\nabla f_i(\Ab_i\yb_i^{(k-1)})-\nabla f_i(\Ab_i\yb_i^{(k)}))^T\Ab_i(\yb_i^{(k)}-\yb^\star)\notag \\
 &\textstyle \geq \frac{-1}{2\rho_i} \|\nabla f_i(\Ab_i\yb_i^{(k-1)})-\nabla f_i(\Ab_i\yb_i^{(k)})\|^2_2
 \notag \\
 &~~~~~~~~~~~~~~~~~~~~~~ \textstyle~~~~~-\frac{\rho_i}{2}\|\yb_i^{(k)}-\yb^\star\|_{\Ab_i^T\Ab_i}^2 \notag\\
 &\textstyle\geq \!\frac{-L_{f,i}^2}{2\rho_i} \|\yb_i^{(k-1)}\!-\yb_i^{(k)}\|_{\Ab_i^T\Ab_i}^2 \!-\! \frac{\rho_i}{2}\|\yb_i^{(k)}-\yb^\star\|_{\Ab_i^T\Ab_i}^2
\end{align}
for any $\rho_i>0$, where the second inequality is due to \eqref{eq: lipschitz gradient of f} in Assumption \ref{assumption strongly convex fi}. By the strong convexity of $f_i$ and convexity of $g_i$, the third and fourth terms of \eqref{eq: proof 2} can respectively be lower bounded as
\begin{align}\label{eq: proof 4}
&\textstyle (\nabla f_i(\Ab_i\yb_i^{(k)})-\nabla f_i(\Ab_i\yb^\star))^T\Ab_i(\yb_i^{(k)}-\yb^\star)\notag \\
  &\textstyle ~~~~~~~~~~~~~~~~~~~~~~~~~~~~~~~~\geq \sigma_{f,i}^2 \|\yb_i^{(k)}-\yb^\star\|_{\Ab_i^T\Ab_i}^2,
\\
&\textstyle (\partial g_i(\yb_i^{(k)}) - \partial g_i(\yb^\star))^T(\yb_i^{(k)}-\yb^\star) \geq 0.
\label{eq: proof 5}
\end{align}
Moreover, it follows from \eqref{eqn: admm steps for P1 u} and \eqref{eqn: admm steps for P1 v} that the fifth term of
\eqref{eq: proof 2} can be expressed as
\begin{align}
&\textstyle\sum_{j\in \mathcal{N}_i} (\ub_{ij}^{(k)} - \ub_{ij}^\star+ \vb_{ji}^{(k)} - \vb_{ji}^\star)^T(\yb_i^{(k)}-\yb^\star)
\notag
\end{align}
\begin{align}\label{eq: proof 6}
&\textstyle =\sum_{j\in \mathcal{N}_i} (\ub_{ij}^{(k+1)} - \ub_{ij}^\star + \vb_{ji}^{(k+1)} - \vb_{ji}^\star)^T(\yb_i^{(k)}-\yb^\star)
 \notag \\
 &\blue~~~~~~\textstyle-2c \sum_{j\in \mathcal{N}_i} \big(\yb_i^{(k)}-\frac{\yb_i^{(k)}+\yb_j^{(k)}}{2}  \big)^T(\yb_i^{(k)}-\yb^\star).
\end{align}
By substituting \eqref{eq: proof 3} to \eqref{eq: proof 6} into \eqref{eq: proof 2} and summing over $i=1,\ldots,N$, we obtain
\begin{align}\label{eq: proof 7}
  &{\small  \|\yb^{(k)}-\tilde \yb^\star\|_{\Mb}^2
  -\frac{1}{2} \|\yb^{(k-1)}-\yb^{(k)}\|^2_{\tilde \Ab^T \Db_{L_f}\Db_{\rho}^{-1} \tilde \Ab }}
  \notag \\
  &~~~~~~~~~~~~~~~~~~{\small +(\yb^{(k)} - \yb^{(k-1)})^T\Db_{\beta}(\yb^{(k)}-\tilde \yb^\star)}
  \notag \\
  &+\sum_{i=1}^N\sum_{j\in \mathcal{N}_i} (\ub_{ij}^{(k+1)} - \ub_{ij}^\star)^T(\yb_i^{(k)}-\yb^\star)
   \notag \\
  &+ \sum_{i=1}^N\sum_{j\in \mathcal{N}_i} (\vb_{ji}^{(k+1)} - \vb_{ji}^\star)^T(\yb_i^{(k)}-\yb^\star)
  \notag \\
  &+ 2c \sum_{i=1}^N\sum_{j\in \mathcal{N}_i}\!\!
  \bigg(\!\frac{\yb_i^{(k)}+\yb_j^{(k)}}{2}-\frac{\yb_i^{(k-1)}+\yb_j^{(k-1)}}{2} \! \bigg)^T\!\!(\yb_i^{(k)}-\yb^\star)
  \notag \\
  & \leq 0,
\end{align}
where $\yb^{(k)}=[(\yb^{(k)}_1)^T,\ldots,(\yb^{(k)}_N)^T]^T$, $\tilde \Ab=\blkdiag\{\Ab_1,\ldots,\Ab_N\}$, $\Db_{L_f}=\diag\{L_{f,1}^2,\ldots,L_{f,N}^2\}\otimes \Ib_K$, $\Db_{\beta}=\diag\{\beta_1,\ldots,\beta_N\}\otimes \Ib_K$, {
$\Db_{\rho}=\diag\{\rho_1,\ldots,\rho_N\}\otimes \Ib_K$, and as defined in \eqref{eq: M},} $$\Mb= \tilde \Ab^T(\Db_{\sigma_f}-\frac{1}{2}\Db_{\rho})\tilde \Ab.$$
It can be observed from \eqref{eq: KKT3} and also \eqref{eqn: admm steps for P1 u} and \eqref{eqn: admm steps for P1 v} that
\begin{align}\label{eq: proof 7.5}
    &\ub_{ij}^\star + \vb_{ij}^\star =\zerob~\forall j,i, \\
    &\ub_{ij}^{(k)} + \vb_{ij}^{(k)} =\zerob~ \forall j,i,k,    \label{eq: proof 7.55}
\end{align} given the initial $\ub_{ij}^{(0)} + \vb_{ij}^{(0)} =\zerob~\forall j,i,k$ {\blue which is equivalent to setting $\pb_i^{(k)}=\zerob~\forall i\in V$ (See Step 1 of Algorithm \ref{table: IC-ADMM}).}
Besides, due to the symmetric property of $\Wb$, for any $\{\alpha_{ij}\}$, we have
{\small \begin{align}\label{eq: proof 7.6}
  \sum_{i=1}^N\sum_{j\in \mathcal{N}_i}  \alpha_{ij} &= \sum_{i=1}^N\sum_{j=1}^N [\Wb]_{i,j}\alpha_{ij}  \notag \\
  &= \sum_{i=1}^N\sum_{j=1}^N [\Wb]_{i,j}\alpha_{ji}= \sum_{i=1}^N\sum_{j\in \mathcal{N}_i}  \alpha_{ji}.
\end{align}}
\hspace{-0.3cm} By the above two properties, the fourth and fifth terms in the LHS of \eqref{eq: proof 7} can be written as
\begin{align}
&\textstyle \sum_{i=1}^N\sum_{j\in \mathcal{N}_i} (\ub_{ij}^{(k+1)} - \ub_{ij}^\star)^T(\yb_i^{(k)}-\yb^\star)
\notag\\
&~~\textstyle ~~~~~~~~~~+ \sum_{i=1}^N\sum_{j\in \mathcal{N}_i} (\vb_{ji}^{(k+1)} - \vb_{ji}^\star)^T(\yb_i^{(k)}-\yb^\star) \notag \\
& ~~\textstyle  ={\sum_{i=1}^N\sum_{j\in \mathcal{N}_i} (\ub_{ij}^{(k+1)} - \ub_{ij}^\star)^T(\yb_i^{(k)}-\yb^\star) }
\notag
\\
&~~\textstyle {~~~~~~~~~~+ \sum_{i=1}^N\sum_{j\in \mathcal{N}_i} (\vb_{ij}^{(k+1)} - \vb_{ij}^\star)^T(\yb_j^{(k)}-\yb^\star) } \
\notag
\\
&~~\textstyle =
\sum_{i=1}^N\sum_{j\in \mathcal{N}_i} (\ub_{ij}^{(k+1)} - \ub_{ij}^\star)^T(\yb_i^{(k)}-\yb_j^{(k)})
\notag
\end{align}
\begin{align}\label{eq: proof 8}
&~~\textstyle =\frac{2}{c}\sum_{i=1}^N\sum_{j\in \mathcal{N}_i} (\ub_{ij}^{(k+1)} - \ub_{ij}^\star)^T(\ub_{ij}^{(k+1)}-\ub_{ij}^{(k)})\notag \\
&~~\textstyle\triangleq \frac{2}{c}(\ub^{(k+1)} - \ub^\star)^T(\ub^{(k+1)}-\ub^{(k)}),
\end{align}
{where the first equality is owing to \eqref{eq: proof 7.6}, the second equality is by \eqref{eq: proof 7.5} and \eqref{eq: proof 7.55}, and the third equality is due to \eqref{eqn: admm steps for P1 u}.} In \eqref{eq: proof 8}, $\ub^{(k)}$ ($\ub^\star$) is a vector that stacks $\ub_{ij}^{(k)}$ ($\ub_{ij}^\star$) for all $j\in \mathcal{N}_i$, $i=1,\ldots,N$.
The sixth term in the LHS of \eqref{eq: proof 7} can be rearranged as follows
\begin{align}\label{eq: proof 9}
&\textstyle c \sum_{i=1}^N\sum_{j\in \mathcal{N}_i}
  (\yb_i^{(k)}-\yb_i^{(k-1)}  )^T(\yb_i^{(k)}-\yb^\star)
  \notag \\
&~~~~~~~~~~~\textstyle+c \sum_{i=1}^N\sum_{j\in \mathcal{N}_i}
  (\yb_j^{(k)}-\yb_j^{(k-1)}  )^T(\yb_i^{(k)}-\yb^\star) \notag
  \\
&\textstyle= c \sum_{i=1}^N |\mathcal{N}_i|(\yb_i^{(k)}-\yb_i^{(k-1)}  )^T(\yb_i^{(k)}-\yb^\star)
  \notag \\
  &~~~~~~~~~~~\textstyle+c \sum_{i=1}^N\sum_{j=1}^N
  [\Wb]_{i,j}(\yb_j^{(k)}-\yb_j^{(k-1)}  )^T(\yb_i^{(k)}-\yb^\star) \notag \\
&{=  c (\yb^{(k)}-\yb^{(k-1)}  )^T(\Db \otimes \Ib_K)(\yb^{(k)}-\tilde \yb^\star)}
\notag \\
&~~~~~~~~~~~{+c (\yb^{(k)}-\yb^{(k-1)}  )^T(\Wb \otimes \Ib_K)(\yb^{(k)}-\tilde \yb^\star)} \notag \\
&=  c (\yb^{(k)}-\yb^{(k-1)}  )^T[(\Db+\Wb) \otimes \Ib_K](\yb^{(k)}-\tilde \yb^\star).
\end{align}
{ \blue Note that,} by the graph theory \cite{BK:Chung96}, the normalized Laplacian matrix, i.e., $\Db^{-\frac{1}{2}}\Lb\Db^{-\frac{1}{2}}$, have
$\lambda_{\max}(\Db^{-\frac{1}{2}}\Lb\Db^{-\frac{1}{2}})\leq 2$. Thus, in \eqref{eq: proof 9},
\begin{align*}
 \Db+\Wb = 2 \Db - \Lb =\Db^{\frac{1}{2}}(2\Ib_N- \Db^{-\frac{1}{2}}\Lb\Db^{-\frac{1}{2}} )\Db^{\frac{1}{2}}\succeq \zerob.
\end{align*}

{By substituting \eqref{eq: proof 8} and \eqref{eq: proof 9} into \eqref{eq: proof 7}, we obtain
\begin{align}\label{eq: proof 9.5}
  &{\small  \|\yb^{(k)}-\tilde \yb^\star\|_{\Mb}^2
  -\frac{1}{2} \|\yb^{(k-1)}-\yb^{(k)}\|_{\tilde \Ab^T \Db_{L_f}\Db_{\rho}^{-1} \tilde \Ab }^2}
  \notag \\
  &~~~~~~~~~~~~~~~~~~{\small +(\yb^{(k)} - \yb^{(k-1)})^T\Gb(\yb^{(k)}-\tilde \yb^\star)}
  \notag \\
  &+\frac{2}{c}(\ub^{(k+1)} - \ub^\star)^T(\ub^{(k+1)}-\ub^{(k)}) \leq 0,
\end{align}
where as defined in \eqref{eq: G}, $$\Gb \triangleq \Db_{\beta} + c ((\Db+\Wb)\otimes \Ib_K)\succ \zerob.$$ }
{\blue Note that}
\begin{align}\label{eq: proof 9.50}
 &(\ab^{(k)}-\ab^{(k-1)}  )^T\Qb(\ab^{(k)}-\ab^\star) = \frac{1}{2}\|\ab^{(k)}-\ab^\star\|^2_\Qb
 \notag \\
 &~~~+ \frac{1}{2}\|\ab^{(k)}-\ab^{(k-1)}\|^2_\Qb
 -\frac{1}{2}\|\ab^{(k-1)}-\ab^\star\|^2_\Qb
\end{align}
for any sequence $\ab^{(k)}$ and matrix $\Qb\succeq \zerob$. {\blue By applying \eqref{eq: proof 9.50} to each of the terms in \eqref{eq: proof 9.5}, one obtains that}
\begin{align}\label{eq: proof 10}
 &(\yb^{(k)}-\tilde \yb^\star)^T\bigg[\Mb + \frac{1}{2}\Gb\bigg](\yb^{(k)}-\tilde \yb^\star) + \frac{1}{c}\|\ub^{(k+1)} - \ub^\star\|^2_2
 \notag \\
 &\leq \frac{1}{2}(\yb^{(k-1)}-\tilde \yb^\star)^T \Gb(\yb^{(k-1)}-\tilde \yb^\star) \notag \\
 &~~~~~~~~+ \frac{1}{c}\|\ub^{(k)} - \ub^\star\|^2_2 -  \frac{1}{c}\|\ub^{(k+1)} - \ub^{(k)}\|^2_2 \notag \\
 &-(\yb^{(k)}-\yb^{(k-1)})^T\bigg[\frac{1}{2}\Gb \!-\! {\frac{1}{2}\tilde \Ab^T \Db_{L_f}\Db_{\rho}^{-1} \tilde \Ab} \bigg](\yb^{(k)}\!-\yb^{(k-1)}).
\end{align}
{Now, consider the condition on $\beta_i$ in \eqref{eq: conv condition beta}. It can be easily checked that
\eqref{eq: conv condition beta} implies that
\begin{subequations}\label{eq: proof 11}
\begin{align}
  &\sigma_{f,i}^2 - \frac{\rho_i}{2}> 0,~ \\
  &{\beta_i}\Ib_K + c\lambda_{\min}(\Db+\Wb)\Ib_K - \frac{L_{f,i}^2}{\rho_i}\Ab^T_i\Ab_i \succ \zerob,
\end{align}
\end{subequations}
{\blue for some $\sigma_{f,i}^2\leq \rho_i <2\sigma_{f,i}^2~\forall i\in V$,} and therefore
\begin{align}\label{eq: proof 11.5}
  \Mb\succeq \zerob,~\Gb - \tilde \Ab^T \Db_{L_f}\Db_{\rho}^{-1} \tilde \Ab \succ \zerob.
\end{align}
With \eqref{eq: proof 11.5}, \eqref{eq: proof 10} implies {\blue the following two results  {\sf (R1)}} as $k\to \infty$, the sequence $\frac{1}{2}\|\yb^{(k)}-\tilde \yb^\star\|^2_{\Gb}+\frac{1}{c}\|\ub^{(k+1)} - \ub^\star\|^2_2$ converges
for any pair of optimal $\tilde \yb^\star$ and $\ub^\star$ to problem \eqref{consensus problem equi}; and {\blue {\sf (R2)}}
\begin{align}
  {\yb^{(k)}-\yb^{(k-1)} \rightarrow \zerob, ~\ub^{(k+1)}-\ub^{(k)} \rightarrow \zerob.}
  \label{eq: proof 10.5}
\end{align}

{\blue The result {\sf (R1)}} implies that the sequences of $\{\yb_i^{(k)}\}$ and $\{\ub_{ij}^{(k)}\}$ (so is $\{\vb_{ij}^{(k)}\}$) are bounded.
Let $\tilde{ \hat \yb} =[(\hat \yb_1)^T,\ldots,(\hat \yb_N)^T]^T$, $\hat \ub_{ij}$ and $\hat \vb_{ij}$ be a set of limit points of $\{\yb^{(k)}\}$, $\{\ub_{ij}^{(k)}\}$ and $\{\vb_{ij}^{(k)}\}$, respectively.
Firstly, by the result of $\ub^{(k+1)}-\ub^{(k)} \rightarrow \zerob$ and \eqref{eqn: admm steps for P1 u}, we have
\begin{align}\label{eq: proof 11.6}
  \yb_i^{(k)}-  \yb_j^{(k)}\to \zerob \Longrightarrow \hat \yb \triangleq \hat \yb_i =\hat \yb_j,~\forall j,i.
\end{align}
Secondly, by \eqref{eq: proof 7.55}, we have
\begin{align}\label{eq: proof 11.7}
    &\hat \ub_{ij} + \hat \vb_{ij} =\zerob~ \forall j,i.
\end{align}
Thirdly, by applying the result of $\yb^{(k)}-\yb^{(k-1)} \rightarrow \zerob$ and \eqref{eq: proof 11.6} to
\eqref{eq: proof 1.0}, we have
\begin{align}
  \zerob&=\Ab_i^T\nabla f_i(\Ab_i \hat \yb_i)
   +\partial g_i(\hat \yb_i) +
  \sum_{j\in \mathcal{N}_i} (\hat \ub_{ij} + \hat \vb_{ji} )
  \label{eq: proof 11.8}
\end{align}
for all $i\in V$. So, $\tilde{ \hat \yb}$ and $\{\hat \ub_{ij},\hat \vb_{ij}\}$ are in fact a pair of optimal primal and dual solutions to problem \eqref{consensus problem equi} [see \eqref{eq: KKT1}, \eqref{eq: KKT2} and \eqref{eq: KKT3}]. {\blue Therefore, according to {\sf (R1)},} the sequence
$\frac{1}{2}\|\yb^{(k)}-\tilde {\hat \yb}\|^2_{\Gb}+\frac{1}{c}\|\ub^{(k+1)} - \hat \ub\|^2_2$ converges.
{\blue Furthermore, since $\frac{1}{2}\|\yb^{(k)}-\tilde {\hat \yb}\|^2_{\Gb}+\frac{1}{c}\|\ub^{(k+1)} - \hat \ub\|^2_2$ has a limit value equal to zero, we conclude that $\frac{1}{2}\|\yb^{(k)}-\tilde {\hat \yb}\|^2_{\Gb}+\frac{1}{c}\|\ub^{(k+1)} - \hat \ub\|^2_2$ in fact converges to zero.} This says that
  $\yb_i^{(k)} \rightarrow \hat \yb~ \forall~i\in V$ and  $\ub^{(k+1)} \rightarrow \hat \ub.$ 
}
%
The proof is thus complete. \hfill $\blacksquare$

{\bf Proof of Theorem \ref{thm: conv of inexact pc admm}(b):} 
Let $0<\alpha<1$ be some positive number and rewrite \eqref{eq: proof 10} as
{\small \begin{align}\label{eq: proof 12}
 &\bigg(\!\|\yb^{(k)}-\tilde \yb^\star\|_{\frac{1}{2}\Gb+\alpha\Mb}^2 + \frac{1}{c}\|\ub^{(k+1)} - \ub^\star\|^2_2\!\bigg)
 + \|\yb^{(k)}-\tilde \yb^\star\|_{(1-\alpha)\Mb}^2
 \notag \\
 &~
 + \|\yb^{(k-1)}-\tilde \yb^\star\|_{\alpha\Mb}^2+ \frac{1}{c}\|\ub^{(k+1)} - \ub^{(k)}\|^2_2
 \notag\\
 & +(\yb^{(k)}-\yb^{(k-1)})^T\bigg[\frac{1}{2}\Gb - {\frac{1}{2}\tilde \Ab^T \Db_{L_f}\Db_{\rho}^{-1} \tilde \Ab}\bigg](\yb^{(k)}-\yb^{(k-1)})
 \notag \\
 &\leq \bigg(\|\yb^{(k-1)}-\tilde \yb^\star\|_{\frac{1}{2}\Gb+\alpha\Mb}^2 + \frac{1}{c}\|\ub^{(k)} - \ub^\star\|^2_2\bigg).
\end{align}}\hspace{-0.3cm}
 {Then, in order to prove linear convergence rate, i.e., for some $\delta>0$,
 \begin{align*}
 &\big(\|\yb^{(k)}-\tilde \yb^\star\|_{\frac{1}{2}\Gb+\alpha\Mb}^2 + \frac{1}{c}\|\ub^{(k+1)} - \ub^\star\|^2_2\big)
 \notag \\
 &\leq
 \frac{1}{1+\delta} \big(\|\yb^{(k-1)}-\tilde \yb^\star\|_{\frac{1}{2}\Gb+\alpha\Mb}^2 + \frac{1}{c}\|\ub^{(k)} - \ub^\star\|^2_2\big),
 \end{align*}
  it is sufficient to show that } 
{\small\begin{align}\label{eq: proof 13}
\!\!\!\!\!\!\!\! &
 \|\yb^{(k)}-\tilde \yb^\star\|_{(1-\alpha)\Mb}^2
 + \|\yb^{(k-1)}-\tilde \yb^\star\|_{\alpha\Mb}^2+ \frac{1}{c}\|\ub^{(k+1)} - \ub^{(k)}\|^2_2
 \notag \\
 &~
  +(\yb^{(k)}-\yb^{(k-1)})^T\bigg[\frac{1}{2}\Gb - {\frac{1}{2}\tilde \Ab^T \Db_{L_f}\Db_{\rho}^{-1} \tilde \Ab}\bigg](\yb^{(k)}-\yb^{(k-1)})
 \notag \\
 &\geq \delta \bigg(\|\yb^{(k)}-\tilde \yb^\star\|_{\frac{1}{2}\Gb+\alpha\Mb}^2 + \frac{1}{c}\|\ub^{(k+1)} - \ub^\star\|^2_2\bigg).
\end{align}}\hspace{-0.3cm}
Recall from \eqref{eq: proof 1} and \eqref{eq: proof 6} that
\begin{align}\label{eq: proof 14}
  &\textstyle \Ab_i^T\nabla f_i(\Ab_i\yb_i^{(k-1)})-\Ab_i^T\nabla f_i(\Ab_i\yb^\star) + \beta_i (\yb_i^{(k)} - \yb_i^{(k-1)})
     \notag \\
  &\textstyle +\sum_{j\in \mathcal{N}_i} (\ub_{ij}^{(k+1)} - \ub_{ij}^\star)+ \sum_{j\in \mathcal{N}_i} (\vb_{ji}^{(k+1)} - \vb_{ji}^\star)
  \notag\\
  &\textstyle  + 2c \sum_{j\in \mathcal{N}_i}\!
  \bigg(\!\frac{\yb_i^{(k)}+\yb_j^{(k)}}{2}\!-\!\frac{\yb_i^{(k-1)}+\yb_j^{(k-1)}}{2} \! \bigg)=\zerob.
\end{align}
By applying \eqref{eq: proof 7.5} and \eqref{eq: proof 7.55},
\eqref{eq: proof 14} can be expressed as
\begin{align}\label{eq: proof 15}
  &\Ab_i^T\nabla f_i(\Ab_i\yb_i^{(k-1)})-\Ab_i^T\nabla f_i(\Ab_i\yb^\star) + \beta_i (\yb_i^{(k)} - \yb_i^{(k-1)})
    \notag \\
  &\textstyle+\sum_{j\in \mathcal{N}_i} (\ub_{ij}^{(k+1)} - \ub_{ji}^{(k+1)}- \ub_{ij}^\star +\ub_{ji}^\star  )\notag \\
  &\textstyle+ c \sum_{j\in \mathcal{N}_i}\!\!
  \bigg(\!\yb_i^{(k)}-\yb_i^{(k-1)} +\yb_j^{(k)}-\yb_j^{(k-1)} \! \bigg)=\zerob.
\end{align}
After stacking \eqref{eq: proof 15} for $i=1,\ldots,N$, one obtains
\begin{align}\label{eq: proof 16}
   \tilde \Ab^T(\nabla \fb(\tilde \Ab\yb^{(k-1)})&-\nabla \fb(\tilde \Ab\tilde \yb^\star))+ \Gb(\yb^{(k)}-\yb^{(k-1)})
   \notag \\
   &+ {\Upsilonb} (\ub^{(k+1)}-\ub^\star)
=\zerob.
\end{align}
where {$\nabla \fb(\tilde \Ab\yb^{(k)})\!\triangleq \!\![(\nabla f_1(\Ab_1\yb_1^{(k)}))^T\!,\ldots,\!(\nabla f_1(\Ab_N\yb_N^{(k)}))^T]^T$} and {$\Upsilonb\in \mathbb{R}^{KN \times 2|\mathcal{E}|K}$} is a linear mapping matrix satisfying
\begin{align}\label{eq: Psib}
  \begin{bmatrix}
    \sum_{j\in \mathcal{N}_1} (\ub_{1j}^{(k+1)}-\ub_{j1}^{(k+1)})\\
    \vdots \\
    \sum_{j\in \mathcal{N}_N} (\ub_{Nj}^{(k+1)}-\ub_{jN}^{(k+1)})
  \end{bmatrix} = \Upsilonb \ub^{(k+1)}.
\end{align}
According to \cite{ShiLing2013J}\footnote{\blue Note that the matrix $\Upsilonb$ corresponds to matrix $M_{-}$ in \cite{ShiLing2013J}.}, both $\ub^{(k+1)}$ and $\ub^\star$ lie in the range space of $\Upsilonb^T$.
Hence, one can show that
\begin{align}\label{eq: proof 17}
  \|\Upsilonb (\ub^{(k+1)} - \ub^\star)\|^2 \geq \sigma_{\min}^2(\Upsilonb) \|\ub^{(k+1)} - \ub^\star\|^2_2
\end{align} where $\sigma_{\min}(\Upsilonb)>0$ is the minimum nonzero singular value of $\Upsilonb$.
From \eqref{eq: proof 16}, we have that
{\small \begin{align}\label{eq: proof 18}
   &\|\Gb(\yb^{(k)}-\yb^{(k-1)})\|^2_2
   \notag \\
   &=\|-\tilde \Ab^T(\nabla \fb(\tilde \Ab\yb^{(k-1)})-\nabla \fb(\tilde \Ab\tilde \yb^\star))- \Upsilonb (\ub^{(k+1)}-\ub^\star)\|^2_2 \notag \\
   &\geq (1 -\mu) \|\tilde \Ab^T(\nabla \fb(\tilde \Ab\yb^{(k)})-\nabla \fb(\tilde \Ab\tilde \yb^\star))\|^2_2 \notag \\
   &~~~~~~~~~~~~~~~~~~~~~~~~~~~~~~~~~~+ (1-\frac{1}{\mu})\|\Upsilonb (\ub^{(k+1)}-\ub^\star)\|^2_2 \notag
   \\
   &\geq (1 -\mu) \lambda_{\max}(\tilde\Ab^T\tilde\Ab)\|(\nabla \fb(\tilde \Ab\yb^{(k)})-\nabla \fb(\tilde \Ab\tilde \yb^\star))\|^2_2 \notag
   \\
     &~~~~~~~~~~~~~~~~~~~~~~~~~~~~~~~~~~+ (1-\frac{1}{\mu})\sigma_{\min}^2(\Upsilonb) \|\ub^{(k+1)} - \ub^\star\|_2^2
     \notag
      \\
   &\geq (1 -\mu) \lambda_{\max}(\tilde\Ab^T\tilde\Ab) \|(\yb^{(k-1)}-\yb^\star)\|_{\tilde \Ab^T \Db_{L_f} \tilde \Ab}^2
   \notag \\
   &~~~~~~~~~~~~~~~~~~~~~+ (1-\frac{1}{\mu})\sigma_{\min}^2(\Upsilonb) \|\ub^{(k+1)} - \ub^\star\|_2^2,
\end{align}}
where the first inequality is due to the fact that
\begin{align}\label{eq: inequality }
  \|\ab+\qb\|_2^2 \geq (1 -\mu)\|\ab\|_2^2 + (1-\frac{1}{\mu})\|\qb\|_2^2
\end{align} for any $\ab,\qb$ and $\mu>0,$ the second inequality is obtained by setting $\mu>1$ and \eqref{eq: proof 17}, and the last inequality is by \eqref{eq: lipschitz gradient of f}.
Equation \eqref{eq: proof 18} implies that
\begin{align}\label{eq: proof 19}
\!\!  &\frac{\delta}{c}\|\ub^{(k+1)} - \ub^\star\|_2^2 \leq
  \frac{\delta}{c(1-\frac{1}{\mu})\sigma_{\min}^2(\Upsilonb)}\|\yb^{(k)}-\yb^{(k-1)}\|_{\Gb^T\Gb}^2
  \notag \\
  &+\frac{\delta (\mu-1) \lambda_{\max}(\tilde\Ab^T\tilde\Ab) }{c(1-\frac{1}{\mu})\sigma_{\min}^2(\Upsilonb)}
  \|(\yb^{(k-1)}-\tilde \yb^\star)\|_{\tilde \Ab^T \Db_{L_f} \tilde \Ab}^2.
\end{align}
{According to \eqref{eq: proof 19}, \eqref{eq: proof 13} can hold true if
\begin{align*}
  &\|\yb^{(k)}-\tilde \yb^\star\|_{(1-\alpha)\Mb}^2 \geq
  \delta \|\yb^{(k)}-\tilde \yb^\star\|_{\frac{1}{2}\Gb+\alpha\Mb}^2, \\
  &(\yb^{(k)}-\yb^{(k-1)})^T\bigg[\frac{1}{2}\Gb - {\frac{1}{2}\tilde \Ab^T \Db_{L_f}\Db_{\rho}^{-1} \tilde \Ab}\bigg](\yb^{(k)}-\yb^{(k-1)}) \notag
  \\
    &~~~~\geq \frac{\delta}{c(1-\frac{1}{\mu})\sigma_{\min}^2(\Upsilonb)}\|\yb^{(k)}-\yb^{(k-1)}\|_{\Gb^T\Gb}^2, \\
  &  \|\yb^{(k-1)}-\tilde \yb^\star\|_{\alpha\Mb}^2
  \notag \\
  &~~~~~\geq    \frac{\delta (\mu-1) \lambda_{\max}(\tilde\Ab^T\tilde\Ab) }{c(1-\frac{1}{\mu})\sigma_{\min}^2(\Upsilonb)}
  \|(\yb^{(k-1)}-\tilde \yb^\star)\|_{\tilde \Ab^T \Db_{L_f} \tilde \Ab}^2,
\end{align*}
which are respectively satisfied
if the following three conditions can be satisfied
for some $\delta>0$
{\small\begin{subequations}\label{eq: sufficient for linearly conv}
\begin{align}
&(1-\alpha)\Mb \succeq
\delta \bigg(\frac{1}{2}\Gb + \alpha\Mb\bigg),
\\
& \frac{1}{2}\Gb - {\frac{1}{2}\tilde \Ab^T \Db_{L_f}\Db_{\rho}^{-1} \tilde \Ab} \succeq \frac{\delta}{c(1-\frac{1}{\mu})\sigma_{\min}^2(\Upsilonb)} \Gb^T\Gb,\\
& \alpha(\Db_{\sigma_f}-\frac{1}{2}\Db_{\rho}) \succeq  \delta \frac{(\mu-1) \lambda_{\max}(\tilde\Ab^T\tilde\Ab) }{c(1-\frac{1}{\mu})\sigma_{\min}^2(\Upsilonb)}\Db_{L_f}.
\end{align}
\end{subequations}}
\hspace{-0.11cm}Note that, given $\beta_i$'s in \eqref{eq: conv condition beta}, we have
$\Db_{\sigma_f}-\frac{1}{2}\Db_{\rho}\succ \zerob$ and $\Gb - \frac{1}{\rho}\tilde \Ab^T \Db_{L_f}\Db_{\rho}^{-1} \tilde \Ab \succ \zerob$ (see \eqref{eq: proof 11} and \eqref{eq: proof 11.5}); moreover, since $\Ab_i$'s are full column rank, we have $\Mb\succ \zerob$.} Hence there must exist some $\delta>0$ such that the three conditions in \eqref{eq: sufficient for linearly conv} all hold true.
\hfill $\blacksquare$

\section{Proof of Theorem \ref{thm: conv of inexact dadmm}}
\label{proof of conv of inexact dadmm}
{\bf Proof of Theorem \ref{thm: conv of inexact dadmm}(a):}
                                        %
%
Let {$\xb^\star \triangleq [(\xb_1^\star)^T,\ldots,(\xb_N^\star)^T]^T$ and $\nub^\star$ be a pair of optimal primal and dual solutions} to {\sf (P2)}, and let
$\tilde \nub^\star \triangleq [(\nub_1^\star)^T,\ldots,(\nub_N^\star)^T]^T$ and $\{\ub_{ij}^\star,\vb_{ij}^\star,j\in \mathcal{N}_i\}_{i=1}^N$ be a pair of
optimal primal and dual solutions to problem \eqref{consensus problem equi dual}. Then they respectively satisfy the following optimality conditions
\begin{align}
  &  \Ab_i^T \nabla f_i(\xb_i^\star) + \partial g_i(\xb_i^\star)+\Eb_i^T\nub^\star =\zerob, i\in V,\label{eq: KKT00}\\
  &\textstyle \sum_{i=1}^N \Eb_i\xb_i^\star =\qb, \label{eq: KKT11}
  \\
\!\!\!\!\!  \!\!\!\!\!&\textstyle\partial \varphi_i(\nub_i^\star) + \frac{1}{N}\qb+
  \sum_{j\in \mathcal{N}_i} (\ub_{ij}^\star + \vb_{ji}^\star) =\zerob,~i\in V, \label{eq: KKT22}\\
  &\textstyle \nub_i^\star = \nub_j^\star~\forall j\in \mathcal{N}_i, i\in V, \label{eq: KKT33}\\
  &\textstyle \ub_{ij}^\star + \vb_{ij}^\star =\zerob~\forall j\in \mathcal{N}_i,i\in V. \label{eq: KKT44}
\end{align}
where $\partial \varphi_i(\nub_i^\star)=-\Eb_i \xb_i^\star$  as $\xb_i^\star$ is a maximizer to \eqref{eq: varphi} with $\nub=\nub_i^\star$ \cite{Boydsubgradient}. {Under Assumption \ref{assumption connected graph}, \eqref{eq: KKT2} implies that $\nub^\star\triangleq  \nub_1^\star=\cdots=\nub_N^\star$ and $\tilde \nub^\star = \mathbf{1}_N \otimes \nub^\star$.}

Firstly, by recalling that $\pb_i^{(k)}=\textstyle \sum_{j\in \Nc_i} ( \ub_{ij}^{(k)} + \vb_{ji}^{(k)} )$, it follows from \eqref{eq: proof of thm 2 eq1} and \eqref{eq: KKT22} that
\begin{align}
 \zerob=&\textstyle-(\Eb_i\xb_i^{(k)} -\frac{1}{N}\qb ) + \sum_{j\in \mathcal{N}_i}
    (\ub_{ij}^{(k+1)}+\vb_{ji}^{(k+1)}) \notag \\
    &\textstyle+ c\sum_{j\in \mathcal{N}_i} (\nub_i^{(k)}+\nub_j^{(k)}-\nub_i^{(k-1)}-\nub_j^{(k-1)})
   \label{eq: proof of thm 4 eq01}  \\
 =& \textstyle-\Eb_i\xb_i^\star+ \frac{1}{N}\qb+
  \sum_{j\in \mathcal{N}_i} \ub_{ij}^\star + \sum_{j\in \mathcal{N}_i} \vb_{ji}^\star.\label{eq: proof of thm 4 eq1}
\end{align}
By multiplying $\nub_i^{(k)}-\nub^\star$ to the both sides of \eqref{eq: proof of thm 4 eq1}, we obtain
\begin{align}\label{eq: proof of thm 4 eq2}
 & \sum_{j\in \mathcal{N}_i}
    (\ub_{ij}^{(k+1)}+\vb_{ji}^{(k+1)}-\ub_{ij}^\star - \vb_{ji}^\star)^T (\nub_i^{(k)}-\nub^\star)
    \notag \\
    &+ c\sum_{j\in \mathcal{N}_i} (\nub_i^{(k)}+\nub_j^{(k)}-\nub_i^{(k-1)}-\nub_j^{(k-1)})^T(\nub_i^{(k)}-\nub^\star)
    \notag\\
    &-(\xb_i^{(k)} -\xb_i^\star )^T\Eb_i^T (\nub_i^{(k)}-\nub^\star)=\zerob.
\end{align}
Secondly, {from the optimality of \eqref{eq: dual ADMM x linearized}, we have that
\begin{align}
  \zerob & \textstyle= \Ab_i^T\nabla f_i(\Ab_i\xb_i^{(k-1)}) +\partial g(\xb_i^{(k)}) \notag \\
  &~~\textstyle + \frac{1}{2|\mathcal{N}_i|}\Eb_i^T
    \bigg[\frac{1}{c}(\Eb_i\xb_i^{(k)}-\frac{1}{N}\qb)   - \frac{1}{c}\sum_{j\in \mathcal{N}_i}
    (\ub_{ij}^{(k)}+\vb_{ji}^{(k)}) \notag \\
    &~~ \textstyle + \sum_{j\in \mathcal{N}_i} (\nub_i^{(k-1)}+\nub_j^{(k-1)}) \bigg] +\Pb_i(\xb_i^{(k)}-\xb_i^{(k-1)})
    \notag \\
    &=\Ab_i^T\nabla f_i(\Ab_i\xb_i^{(k-1)}) +\partial g(\xb_i^{(k)}) + \Eb_i^T\nub_i^{(k)}
     \notag \\
    &~~~~~~~~~~~~~~~~~~~~~~~~~~+\Pb_i(\xb_i^{(k)}-\xb_i^{(k-1)})
    \label{eq: proof of thm 4 eq2.5} \\
    &\textstyle =\Ab_i^T(\nabla f_i(\Ab_i\xb_i^{(k-1)})\!-\!\nabla f_i(\Ab_i\xb_i^{(k)}))\! +\! \Ab_i^T \nabla f_i(\Ab_i\xb_i^{(k)})
     \notag \\
    &~~~~~~~+ \partial g(\xb_i^{(k)})+ \Eb_i^T\nub_i^{(k)}+\Pb_i(\xb_i^{(k)}-\xb_i^{(k-1)}) \label{eq: proof of thm 4 eq3}
    \\
    &=\Ab_i^T\nabla f_i(\Ab_i\xb_i^\star) +\partial g(\xb_i^\star) + \Eb_i^T\nub^\star, \label{eq: proof of thm 4 eq4}
\end{align}
where, in the first equality, we have added and subtracted $\frac{1}{2c|\Nc_i|}\Eb_i^T\Eb_i\xb_i^{(k)}$ {\blue and defined}
\begin{align}
 \textstyle \Pb_i\triangleq \beta_i\Ib_K - \frac{1}{2c|\Nc_i|}\Eb_i^T\Eb_i;
\end{align}
the second equality is due to \eqref{eq: dual ADMM lambda 2}; and the last equality is because $\xb_i^\star$ is a maximizer to \eqref{eq: varphi} with $\nub=\nub_i^\star$.}
%
\ifconfver
{Multiplying both \eqref{eq: proof of thm 4 eq3} and \eqref{eq: proof of thm 4 eq4} with $\xb_i^{(k)}-\xb_i^\star$, combining with \eqref{eq: proof of thm 4 eq2}, and summing for $i=1,\ldots,N$, yields \eqref{eq: proof of thm 4 eq5} on the top of the next page.
\begin{figure*}[t]
\begin{align}\label{eq: proof of thm 4 eq5}
  &{\small \sum_{i=1}^N(\nabla f_i(\Ab_i\xb_i^{(k-1)})-\nabla f_i(\Ab_i\xb_i^{(k)}))^T\Ab_i(\xb_i^{(k)}-\xb_i^\star)+
  \sum_{i=1}^N(\xb_i^{(k)}-\xb_i^{(k-1)})^T\Pb_i(\xb_i^{(k)}-\xb_i^\star)}
   \notag \\
  &~+\sum_{i=1}^N(\nabla f_i(\Ab_i\xb_i^{(k)})-\nabla f_i(\Ab_i\xb_i^\star))^T\Ab_i(\xb_i^{(k)}-\xb_i^\star)+\sum_{i=1}^N(\partial g_i(\xb_i^{(k)}) - \partial g_i(\xb_i^\star))^T(\xb_i^{(k)}-\xb_i^\star)
    \notag \\
    &~+
  \sum_{i=1}^N\sum_{j\in \mathcal{N}_i}
    (\ub_{ij}^{(k+1)}+\vb_{ji}^{(k+1)}-\ub_{ij}^\star - \vb_{ji}^\star)^T (\nub_i^{(k)}-\nub_i^\star)+ c\sum_{i=1}^N\sum_{j\in \mathcal{N}_i} (\nub_i^{(k)}+\nub_j^{(k)}-\nub_i^{(k-1)}-\nub_j^{(k-1)})^T(\nub_i^{(k)}-\nub_i^\star) =\zerob.
\end{align}
\hrulefill
\end{figure*}}
\else
{Multiplying both \eqref{eq: proof of thm 4 eq3} and \eqref{eq: proof of thm 4 eq4} with $\xb_i^{(k)}-\xb_i^\star$, combining with \eqref{eq: proof of thm 4 eq2}, and summing for $i=1,\ldots,N$, yields
\begin{align}\label{eq: proof of thm 4 eq5}
  &\sum_{i=1}^N(\nabla f_i(\Ab_i\xb_i^{(k-1)})-\nabla f_i(\Ab_i\xb_i^{(k)}))^T\Ab_i(\xb_i^{(k)}-\xb_i^\star)+
  \sum_{i=1}^N(\xb_i^{(k)}-\xb_i^{(k-1)})^T\Pb_i(\xb_i^{(k)}-\xb_i^\star)+
  \notag \\
  &
  \sum_{i=1}^N(\nabla f_i(\Ab_i\xb_i^{(k)})-\nabla f_i(\Ab_i\xb_i^\star) )^T\Ab_i(\xb_i^{(k)}-\xb_i^\star) +
  \sum_{i=1}^N(\partial g(\xb_i^{(k)}) -\partial g(\xb_i^\star))^T(\xb_i^{(k)}- xb_i^\star)\notag \\
  & ~~~~~~+
  \sum_{i=1}^N\sum_{j\in \mathcal{N}_i}
    (\ub_{ij}^{(k+1)}+\vb_{ji}^{(k+1)}-\ub_{ij}^\star - \vb_{ji}^\star)^T (\nub_i^{(k)}-\nub_i^\star)
    \notag \\
    &~~~~~~+ c\sum_{i=1}^N\sum_{j\in \mathcal{N}_i} (\nub_i^{(k)}+\nub_j^{(k)}-\nub_i^{(k-1)}-\nub_j^{(k-1)})^T(\nub_i^{(k)}-\nub_i^\star)=\zerob.
\end{align}}
\fi

{Similar to \eqref{eq: proof 8} and by \eqref{eq: inexact ADMM u v}, the fifth term in the LHS of \eqref{eq: proof of thm 4 eq5} can be expressed as
\begin{align}\label{eq: proof of thm 4 eq5.1}
  &\textstyle \sum_{i=1}^N\sum_{j\in \mathcal{N}_i}
    (\ub_{ij}^{(k+1)}+\vb_{ji}^{(k+1)}-\ub_{ij}^\star - \vb_{ji}^\star)^T (\nub_i^{(k)}-\nub_i^\star) \notag \\
    =& \frac{2}{c}(\ub^{(k+1)} - \ub^\star)^T(\ub^{(k+1)}-\ub^{(k)}).
\end{align}
Moreover, the sixth term in the LHS of \eqref{eq: proof of thm 4 eq5} can be shown as
\begin{align}\label{eq: proof of thm 4 eq5.2}
&\textstyle c \sum_{i=1}^N\sum_{j\in \mathcal{N}_i}
  (\nub_i^{(k)}-\nub_i^{(k-1)}  )^T(\nub_i^{(k)}-\nub^\star_i)
  \notag\\
&~~~~~~~~~~~\textstyle +c \sum_{i=1}^N\sum_{j\in \mathcal{N}_i}
  (\nub_j^{(k)}-\nub_j^{(k-1)}  )^T(\nub_i^{(k)}-\nub_i^\star) \notag
  \\
&\textstyle = c \sum_{i=1}^N |\mathcal{N}_i|(\nub_i^{(k)}-\nub_i^{(k-1)}  )^T(\nub_i^{(k)}-\nub_i^\star)
\notag \\
  &\textstyle ~~~~~~~  +c \sum_{i=1}^N\sum_{j=1}^N
     [\Wb]_{i,j}(\nub_j^{(k)}-\nub_j^{(k-1)})^T(\nub_i^{(k)}-\nub_i^\star) \notag \\
&\textstyle =  c (\nub^{(k)}-\nub^{(k-1)}  )^T\Qb(\nub^{(k)}-\tilde\nub^\star),
 \end{align}
where $\Qb\triangleq (\Db+\Wb)\otimes \Ib_M$. By applying \eqref{eq: proof 3}, \eqref{eq: proof 4}, \eqref{eq: proof 5},
\eqref{eq: proof of thm 4 eq5.1} and \eqref{eq: proof of thm 4 eq5.2} to \eqref{eq: proof of thm 4 eq5}, one obtains
\begin{align}\label{eq: proof of thm 4 eq5.3}
  &{\small  \|\xb^{(k)}-\xb^\star\|_{\Mb}^2
  -\frac{1}{2} \|\xb^{(k-1)}-\xb^{(k)}\|_{\tilde \Ab^T \Db_{L_f}\Db_{\rho}^{-1} \tilde \Ab }^2}
  \notag \\
  &~~~~~~~~~~~~~~~~~~{\small +(\xb^{(k)} - \xb^{(k-1)})^T\Pb(\xb^{(k)}- \xb^\star)}
  \notag \\
  &~~~~~~~~~~~~~~~~~~{\small +c (\nub^{(k)}-\nub^{(k-1)}  )^T\Qb(\nub^{(k)}-\tilde\nub^\star)} \notag \\
  &+\frac{2}{c}(\ub^{(k+1)} - \ub^\star)^T(\ub^{(k+1)}-\ub^{(k)}) \leq 0,
\end{align}
where 
$\nub^{(k)}=[(\nub^{(k)}_1)^T,\ldots,(\nub^{(k)}_N)^T]^T$, $\Pb=\blkdiag\{\Pb_1,\ldots,\Pb_N\}\succ \zerob$, and $\Db_{\sigma_f}$, $\Db_{L_f}$, $\Db_{\rho}$, $\tilde \Ab$ and $\Mb$  are all defined below \eqref{eq: proof 7}.
After applying \eqref{eq: proof 9.5} to \eqref{eq: proof of thm 4 eq5.3},} we obtain
\begin{align}\label{eq: proof of thm 4 eq6}
 &\|\xb^{(k)}- \xb^\star\|_{\Mb+\frac{1}{2}\Pb}^2+ \frac{1}{c}\|\ub^{(k+1)} - \ub^\star\|^2_2
 + \frac{c}{2}\|\nub^{(k)} - \tilde \nub^\star\|^2_{\Qb}\notag \\
 &\leq \|\xb^{(k-1)}-\xb^\star\|_{\frac{1}{2}\Pb}^2 + \frac{1}{c}\|\ub^{(k)} - \ub^\star\|^2_2
 +\frac{c}{2}\|\nub^{(k-1)} - \tilde \nub^\star\|^2_{\Qb}
 \notag \\
 &~~~
  -\frac{1}{2}\|\xb^{(k)}-\xb^{(k-1)}\|_{\Pb - \tilde \Ab^T \Db_{L_f}\Db_{\rho}^{-1} \tilde \Ab}^2- \frac{1}{c}\|\ub^{(k+1)} - \ub^{(k)}\|^2_2 \notag \\
 &~~~-\frac{c}{2}\|\nub^{(k)} - \nub^{(k-1)}\|^2_{\Qb}.
\end{align}

It is easy to show that, under \eqref{eq: conv condition beta idcadmm}, it holds true that
\begin{align}\label{eq: proof of thm 4 eq6.5}
  \sigma_{f,i}^2 - \frac{\rho_i}{2}> 0,~\Pb_i- \frac{L_{f,i}^2}{\rho_i}\Ab_i^T\Ab_i \succ \zerob,~\forall~i\in V,
\end{align}
for some {\blue $\sigma_{f,i}^2\leq \rho_i < 2\sigma_{f,i}^2$} $\forall i\in V$, {which implies that
\begin{align*}
  \Pb\succ \zerob,~ \Pb - \tilde \Ab^T \Db_{L_f}\Db_{\rho}^{-1} \tilde \Ab \succ \zerob.
\end{align*} Thus, \eqref{eq: proof of thm 4 eq6} implies that
{\blue \sf (R1)} the sequence $\|\xb^{(k)}- \xb^\star\|_{\Mb+\frac{1}{2}\Pb}^2+ \frac{1}{c}\|\ub^{(k+1)} - \ub^\star\|^2_2
 + \frac{c}{2}\|\nub^{(k)} - \tilde \nub^\star\|^2_{\Qb}$ converges for any optimal $\xb^\star$ to {\sf (P2)}, and optimal $\tilde \nub^\star$ and $\ub^\star$ to problem \eqref{consensus problem equi dual}; and {\blue \sf (R2)}
\begin{align}\label{eq: proof of thm 4 eq7}
 &\xb^{(k)} - \xb^{(k-1)} \rightarrow \zerob, ~\ub^{(k+1)} -\ub^{(k)}\rightarrow\zerob, \\
 &\|\nub^{(k)} - \nub^{(k-1)}\|^2_{\Qb} \to \zerob.\label{eq: proof of thm 4 eq7.5}
\end{align}

Let ${ \hat \xb} =[(\hat \xb_1)^T,\ldots,(\hat \xb_N)^T]^T$, $\tilde{ \hat \nub} =[(\hat \nub_1)^T,\ldots,(\hat \nub_N)^T]^T$, $\hat \ub_{ij}$ and $\hat \vb_{ij}$ be {\blue a set of} limit points of $\{\xb^{(k)}\}$, $\{\nub_1^{(k)},\ldots,\nub_N^{(k)}\}$, $\{\ub_{ij}^{(k)}\}$ and $\{\vb_{ij}^{(k)}\}$, respectively.
Firstly, by applying the fact of $\xb^{(k)} - \xb^{(k-1)}\rightarrow\zerob$ to \eqref{eq: proof of thm 4 eq3}, we have
\begin{align}\label{eq: proof of thm 4 eq8}
    \Ab_i^T \nabla f_i(\Ab_i\hat\xb_i)+ \partial g(\hat\xb_i)+ \Eb_i^T\hat\nub_i = \zerob,~
    \forall~i\in V.
\end{align}
Secondly, by \eqref{eq: proof 7.55}, we have
\begin{align}\label{eq: proof 11.799}
    &\hat \ub_{ij} + \hat \vb_{ij} =\zerob~ \forall j,i.
\end{align}
Thirdly, applying the fact of $\ub_{ij}^{(k+1)}-\ub_{ij}^{(k)}\rightarrow \zerob$ to \eqref{eq: inexact ADMM ub} yields
\begin{align}\label{eq: proof of thm 4 eq9}
 \nub_i^{(k)}-\nub_j^{(k)}\to \zerob \Longrightarrow \hat \nub \triangleq\hat \nub_i = \hat\nub_j~\forall j\in \mathcal{N}_i, i\in V
\end{align}
The result of $\nub_i^{(k)}-\nub_j^{(k)}\to \zerob~\forall j,i$ and Assumption 1 implies that
$       \nub^{(k)} -{\bf 1}_N\otimes \nub^{(k)}_i \rightarrow \zerob$
for any $i\in V$. Since the Laplacian matrix $\Lb{\bf 1}_N=\zerob$ \cite{BK:Chung96}, one obtains
\begin{align}
&\|\nub^{(k)} -  \nub^{(k-1)}\|^2_{\Qb}\notag \\
&\rightarrow ({\bf 1}_N\otimes (\nub^{(k)}_i-\nub^{(k-1)}_i))^T\Qb({\bf 1}_N\otimes (\nub^{(k)}_i-\nub^{(k-1)}_i))\notag\\
&=({\bf 1}_N^T(\Db+\Wb) {\bf 1}_N) \|\nub^{(k)}_i - \nub^{(k-1)}_i\|^2_2 \notag
\\
&=({\bf 1}_N^T(2\Db-\Lb) {\bf 1}_N) \|\nub^{(k)}_i - \nub^{(k-1)}_i\|^2_2
\notag \\
&\textstyle =(2\sum_{j=1}^N|\Nc_j|)\|\nub^{(k)}_i - \nub^{(k-1)}_i\|^2_2,
\end{align}
which, when combined with \eqref{eq: proof of thm 4 eq7.5}, further implies that
\begin{align}\label{eq: proof of thm 4 eq10}
 \nub^{(k)}_i - \nub^{(k-1)}_i \to \zerob~\forall i\in V.
\end{align}
By applying \eqref{eq: proof of thm 4 eq10} to \eqref{eq: proof of thm 4 eq01}, one obtains
\begin{align}\label{eq: proof of thm 4 eq11}
  \zerob & \textstyle =-\Eb_i\hat\xb_i+ \frac{1}{N}\qb+
  \sum_{j\in \mathcal{N}_i} \hat\ub_{ij} + \sum_{j\in \mathcal{N}_i} \hat\vb_{ji}   \\
  &\textstyle=\partial \varphi_i(\hat \nub_i) + \frac{1}{N}\qb+
  \sum_{j\in \mathcal{N}_i} \hat\ub_{ij} +\sum_{j\in \mathcal{N}_i} \hat\vb_{ji},\label{eq: proof of thm 4 eq12}
\end{align}
where $\partial \varphi_i(\hat \nub_i)=-\Eb_i\hat\xb_i$ since \eqref{eq: proof of thm 4 eq8} implies that $\hat\xb_i$ is a maximizer to
\eqref{eq: varphi} with $\nub=\hat\nub_i$ \cite{Boydsubgradient}.
Finally, by summing \eqref{eq: proof of thm 4 eq11} for $i=1,\ldots,N$, followed by applying \eqref{eq: proof 7.6} and \eqref{eq: proof 11.799}, one obtains
\begin{align}\label{eq: proof of thm 4 eq13}
  \textstyle \sum_{i=1}^N\Eb_i\hat\xb_i=\qb.
\end{align}

The results in \eqref{eq: proof of thm 4 eq8}, \eqref{eq: proof 11.799}, \eqref{eq: proof of thm 4 eq9}, \eqref{eq: proof of thm 4 eq12} and \eqref{eq: proof of thm 4 eq13} imply that $\hat \xb$ and $\hat \nub $ are in fact a pair of optimal primal and dual solutions to {\sf (P2)}, and $\tilde {\hat \nub}$ and  $\{\hat \ub_{ij},\hat \vb_{ij}\}$ are a pair of
optimal primal and dual solutions to problem \eqref{consensus problem equi dual} [see \eqref{eq: KKT00} to \eqref{eq: KKT44}].
Thus, {\blue according to {\sf (R1)}, the sequence $\|\xb^{(k)}- \hat \xb\|_{\Mb+\frac{1}{2}\Pb}^2+ \frac{1}{c}\|\ub^{(k+1)} - \hat \ub\|^2_2
 + \frac{c}{2}\|\nub^{(k)} - \tilde {\hat \nub}\|^2_{\Qb}$ in fact }converges to zero and thereby
$\xb^{(k)} \to \hat \xb$, $\ub^{(k+1)} \to \hat \ub$ and $\nub_i^{(k)} \to \hat \nub~\forall i\in V.$}
\hfill $\blacksquare$

{\bf Proof of Theorem \ref{thm: conv of inexact dadmm}(b):} We assume that $\phi_i(\xb_i)=f_i(\Ab_i\xb_i)$, $\Ab_i$ has full column rank and  $\Eb_i$ has full row rank, for all $i\in V$. Denote $\rb^{(k)}\triangleq \|\xb^{(k)}- \xb^\star\|_{\alpha\Mb+\frac{1}{2}\Pb}^2+ \frac{1}{c}\|\ub^{(k+1)} - \ub^\star\|^2_2
 + \frac{c}{2}\|\nub^{(k)} - \tilde \nub^\star\|^2_{\Qb}$ for some $\alpha>0$. One can write
\eqref{eq: proof of thm 4 eq6} as follows
\begin{align*}
 &\rb^{(k)}+\|\xb^{(k)}- \xb^\star\|_{(1-\alpha)\Mb}^2 + \|\xb^{(k-1)}- \xb^\star\|_{\alpha\Mb}^2
 \notag \\
 &\leq \rb^{(k-1)}
  -\frac{1}{2}\|\xb^{(k)}-\xb^{(k-1)}\|_{\Pb - {\frac{1}{2}\tilde \Ab^T \Db_{L_f}\Db_{\rho}^{-1} \tilde \Ab}}^2\notag \\
  &~~~- \frac{1}{c}\|\ub^{(k+1)} - \ub^{(k)}\|^2_2-\frac{c}{2}\|\nub^{(k)} - \nub^{(k-1)}\|^2_{\Qb}.
\end{align*}
Therefore, it suffices to show that, for some $\delta>0$,
\begin{align}\label{eq: proof of thm 4b eq2}
\small  &\|\xb^{(k)}- \xb^\star\|_{(1-\alpha)\Mb}^2 + \|\xb^{(k-1)}- \xb^\star\|_{\alpha\Mb}^2
  \notag \\
  &+\frac{1}{2}\|\xb^{(k)}-\xb^{(k-1)}\|_{\Pb - {\frac{1}{2}\tilde \Ab^T \Db_{L_f}\Db_{\rho}^{-1} \tilde \Ab}}^2+ \frac{1}{c}\|\ub^{(k+1)} - \ub^{(k)}\|^2_2 \notag \\
  &+\frac{c}{2}\|\nub^{(k)} - \nub^{(k-1)}\|^2_{\Qb}
  \geq \delta \rb^{(k)}.
\end{align}
Firstly, from \eqref{eq: proof of thm 4 eq2.5} and \eqref{eq: proof of thm 4 eq4}, we have that (without $g_i$'s)
\begin{align}\label{eq: proof of thm 4b eq3}
  \Ab_i^T(\nabla f_i(\Ab_i\xb_i^{(k-1)}) &-\nabla f_i(\Ab_i\xb_i^\star)) + \Eb_i^T(\nub_i^{(k)} - \nub^\star) \notag \\
  &+\Pb_i(\xb_i^{(k)}-\xb_i^{(k-1)}) =\zerob.
\end{align}
By applying \eqref{eq: inequality } to \eqref{eq: proof of thm 4b eq3}, we have, for some $\mu_1>1$,
\begin{align}\label{eq: proof of thm 4b eq4}
  &\|\Pb_i(\xb_i^{(k)}-\xb_i^{(k-1)})\|^2
  \notag \\
  &\geq \!(1-\mu_1)\|\Ab_i^T(\nabla f_i(\Ab_i\xb_i^{(k-1)})\! -\!\nabla f_i(\Ab_i\xb_i^\star))\|^2\!\notag \\
  &~~~~~~~~~~~~~~+\!(1\!-\!\frac{1}{\mu_1})\|\Eb_i^T(\nub_i^{(k)}\! - \! \nub^\star)\|^2_2 \notag \\
  &\geq (1-\mu_1)L_{f,i}\lambda_{\max}^2(\Ab_i^T\Ab_i)\|\xb_i^{(k-1)} -\xb_i^\star)\|^2_2
  \notag \\
  &~~~~~~~~~~~~~~+(1-\frac{1}{\mu_1})\lambda_{\min}(\Eb_i\Eb_i^T)\|\nub_i^{(k)} - \nub^\star\|^2_2,
\end{align} where the second inequality is obtained by \eqref{eq: lipschitz gradient of f}.
Note that $\Db+\Wb=2\Db-\Lb \preceq 2\Db$ as $\Lb\succeq \zerob$ \cite{BK:Chung96}. Hence, we have
\begin{align}\label{eq: proof of thm 4b eq5}
\!\!\!  &\frac{c\delta}{2}\|\nub^{(k)} - \tilde \nub^\star\|^2_{\Qb}
  \leq {c\delta}\|\nub^{(k)} - \tilde \nub^\star\|^2_{\Db\otimes \Ib_M}\notag \\
  &\leq {c\delta} \tau_1
  \|(\xb^{(k)}-\xb^{(k-1)})\|^2_{\Pb^T\Pb}
  +{c\delta} \tau_2
  \|\xb^{(k-1)} -\xb^\star)\|^2_2,
\end{align}
where the second inequality is due to \eqref{eq: proof of thm 4b eq4},
 $\tau_1=\max_{i\in V}\big\{\frac{|\Nc_i|}{(1-\frac{1}{\mu_1})\lambda_{\min}(\Eb_i\Eb_i^T)}\big\}>0$ and
$\tau_2=\max_{i\in V}\big\{\frac{(\mu_1-1)\lambda_{\max}^2(\Ab_i^T\Ab_i)|\Nc_i|L_{f,i}}{(1-\frac{1}{\mu_1})\lambda_{\min}(\Eb_i\Eb_i^T)}\big\}>0$ are finite given that
$\Eb_i$'s have full row rank.

Secondly, upon stacking \eqref{eq: proof of thm 4 eq1} for all $i\in V$ and applying \eqref{eq: KKT3} and \eqref{eq: proof 7.5}, one obtains
\begin{align}\label{eq: proof of thm 4b eq6}
  \Upsilonb(\ub^{(k+1)}-\ub^\star) &+ c \Qb(\nub^{(k)}-\nub^{(k-1)}) \notag \\
   &-\Eb(\xb^{(k)}-\xb^\star)=\zerob,
\end{align}
where $\Eb=\blkdiag\{\Eb_1,\ldots,\Eb_N\}$ and $\Upsilonb$ is given in \eqref{eq: Psib}.
Analogously, by applying \eqref{eq: inequality } to
\eqref{eq: proof of thm 4b eq6} and by \eqref{eq: proof 17}, one can show that, for some  $\mu_2>1$,
\begin{align}\label{eq: proof of thm 4b eq7}
  \frac{\delta}{c}\|\ub^{(k+1)}-\ub^\star\|^2_2 &\leq \frac{\delta}{c\tau_3}\|\xb^{(k)}-\xb^\star\|^2_{\Eb^T\Eb}
  \notag \\
  &+ \frac{\delta(\mu_2-1)c}{\tau_3}\|\nub^{(k)}-\nub^{(k-1)}\|^2_{\Qb},
\end{align}
where
$\tau_3={(1-\frac{1}{\mu_2})\sigma_{\min}^2(\Upsilonb)}>0$.
By \eqref{eq: proof of thm 4b eq5} and \eqref{eq: proof of thm 4b eq7}, sufficient conditions for satisfying \eqref{eq: proof of thm 4b eq2}
are therefore given by: $\forall i\in V,$
\vspace{-0.2cm}
{\small \begin{subequations}\label{eq: proof of thm 4b eq8}
\begin{align}
\!\!\!  (1-\alpha-\delta\alpha)(\sigma_{f,i}^2-\frac{\rho_i}{2})\Ab_i^T\Ab_i   &\succeq \frac{\delta}{2}\Pb_i +
  \frac{\delta}{c\tau_3}\Ab_i^T\Ab_i, \\
  \alpha(\sigma_{f,i}^2-\frac{\rho_i}{2})\Ab_i^T\Ab_i &\succeq c\delta \tau_2 \Ib_{K}, \\
  \frac{1}{2}\Pb_i - \frac{L_{f,i}^2}{2\rho_i}\Ab_i^T\Ab_i &\succeq c\delta \tau_1 \Pb_i^T\Pb_i, \\
  \frac{1}{2} &\geq \frac{\delta(\mu_2-1)}{\tau_3}.
\end{align}
\end{subequations}}\hspace{-0.1cm}
Under \eqref{eq: proof of thm 4 eq6.5} and full column rank $\Ab_i$'s, we see that \eqref{eq: proof of thm 4b eq8} {\blue is true} for some $\delta>0$.
The proof is complete. \hfill $\blacksquare$
\vspace{-0.1cm}
\footnotesize
\bibliography{distributed_opt,smart_grid}
\end{document}